\documentclass[usenatbib]{mn2e}
\usepackage{graphicx}
\usepackage{epstopdf}
\usepackage{natbib}
\usepackage{verbatim}
\usepackage{amssymb,amsmath}
\usepackage{ulem}
\usepackage[bookmarks,bookmarksnumbered,colorlinks=true, citecolor=blue, linkcolor=black, draft=false]{hyperref}

\usepackage{color}
\usepackage[mathscr]{eucal}

\newcommand\aj{AJ}
\newcommand\apj{ApJ}
\newcommand\apjl{ApJ}
\newcommand\aap{A\&A}
\newcommand\mnras{MNRAS}
\newcommand\pasp{PASP}
\newcommand\nat{Nature}
\newcommand\fcp{Fund.~Cosmic~Phys.}

\title[IllustrisTNG MZR]{The evolution of the mass-metallicity relation in IllustrisTNG}

\author[P. Torrey et al.]
       {\parbox{18cm}{Paul~Torrey$^{1}$\thanks{ptorrey@mit.edu}\thanks{Hubble Fellow},
       Mark Vogelsberger$^{1}$\thanks{Alfred P. Sloan Fellow},
       Federico Marinacci$^{1}$,
       R\"udiger Pakmor$^{2}$, \\
       Volker Springel$^{2,3,4}$,
       Dylan Nelson$^{4}$,
       Jill Naiman$^{5}$,
       Annalisa Pillepich$^{6}$,
       Shy Genel$^{7,8}$,
       Rainer Weinberger$^{2}$, 
       Lars Hernquist$^{5}$
       }\vspace{0.3cm}\\ 
         $^1$ MIT Kavli Institute for Astrophysics \& Space Research, Cambridge, MA, 02139, USA\\
         $^2${Heidelberg Institute for Theoretical Studies, Schloss-Wolfsbrunnenweg 35, D-69118 Heidelberg, Germany}\\
         $^3${Zentrum f{\"u}r Astronomie der Universit{\"a}t Heidelberg, ARI, M{\"o}nchhofstr. 12-14, D-69120 Heidelberg, Germany}\\
	$^4${Max-Planck-Institut f{\"u}r Astrophysik, Karl-Schwarzschild-Str. 1, 85741 Garching, Germany}\\
         $^5${Harvard--Smithsonian Center for Astrophysics, 60 Garden Street, Cambridge, MA 02138}\\
	$^6${Max-Planck-Institut f{\"u}r Astronomie, K{\"o}nigstuhl 17, 69117 Heidelberg, Germany}\\
	$^{7}$Center for Computational Astrophysics, Flatiron Institute, 162 Fifth Avenue, New York, NY 10010, USA\\
	$^{8}$Columbia Astrophysics Laboratory, Columbia University, 550 West 120th Street, New York, NY 10027, USA\\
         }

\begin{document}

\maketitle

\begin{abstract}
The coevolution of galaxies and their metal content serves as an important test for galaxy feedback models.
We analyze the distribution and evolution of metals within the IllustrisTNG simulation suite with a focus on the gas-phase mass-metallicity relation (MZR).
We find that the IllustrisTNG model broadly reproduces the slope and normalization evolution of the MZR across the redshift range $0<z<2$ and mass range $10^9 < M_*/\mathrm{M}_\odot < 10^{10.5}$.
We make predictions for the high redshift ($2<z<10$) metal content of galaxies which is described by a gradual decline in the normalization of the metallicity with an average high redshift ($z>2$) evolution fit by $\mathrm{d\;log(Z)}/\mathrm{dz} \approx - 0.064$.
Our simulations indicate that the metal retention efficiency of the interstellar medium (ISM) is low:  a majority of gas-phase metals ($\sim$ 85 per cent at $z=0$) live outside of the ISM, either in an extended gas disk, the circumgalactic medium, or outside the halo.
Nevertheless, the redshift evolution in the simulated MZR normalization is driven by the higher gas fractions of high redshift galaxies, not by changes to the metal retention efficiency.
The scatter in the simulated MZR contains a clear correlation with the gas-mass or star formation rate of the system, in agreement with the observed fundamental metallicity relation.
The scatter in the MZR is driven by a competition between periods of enrichment- and accretion-dominated metallicity evolution.
We expect that while the normalization of the MZR declines with redshift, the strength of the 
correlation between metallicity and gas-mass at fixed stellar mass
is not a strong function of redshift.
Our results indicate that the ``regulator" style models are best suited for simultaneously explaining the shape, redshift evolution, and existence of correlated scatter with gas fraction about the MZR.
\end{abstract}


\renewcommand{\thefootnote}{\fnsymbol{footnote}}

\section{Introduction}

Aging stellar populations within galaxies synthesize new heavy elements -- or metals -- which are redistributed within the interstellar medium (ISM).
These metals enrich the ISM and are redistributed via gas motion which imprints implicit information about each galaxy's star formation history and baryon cycle.
The distribution of metals within galaxies plays an important role in constraining our understanding of 
when galaxies acquire their fuel for star formation, 
where stars are formed, 
how the ejecta from the aging stellar populations is returned to the ISM, 
and the role that outflows may have in removing both mass and metals from the galaxy.

Perhaps the most fundamental and early recognized relationship between galaxies and their metal content is the stellar-mass versus gas-phase metallicity relationship~\citep[hereafter, the  mass metallicity relationship, or MZR;][]{Lequeux1979, Tremonti2004, KewleyEllison}.
The MZR forms a tight correlation (i.e. $\sim0.1$ dex) that has been observed over several orders of magnitude in stellar mass, over an order of magnitude in metallicity, and across a wide redshift range out to $z\sim2$.
The MZR shows that metallicity generally correlates with galaxy stellar mass such that low-mass galaxies tend to have lower metallicities then their higher mass counterparts.
The MZR asymptotes around $M_*\approx10^{10.5}\mathrm{M}_\odot$~\citep[e.g.,][]{Tremonti2004} yielding a nearly flat relationship between mass and metallicity for higher mass systems.
The normalization of the MZR is observed to evolve slightly with time, with higher redshift galaxies having somewhat lower metallicities than their lower redshift counterparts~\citep[e.g.,][]{Savaglio2005, Erb2006, Maiolino2008, LaraLopez2010, Zahid2014}.

What remains missing is a clear physical understanding of why the metal content of galaxies tracks the stellar mass of galaxies as observed.
Understanding the shape and evolution of the MZR alone has proven to be a challenging task.
The first analytic, one-zone, simple chemical-evolution model that was developed to shed light on the driving forces behind the MZR and its evolution was the closed box model~\citep[][]{Searle1972, Tinsley1980}.
This model was made analytically tractable by assuming that 
(i) the total baryon content of the system is fixed (hence the `closed box' title) and 
(ii) the evolution of the gas-mass, stellar mass, gas-phase metal-mass, and stellar-phase metal-mass is completely determined by star formation and the associated metal yield.
Unfortunately, while the closed box model gives physical insight and yields clear predictions for the coevolution of galactic metals and galactic gas/stellar mass, it has now been widely established that the observed MZR cannot be directly explained with the fiducial closed box model alone~\citep[e.g.,][]{Chiappini1997, Tremonti2004}.
Numerous attempts have been made to extend and improve the closed box model's ability to explain the MZR by accounting for inflows of pristine gas~\citep[e.g., going back to][]{Larson1972} or outflows of metal enriched gas~\citep[e.g.,][]{Larson1974}.
Similar incarnations of the analytic gas and metal evolution equations first laid out in~\citet{Larson1972} and~\citet{Larson1974} continue to feature prominently in the literature as possible explanations for the shape, normalization, and evolution of the MZR~\citep[e.g.,][]{Dalcanton2007, Finlator2008, Lilly2013, Zahid2014}.

Alternative modern interpretations of the MZR cast galaxies as more dynamic gas and metal reservoirs, subject to (and possibly dominated by) pristine gas inflows and enriched gas outflows~\citep[as in, e.g.,][]{Finlator2008, Dave2012}.
In stark contrast to the closed box type models where galactic metallicity is a reflection of the integrated formation history of the galaxy, \citet{Finlator2008} argued that the impact of inflows, outflows, and mixing implies that galactic metallicity is more likely set by the \textit{current} galaxy properties.
Specifically,  the \citet{Finlator2008} model defines an equilibrium metallicity as $Z_{\mathrm{eq}} = \dot M_{\mathrm{Z}}/\dot M_{\mathrm{gas}}$ which is set by the {\it current rate} of new metal production and pristine gas inflow.
\citet{Finlator2008} and later \cite{Dave2012} showed that this reasonably simple and intuitive model is able to explain a number of features of the MZR including the slope of the low-mass end of the MZR, the turnover at higher masses, and the modified slope/normalization found in simulations with varied feedback efficiencies.

These analytic models act as broad tools that can be used to understand the average relationship between the gas-mass growth of a galaxy, production of stars and associated metal synthesis, metal retention, and ultimately galaxy metallicity evolution.
However, these same models do not accommodate the varied formation tracks that real galaxies follow.
As a result, the previously described simple analytic models do not shed light on the scatter in the MZR, and while these models do seem to describe the shape and normalization evolution of the MZR, they do not necessarily give insight into the actual evolutionary tracks along which individual galaxies evolve.

Addressing the scatter in the MZR is important because it has been argued that a galaxy's position with respect to the MZR correlate with secondary galactic parameters, including the star formation rate~\citep{Ellison_FMR}.
The correlations between the scatter in the MZR and star formation rate have been cast as a fundamental metallicity relationship that exists not just between metallicity and stellar mass, but also with star formation rate~\citep[e.g.,][]{LaraLopez2010, Mannucci_FMR} or gas-mass~\citep{Bothwell2013}.
The existence of this correlated scatter about the MZR likely indicates a non-separarable link between stellar mass, metallicity, and gas-mass.
However, the strength of the correlation between gas-mass or star formation rate and offset from the MZR at a given stellar mass is not constant, and has been shown to depend on galaxy mass~\citep{Bothwell2013}.

A different class of models have attempted to address the existence of correlated scatter about the MZR by allowing for deviations from the MZR~\citep[e.g.,][]{Lilly2013, Forbes2014}.
The key feature of these ``regulator" models is that the gas-mass in a system is allowed to vary, meaning that the metallicity of a system is dependent either explicitly on the current gas-mass~\citep{Lilly2013} or implicitly dependent on the current gas-mass via the recent accretion history of the system~\citep{Forbes2014}.
These ``regulator" models make concrete predictions for the existence of the correlated scatter about the MZR.
Specifically, the \citet{Forbes2014} model predicts a correlation between galactic metallicity and star formation rate at a fixed stellar mass, where the strength of the correlation depends on the relative timescale over which the accretion rates remain coherent against the timescale over which a galaxy loses mass.
The  \citet{Forbes2014} model therefore predicts weaker correlated scatter about the MZR for higher mass systems, as is observed~\citep[e.g.,][]{Bothwell2013}.

Despite their ability to match the observed nature of the correlated scatter in the MZR, the ``regulator" models have not been rigorously tested or validated.
Numerical simulations can be leveraged to not only examine the emergence of the MZR~\citep[e.g.,][]{Dave2012, Torrey2014, DeRossi2017, Dave2017}, but also inspect how individual galaxies evolve and to discriminate between the varied simple analytic models.
Even with dozens of numerical simulation papers focused on the origin of the MZR and the correlated scatter about the MZR, comparatively little attention has been paid to individual galaxy evolution and scrutinizing the extent to which regulator models capture the physical processes that drive galaxies in their stellar mass, gas-mass, and metallicity evolution.  
Addressing this point seems particularly important, given that, e.g., even the best fit \citet{Forbes2014} regulator model requires scatter in accretion rates that is smaller than what is expected from N-body simulations.

In this paper, we explore the nature of the MZR as formed in the IllustrisTNG simulations.
We use the large simulated galaxy population to compare the MZR against observations and quantify the existence of correlated scatter.
Using the high-time-frequency snapshot output of the sub-boxes within the full IllustrisTNG volume, we then consider the detailed time evolution of galaxies to identify the physical processes  that drive the gas-mass, stellar mass, and metallicity evolution of the simulated galaxies.
We compare our results with analytic MZR models in an attempt to validate the existing equilibrium and regulator model frameworks to better describe the MZR evolution seen in IllustrisTNG.

The structure of this paper is as follows:
In Section~\ref{sec:Methods} we describe the numerical simulations used in this paper (the IllustrisTNG simulation) with a focus on the metal enrichment and evolution methods.
In Section~\ref{sec:Results1} we present a global overview of the metal distribution within the IllustrisTNG simulation, including a breakdown of the metal retention efficiencies in various phases.
In Section~\ref{sec:Results2} we present the MZR across the redshift range $0 \leq z \leq 6$ along with a detailed description of the relationship between metallicity and galactic gas-mass.
In Section~\ref{sec:Results3} we consider the time evolution of individual galaxies to quantify the duration of galaxy deviations from the MZR and to address the physical processes that drive galaxies off the MZR.
In Section~\ref{sec:Discussion} we give a discussion focused on the implications of our results for interpreting observations and understanding how our results compare with previous models.
In Section~\ref{sec:Conclusions} we summarize and conclude.

\section{Methods}
\label{sec:Methods}
The analysis presented in this paper is based on the IllustrisTNG simulation suite~\citep{Marinacci2017, Pillepich2017b, Naiman2017, Springel2017, Nelson2017}.
The IllustrisTNG project is an extension of the Illustris simulation project~\citep{vogelsberger2014a, vogelsberger2014b, genel2014, sijacki2015}, which includes a number of targeted improvements to the included galaxy formation model.
The physical model employed in IllustrisTNG simulations builds on the original Illustris model~\citep{Vogelsberger2013, Torrey2014} with important updates and modifications described in~\citet{Weinberger2017} and~\citet{Pillepich2017}.
The IllustrisTNG simulation suite consists of three simulation volumes: TNG50, TNG100, and TNG300.
The IllustrisTNG simulations have been used to study the size evolution of galaxies~\citep{Genel2017}, 
the enrichment of the intercluster medium~\citep{Vogelsberger2017}, and 
the impact of AGN feedback on galaxy quenching~\citep{Weinberger2017b}.
For the analysis presented in this paper, we use the TNG100 simulation which is an analog to the Illustris-1 simulation, with the updated physical model and cosmology consistent with the 2015 Plank collaboration results~\citep{Plank2015}.
We refer interested readers to the introductory papers for further details beyond the brief description provided here.

\subsection{Metal Enrichment}
Here we briefly summarize the metal enrichment procedures that are employed in the IllustrisTNG model.
Stars are formed from dense gas ($n_\mathrm{H}\gtrsim0.13\;\mathrm{cm^{-3}}$) using a star formation prescription that is designed to reproduce the Kennicutt-Schmidt relation~\citep{SH03}.
Simulation star particles record their birth time and are assigned the same metallicity as the ISM gas from which they were born.
Simulation star particles in the high resolution TNG100 simulation (which we study exclusively in this paper) have masses of $m_* \sim 10^6 \mathrm{M}_\odot$ and therefore represent an unresolved large population of real stars.
Each unresolved stellar population is assumed to be comprised of a~\citet{ChabrierIMF} initial mass function (IMF).

As star particles age, mass and metals from the aging stellar populations are returned to the ISM.  
At any timestep, stellar lifetime tables~\citep{Portinari1998} are used to determine which stars within the unresolved full stellar population are expected to be moving off the main sequence.
Mass return and metal yield tables for SNIa~\citep{Nomoto1997}, SNII~\citep{Portinari1998, Kobayashi2006}, and AGB stars~\citep{Karakas2010, Doherty2014, Fishlock2014} are then used to determine the amount of mass and metals that should be returned to the ISM.
The mass and metal return is carried out by finding 64 nearest gas cells and spreading the returned gas-mass and metal-mass among them.
This mass and metal return procedure leads to a time- and spatially-dependent enrichment of the ISM based on the distribution and formation history of the stellar population in any galaxy.
After metals are deposited in the ISM, they are passively advected with the fluid flow.

Galactic winds are driven in the IllustrisTNG model based on the local SFR~\citep[for full details, see][]{Pillepich2017}.
Importantly, however, the winds are assigned a metallicity which is lower than the ambient ISM (specifically, $Z_{\mathrm{wind}} = 0.4 Z_{\mathrm{ISM}}$).
Metal mass not launched in the wind is left in the ISM, such that metal mass is conserved.
The reduction in wind metallicity is motivated by the fact that winds are expected entrain a significant amount of material as they propagate away from their launch site.
Entrainment of low metallicity gas (as is likely to be the case for off-disk-plane material) will naturally dilute the wind metallicity.
Since the IllustrisTNG SF driven winds are hydrodynamically decoupled for a brief period, we adopt a lowered metallicity to account, in part, for any low metallicity gas entrainment which may have occurred.
The reduction in wind metallicity was tuned to encourage a better match to the MZR in the Illustris model~\citep{Vogelsberger2013, Torrey2014} and remains unchanged in the IllustrisTNG setup.

\begin{figure*}
\centerline{\vbox{\hbox{ 
\includegraphics[width=0.5\textwidth]{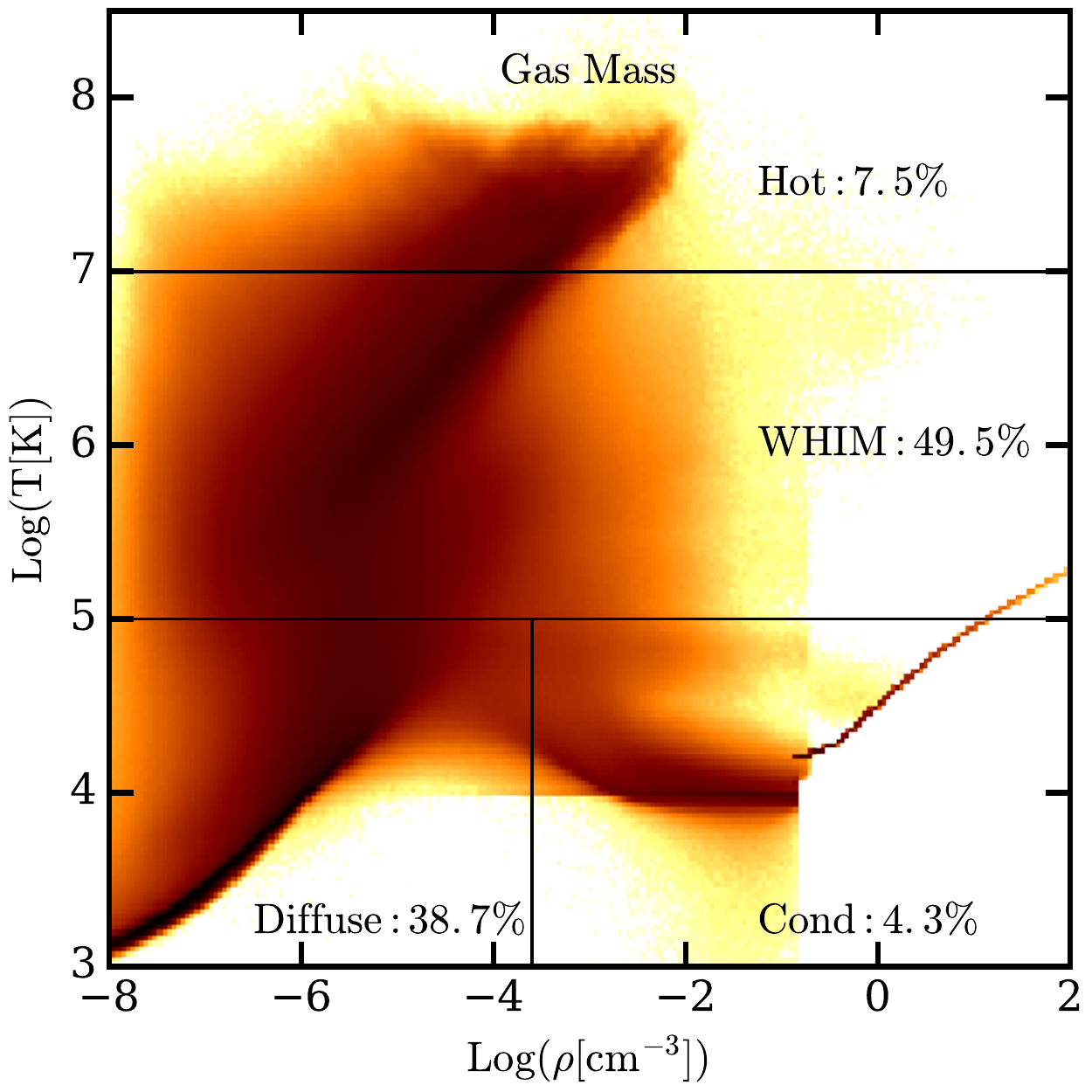}
\includegraphics[width=0.5\textwidth]{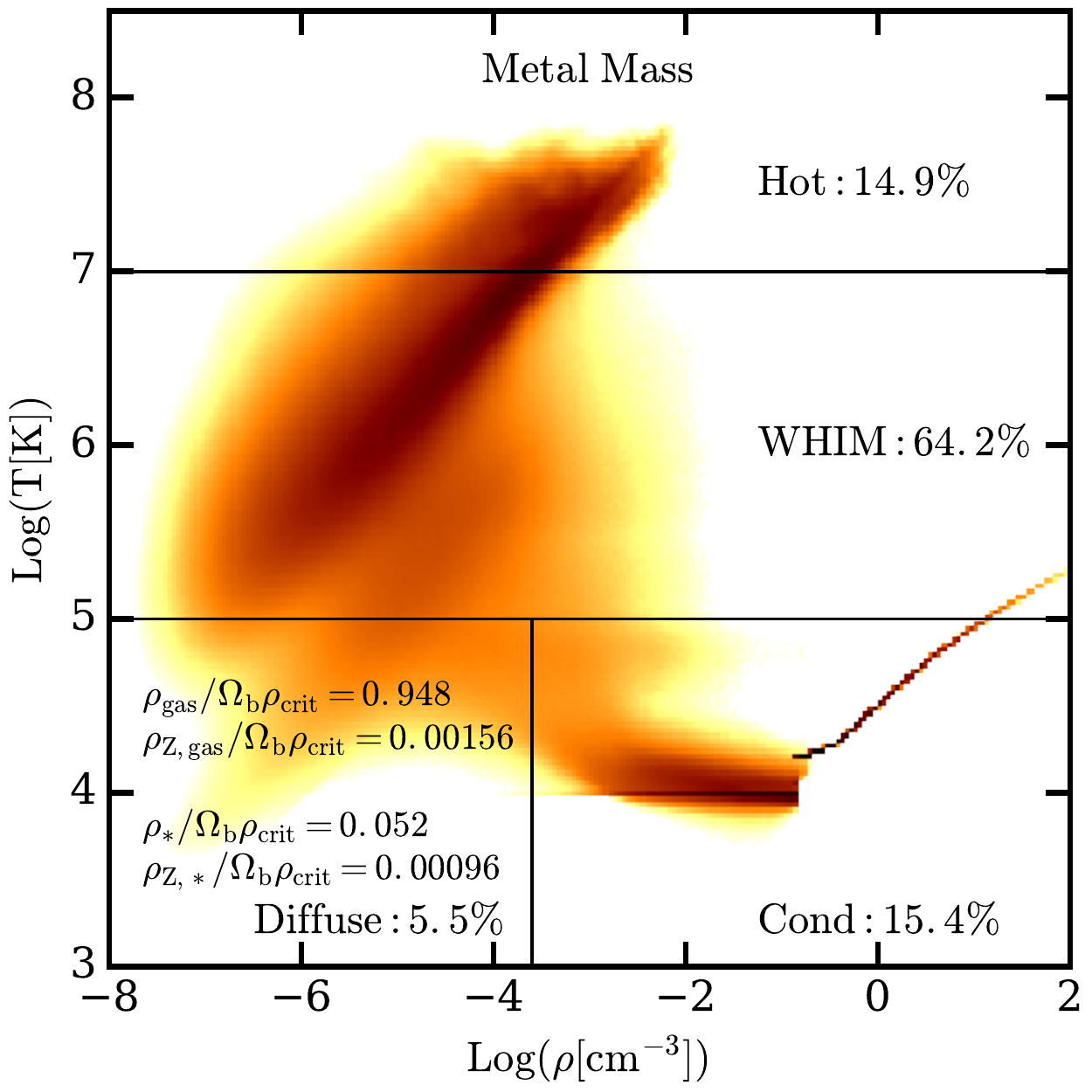}
 }}}
\caption{Phase diagrams showing the redshift $z=0$ global distribution of gas-mass (left) and gas-phase metal-mass (right) as a function of density and temperature.
Thin black lines indicate boundaries used to separate the hot, warm-hot intergalactic medium (WHIM), diffuse, and condensed material~\citep{Dave2001, Haider2016}.
The fractions of gas-mass (or gas-phase metal-mass) within each region is indicated within the plot.
There is a significantly higher relative fraction of metals in the condensed region of phase space, which is where star formation occurs and metals are produced.
However, there is also significant pollution of the WHIM and hot regions of phase space, with $\sim85\%$ of gas-phase metals being outside of the condensed region.
}
\label{fig:global_phase_diagrams}
\end{figure*}

\subsection{Definition of Metallicity and ISM mass}
In this paper we focus our attention on the evolution of galactic gas-phase metallicity, with a further emphasis on comparisons with observations of extragalactic nebular emission line based metallicity measurements.
To make simple comparisons with nebular emission line metallicity measurements, gas-phase metallicity values quoted in this paper are star formation rate weighted metallicity values.
Using star formation rate weighted metallicity measurements limits the sample of galaxies that have assigned metallicities to those with star forming gas.
Throughout this paper, we therefore require non-zero star formation rates in order to quote metallicity values.
In some portions of the paper (where explicitly noted) we require higher star formation threshold values be met in order to further remove nearly quenched galaxies that would likely not be observationally detectable.

We quote two distinct metal abundance values in this paper.  
When discussing ``metallicity", we adopt the scalar metallicity value from the simulation, $Z=M_{Z,\mathrm{gas}}/M_{\mathrm{gas}}$, which we define in the traditional way as the fraction of gas-mass that is composed of metals.

When making comparisons with nebular emission line determined metallicities (quoted as $\mathrm{Log(O/H)+12}$ values) we also use the global metallicity scalar, assuming that Oxygen is 35 per cent of the metal-mass.
In addition, we use SFR-weighted metallicities for each galaxy, in order to provide a more even handed comparison with nebular emission line measurements.
We do not place any emphasis on the absolute normalization of the simulated or observed MZR values since it is well established that nebular emission line metallicity diagnostics have factor of $>2$ uncertainty in their absolute normalization~\citep{KewleyEllison}.
Instead we focus our attention on the shape and normalization evolution of the MZR.

Additionally, in several places in this paper we discuss the co-evolution of galactic metallicity and ISM mass.
We adopt an ISM mass definition as being the total mass of gas over which the metallicity was calculated.
Since in this paper we use SFR-weighted metallicities, our adopted ISM mass definition is therefore the total mass of gas above our employed star formation density threshold of $n_{\mathrm{H}} \gtrsim 0.13 \mathrm{cm}^{-3}$.

\section{Results: Global Metal Distribution}
\label{sec:Results1}
In this section we explore the broad characteristics of the gas-mass and gas-phase metal-mass distribution in the TNG100 simulation.
We first consider global density-temperature phase diagrams of both the gas- and metal-mass distribution, and then consider the contributions to these global phase diagrams from galaxies of varied masses.
Finally, we show the evolution of the metal retention efficiencies for the ISM, circumgalactic medium (CGM), and stars.

\subsection{Global Gas-Mass and Metal-Mass Density-Temperature Phase Diagrams}
Figure~\ref{fig:global_phase_diagrams} shows a density-temperature phase diagram of the global distribution of gas-mass (left) and gas-phase metal-mass (right).
Thin black lines have been placed to mark boundaries between four loosely defined regions within this phase diagram: hot gas ($T>10^{7}\mathrm{K}$), warm-hot gas ($10^{5}\mathrm{K} <T<10^{7}\mathrm{K}$), diffuse material ($\rho <  1000 \rho_c \Omega_\mathrm{b}$; $T<10^{5}\mathrm{K}$), and condensed material ($\rho >  1000 \rho_c \Omega_\mathrm{b}$; $T<10^{5}\mathrm{K}$).
These are the same phase boundaries originally employed in~\citet{Dave2001} which were also used more recently to analyze the original Illustris simulation in~\citet{Haider2016}.
The fraction of gas-mass (left) or metal-mass (right) in each region is indicated within the figure.

\begin{figure*}
\centerline{\vbox{\hbox{ 
\includegraphics[width=0.333\textwidth]{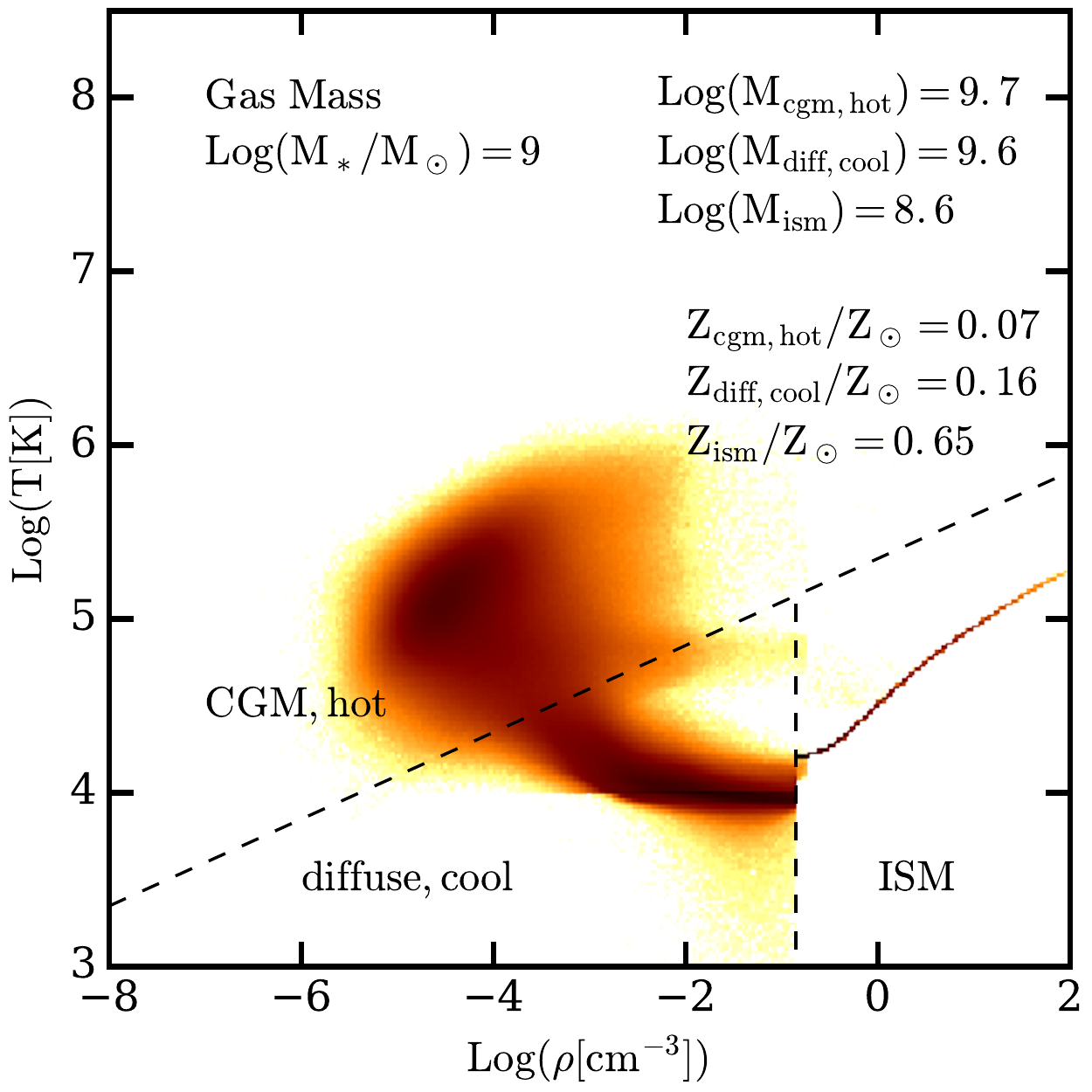}
\includegraphics[width=0.333\textwidth]{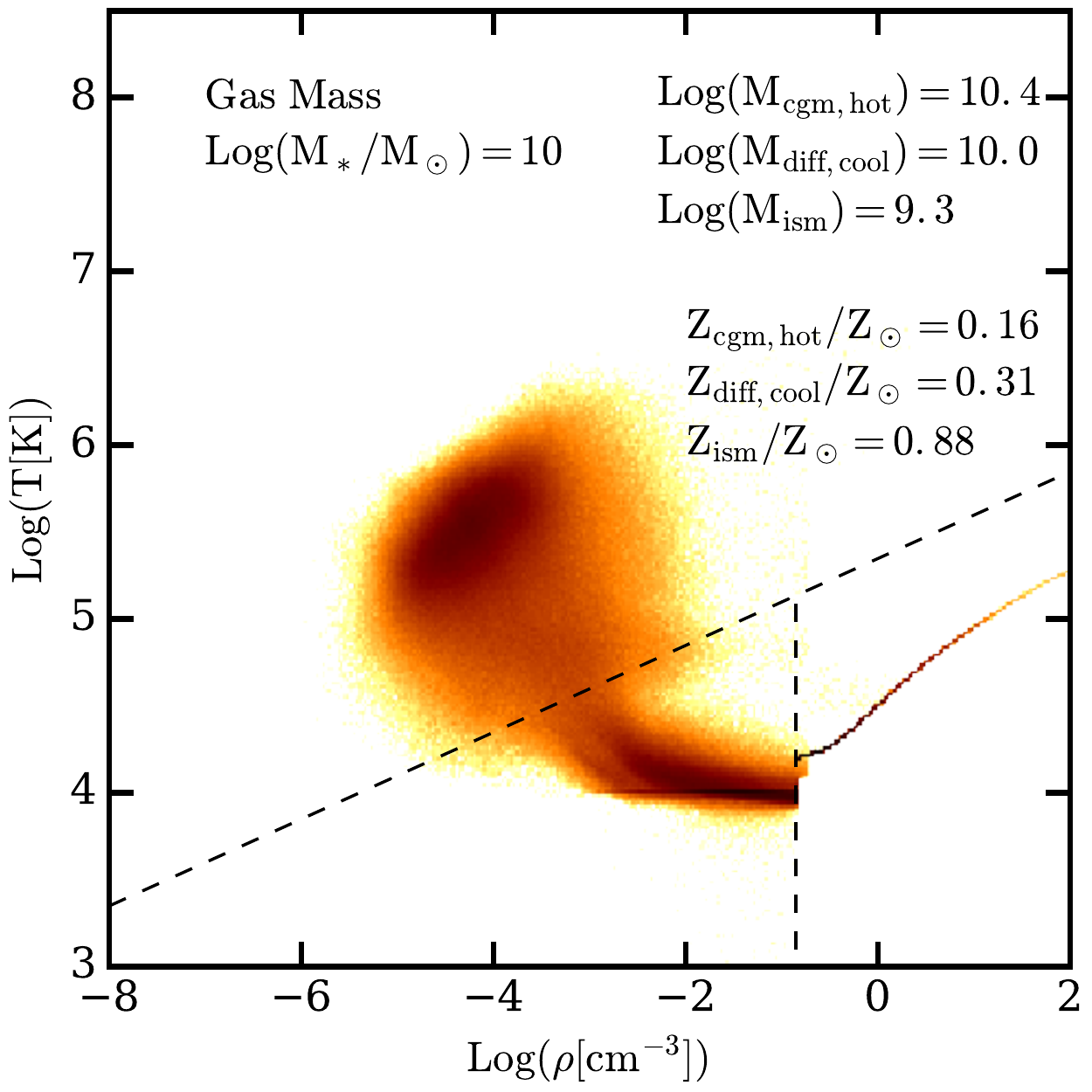}
\includegraphics[width=0.333\textwidth]{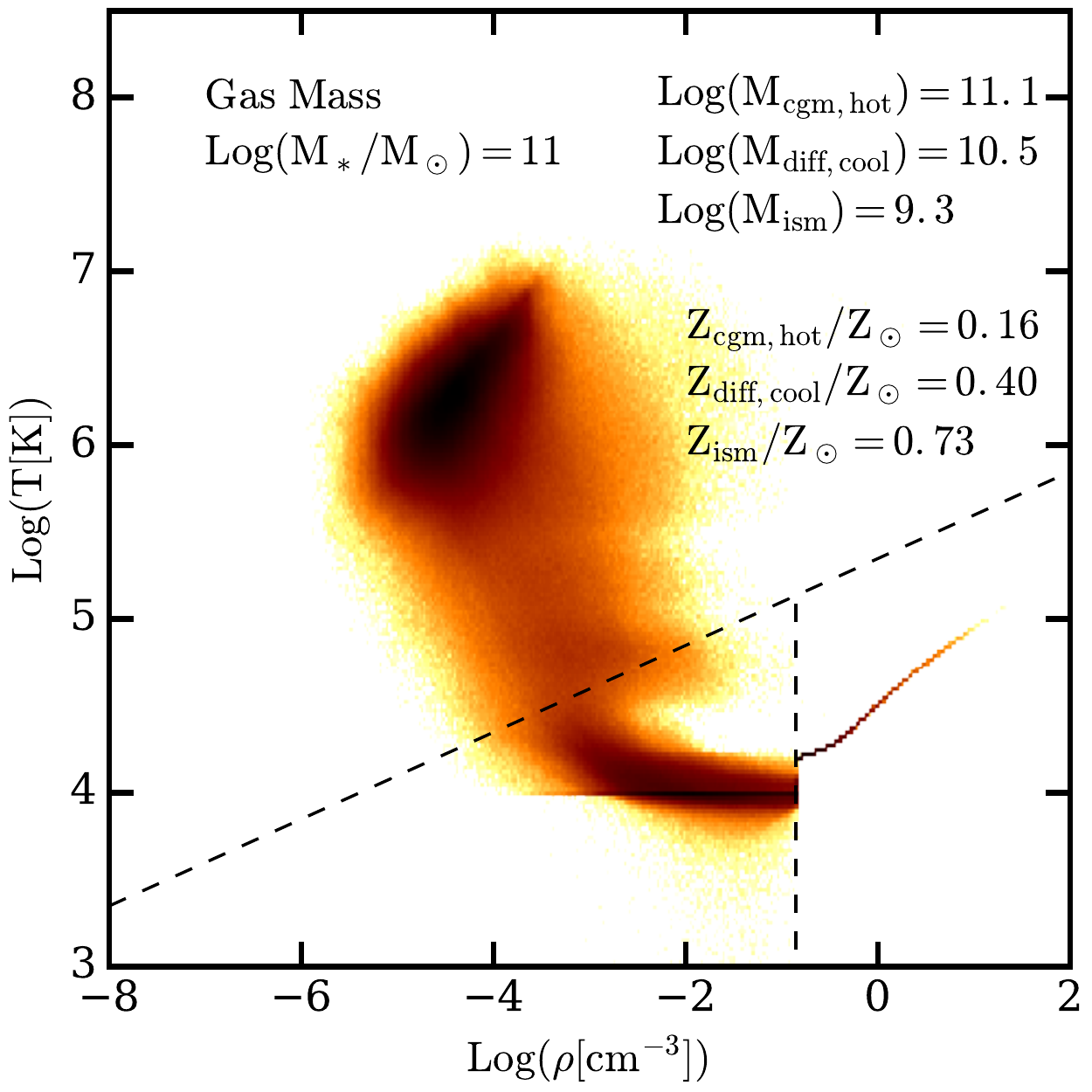}
 }}}
 \centerline{\vbox{\hbox{ 
\includegraphics[width=0.333\textwidth]{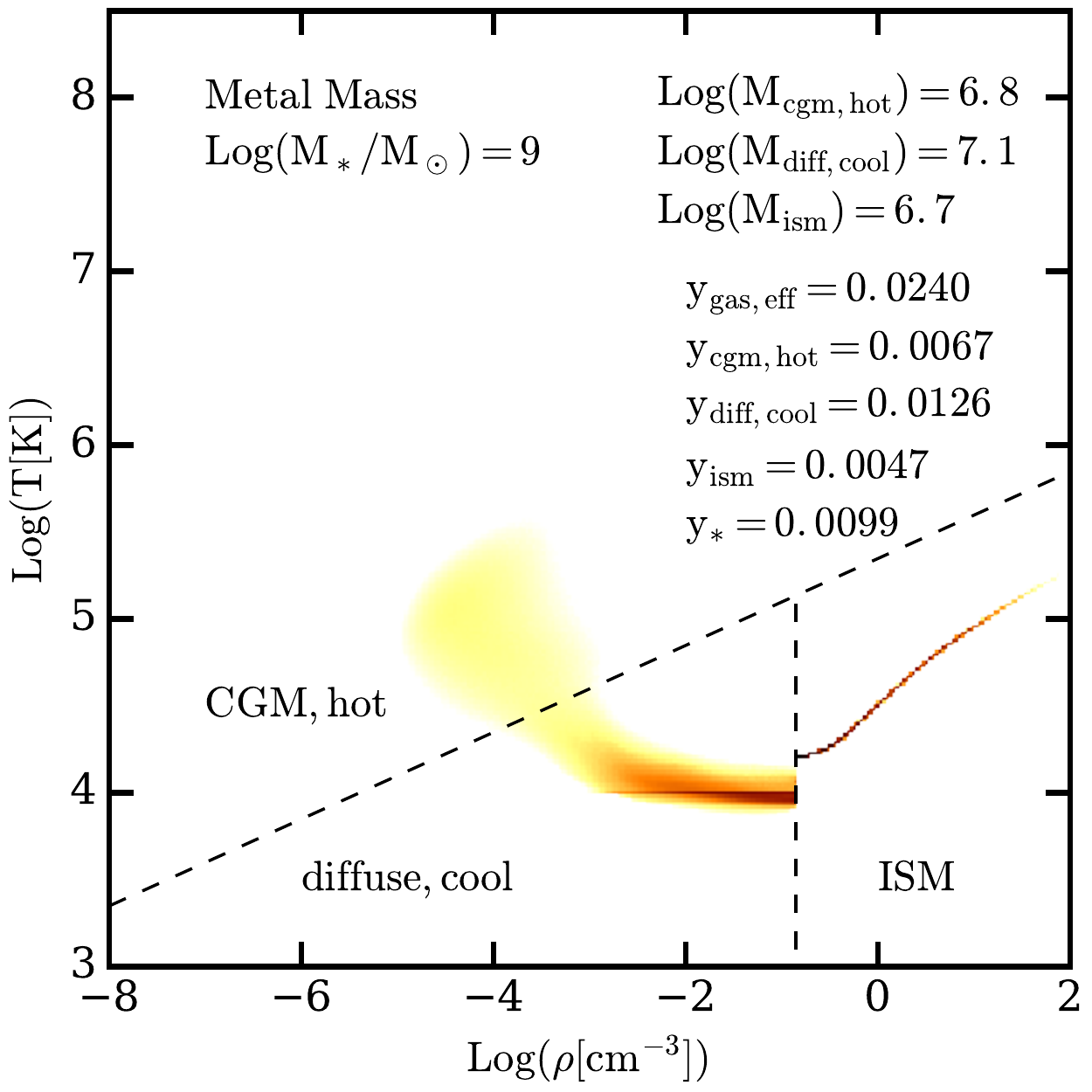}
\includegraphics[width=0.333\textwidth]{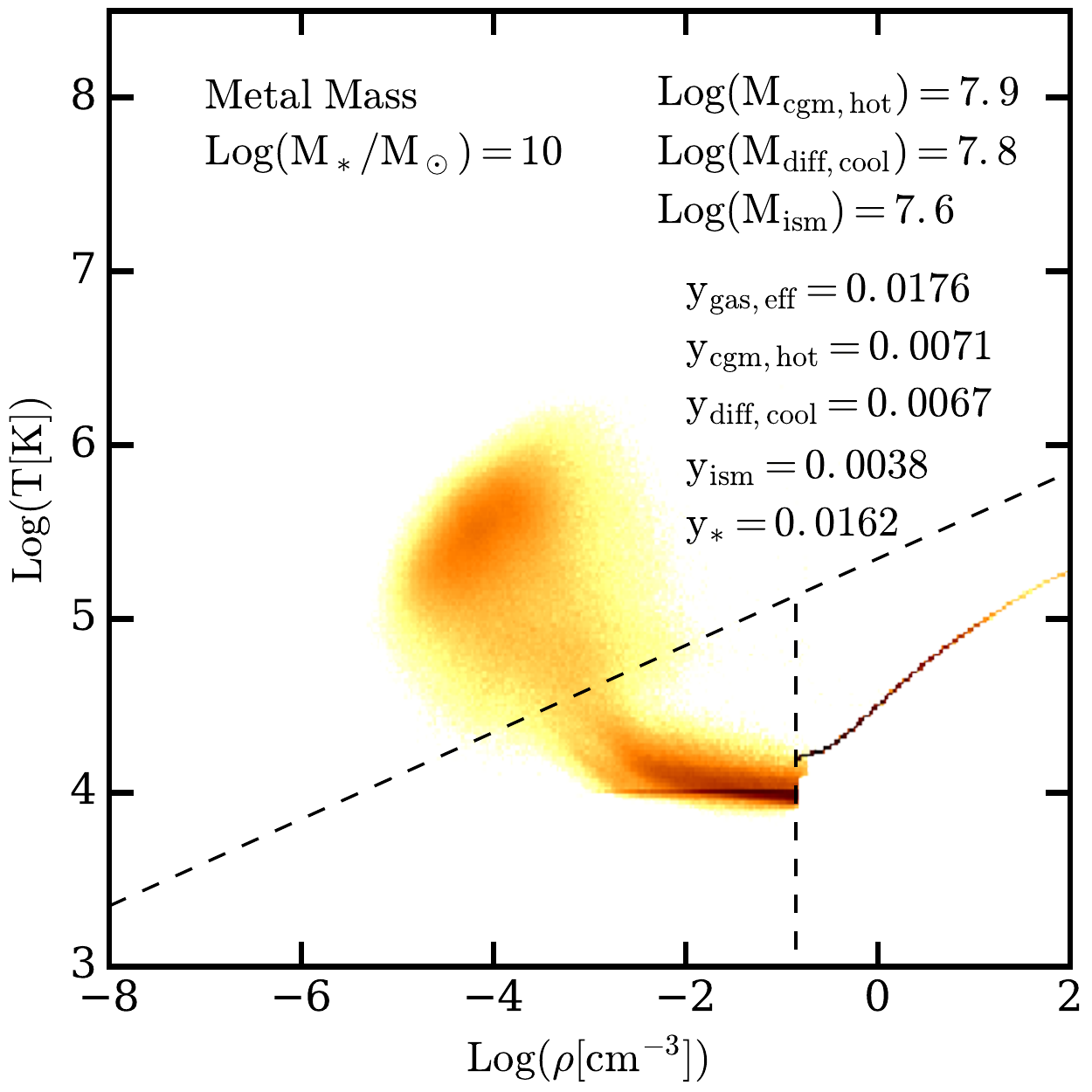}
\includegraphics[width=0.333\textwidth]{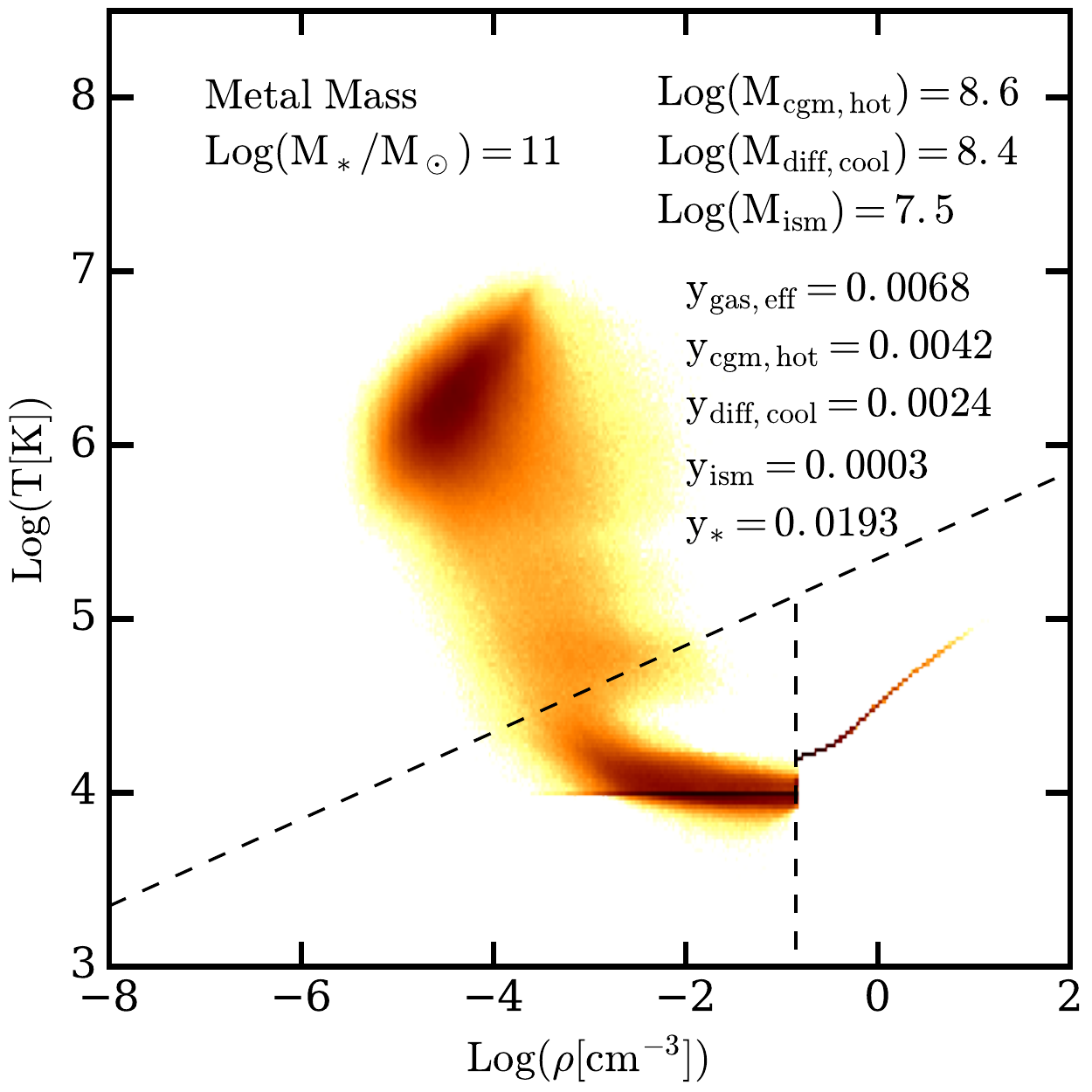}
 }}}
\caption{ Density-temperature phase diagrams indicating the distribution of gas-mass (top row) and gas-phase metal-mass (bottom row) for galaxies with stellar masses of $M_*\approx10^9 \mathrm{M}_\odot$ (left column), $M_*\approx10^{10} \mathrm{M}_\odot$ (center column), and $M_*\approx10^{11} \mathrm{M}_\odot$ (right column).
We split the gas distribution into three regions indicated by thin dashed lines:  the hot CGM material (top region), the dense/cool ISM gas (bottom right), and the diffuse/cool gas (bottom left).
The total gas-mass -- or gas-phase metal-mass -- in each region is indicated within each figure.
The average metallicity of gas in each region, or the effective yield for each region ($y_i = M_{Z,i} / M_*$; see main text for further details) is indicated within the plot.    }
\label{fig:galaxy_phase_diagrams}
\end{figure*}

\begin{table}
\begin{center}
\caption{Effective yields at $z=0$ for the defined gas phases in three mass bins.}
\label{table:effective_yields}
\begin{tabular}{ l c c c  }
\hline
		$M_* $  				& $ 10^9 \mathrm{M}_\odot$		& $10^{10} \mathrm{M}_\odot$	&  $ 10^{11} \mathrm{M}_\odot$	 	\\ 
\hline
\hline
$y_{\mathrm{eff,all\;baryons\;in\;halo}}$		& 0.0348				& 0.0338						& 	0.0271					 	\\
$y_{\mathrm{eff,all\;gas\;in\;halo}}$		& 0.0240				& 0.0176						& 	0.0068					 	\\
$y_{\mathrm{eff,ISM}}$  			& 0.0047  				& 0.0038 						& 	0.0003						\\
$y_{\mathrm{eff,cool,diff}}$		& 0.0126  				& 0.0067 						& 	0.0024						\\
$y_{\mathrm{eff,hot,CGM}}$		& 0.0067  				& 0.0071 						& 	0.0042						\\
$y_{\mathrm{eff,stars}}$			& 0.0099  				& 0.0162 						& 	0.0193						\\
\hline
\hline
\end{tabular}
\end{center}
\end{table}


Generally, the phase-diagram mass distribution for IllustrisTNG is qualitatively identical to what was found in the original Illustris simulation~\citep[e.g.,][]{Haider2016}.
The majority of the redshift $z=0$ IllustrisTNG gas (49.5 per cent) is in the WHIM intermediate temperature range, which is a similar fraction to what was found in the original Illustris (57.8 per cent).
The hot phase gas contains 7.5 per cent of the $z=0$ IllustrisTNG gas, which is a marginal increase from the original Illustris (7.0 per cent). 
There are, however, some more notable changes in the content of the diffuse and condensed regions.
While the diffuse gas increased from 23.2 per cent in Illustris to 38.7 per cent in IllustrisTNG, the condensed gas decreased from 12.0 per cent in Illustris to 4.3 per cent in IllustrisTNG.
The modification to the condensed gas fraction is likely a result of the modified stellar and AGN feedback used in IllustrisTNG compared against Illustris.
Additionally, the equation of state for the low density/temperature IGM material shows some curvature in the IllustrisTNG results, which was not present in Illustris.
This curvature in the IGM equation of state is likely the result of numerical heating, but should not impact the results presented in this paper.
In this Section and throughout this paper, we focus on comparing the gas-mass distribution against the metal-mass distribution as a tool for understanding the metallicity evolution of IllustrisTNG galaxies.

The distribution of gas-mass within the density-temperature phase diagram shown in the left panel of Figure~\ref{fig:global_phase_diagrams} can be contrasted against the gas-phase metal-mass distribution shown in the right panel of the same Figure.
As with the gas-mass, the majority of the gas-phase metals are in the WHIM (64.2 per cent), with 15.4 per cent being found in the condensed region and 14.9 per cent being found in hot gas.
In stark contrast with the gas-mass distribution, a very small fraction (5.5 per cent) of metals are found in diffuse gas.
This is consistent with a basic picture of galaxy formation and metal production where metals are produced deep within galactic potentials and mostly pollute the gas in and immediately around the galaxies where they form.
However, importantly, it is worth noting that while all star formation is associated with gas in the `condensed' region of phase space, the majority of gas-phase metals ($\sim85$ per cent) are found outside of this region.
Feedback from stellar winds and AGN are critical components of the baryon cycle in our simulations, with the resulting metal distribution being spread over a wide range of the density-temperature phase diagram.

\subsection{Distribution of Metals around Galaxies}
To further investigate the distribution of mass and metals around IllustrisTNG galaxies, Figure~\ref{fig:galaxy_phase_diagrams} shows density-temperature phase diagrams for the gas-mass (top row) and gas-phase metal-mass (bottom row) for gas that is bound to galaxies with average masses of $M_* = 10^{9} \mathrm{M}_\odot$ (left column), $M_* = 10^{10} \mathrm{M}_\odot$ (center column), and $M_* = 10^{11} \mathrm{M}_\odot$ (right column).
The phase diagrams were constructed by taking all gravitationally-bound gas in the {\small SUBFIND} catalog~\citep{SUBFIND, Dolag2009} for the 100 galaxies most closely matched in mass to the three respective target masses.
Within each figure, dashed lines indicate boundaries that are used to separate the hot CGM (top region), cool-diffuse gas (lower left), and dense ISM (lower right).
We define here the hot CGM as being gas above the line $\mathrm{log(}T\mathrm{/10^6 K )} = 0.25\; \mathrm{log(}n / 405 \;\mathrm{cm^{-3})}$, which is the same boundary employed in~\citet{Torrey2012b}.
The hot CGM gas boundary used here cuts orthogonally through the gas-mass distribution within the phase diagram at a location of low gas density, which makes our quoted mass results reasonably insensitive to the exact normalization adopted.
As outlined in Section~\ref{sec:Methods}, we define the ISM gas as being material above our employed star formation density threshold of $n_{\mathrm{H}} \gtrsim 0.13 \; \mathrm{cm^{-3}}$.
The amount of material (gas-mass, or metal-mass) within each region is printed in the Figure, as is the average metallicity of gas in each region.

Examining first the gas-mass distribution within these phase diagrams, we find that there is a significant and well defined population of gas in each of the regions at all of the galaxy mass scales.
The ISM mass is described by a characteristically thin distribution that is set by the~\citet{SH03} equation of state model used in the IllustrisTNG simulations.
At redshift $z=0$ (where the phase diagrams were constructed) the ISM is spatially confined to a region in the center of each halo: namely a central star forming gas disk extending between one and a few kiloparsecs in length.
The cool-diffuse gas acts as a continuation of the dense ISM gas, extending down 2-2.5 orders of magnitude in density.
While a fraction of it is associated with the cool-diffuse gas is an extended gas disk, some of the material is distributed further out into the CGM and can be considered a cool CGM component.
The hot CGM gas is clearly distinct from the other two populations of gas, is generally hotter for more massive galaxies, and spatially occupies the entire halo.

The gas-phase metal-mass distribution broadly traces the gas-mass distribution, but is biased toward the cool-diffuse gas and ISM.
This is seen quantitatively by the higher metallicities obtained for the ISM compared against the CGM gas.
The distinction between ISM and CGM metal content and metallicity is most pronounced for the two lower mass systems where the ISM metallicity is $\sim$5-6 times higher ($Z_{\mathrm{cgm}}=0.1Z_\odot$; $Z_{\mathrm{ism}}=0.6Z_\odot$ for the lowest mass bin).
The higher mass galaxies still have enhanced ISM metallicities, but with a somewhat smaller offset (a factor of $\sim$3-4).

Importantly, we note that the IllustrisTNG simulated galaxies host a significant fraction of their gas-mass and gas-phase metal-mass outside of the ISM in either the cool-diffuse gas, or in the hot CGM.
This point is important when considering the applicability of, e.g., closed box models in describing the metallicity evolution of galaxies.
Tracking the metallicity evolution of the ISM requires accounting for both the production of new metals associated with star formation as well as the constant shifting of those metals between different phases.
The enrichment of ISM, cool-diffuse gas, or the CGM relies on a competition between pristine gas inflow naturally associated with cosmological galaxy growth and the injection of new metals either directly associated with star formation or associated with enriched gas accretion/outflows.
For the galaxies considered here, there is a trend where the highest metallicity gas resides in the ISM at the center of the halo.
However, the metal budget in the cool-diffuse gas and CGM is generally larger than the ISM metal-mass budget itself~\citep{Peeples2014}.

To further explore galaxy metal retention and the relative distribution of metals amongst the gas phases, the effective metal yields are printed within the bottom row of Figure~\ref{fig:galaxy_phase_diagrams}.
The quoted effective metal yields are averages across all galaxies falling into each mass bin. 
Generally, the metal yield, $y$, is used to calculate the total amount of metal-mass produced from stars according to $M_{\mathrm{Z}} = y M_*$.
The metal yields printed in Figure~\ref{fig:galaxy_phase_diagrams} are given for several gas phases as well as the `stellar metal yield' where in every case $y$ is defined as
\begin{equation}
y_i = \frac{M_{Z,i}}{M_*} 
\end{equation}
where $M_{z,i}$ is either the total gas-phase metal-mass for $y_{\mathrm{eff}}$, or the metal-mass in one phase, and $M_*$ is the total stellar mass of the galaxy.\footnote{We note that while we discuss the metal yield for stars to retain consistency with the gas phases, the `stellar metal yield' as defined here is simply the stellar metallicity $y_*=M_{Z,*}/M_* = Z_*$.}
These effective metal yields are useful because they indicate the metal retention efficiency of each phase, and we refer to these factors interchangeably as effective metal yields and metal retention efficiencies in the subsequent text.
The total metal yield $y_{\mathrm{eff}}$ gives a direct measure of the metal retention within all gas in the galaxy, with the phase-separated metal yields indicating how those metals are partitioned between the various gas phases or stars within the galaxy.
Table~\ref{table:effective_yields} summarizes the average effective yields calculated for three mass bins at redshift $z=0$.

\begin{figure*}
\centerline{\vbox{\hbox{ 
\includegraphics[width=0.25\textwidth]{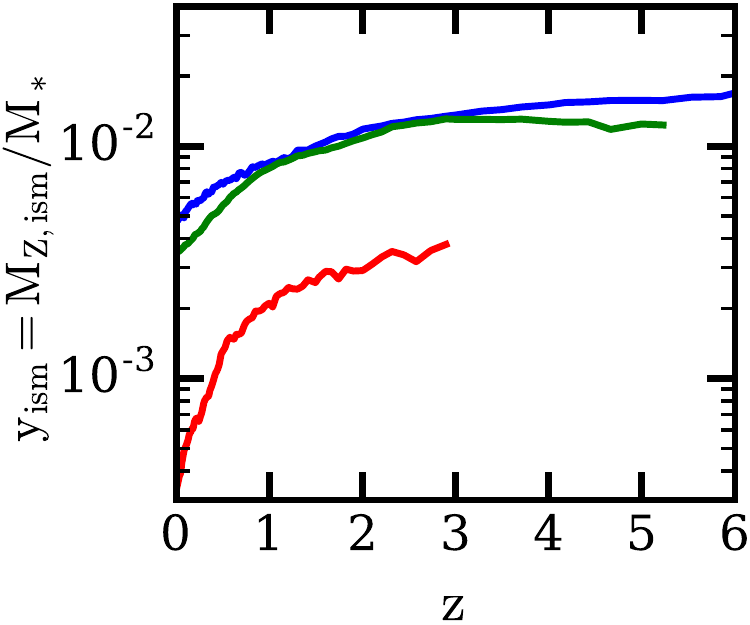}
\includegraphics[width=0.25\textwidth]{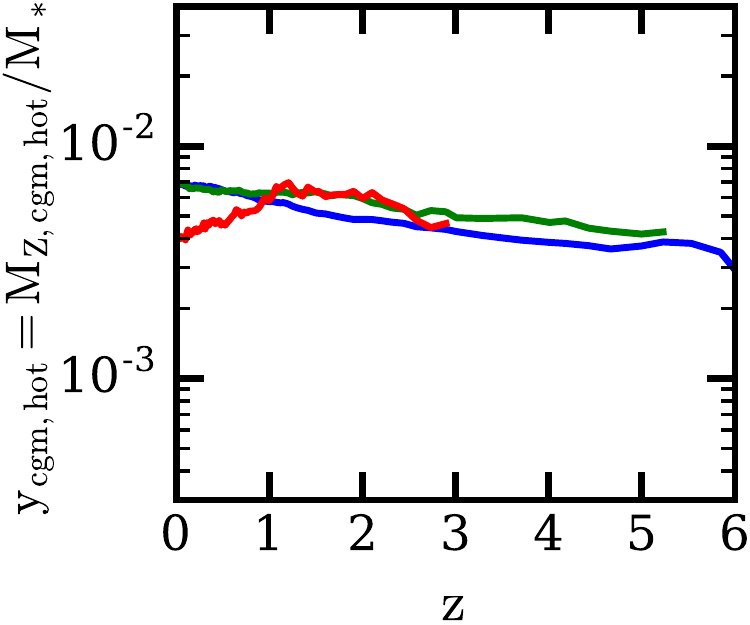}
 \includegraphics[width=0.25\textwidth]{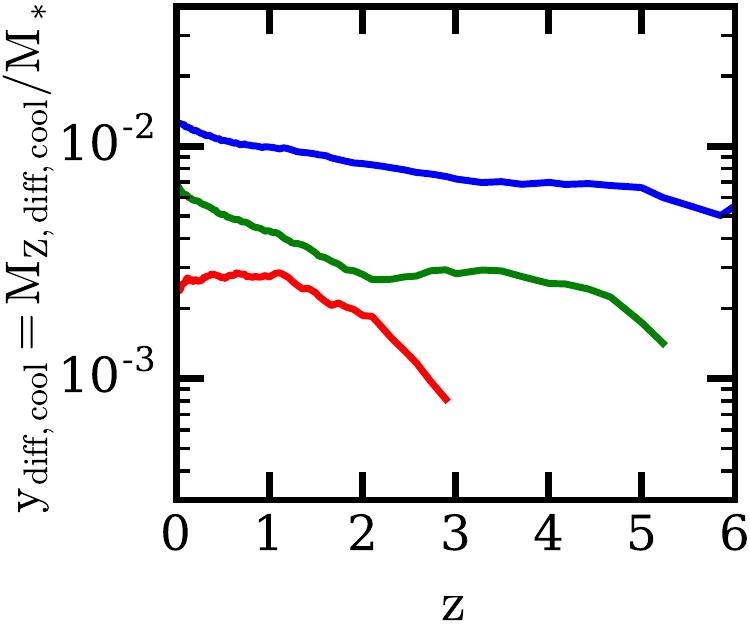}
\includegraphics[width=0.25\textwidth]{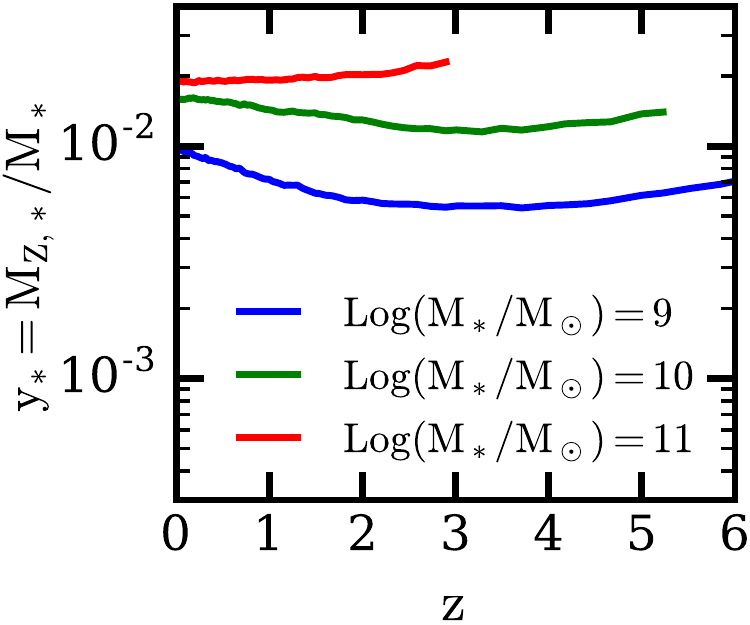}
 }}}
\caption{  The effective yield ($y_i = M_{Z,i} / M_*$; see text for further details) for the ISM (left), the hot CGM (center-left), the diffuse cool gas (center-right), and stars (right) as a function of redshift. 
The effective yields are a normalized parameter indicating the galactic metal retention within each phase.
The effective yield shown here has been calculated for three mass bins in each panel based on the 100 galaxies with masses closest to the target mass.  
Lines are omitted when the bin size becomes larger than $0.1$ dex.
Although metals are produced from star formation in dense ISM gas, the effective yield for the ISM is low.
There is a clear trend where galaxies decrease (increase) the amount of metals they store in the diffuse-cool gas (stars) as their stellar mass increases.
The largest reservoir of metals for low-mass galaxies is the diffuse cool gas, whereas the largest reservoir of metals for high-mass galaxies is locked into stars. 
}
\label{fig:yield_evo}
\end{figure*}

For the IllustrisTNG galaxies, the total metal yield \textit{decreases} with increasing stellar mass from $y_{\mathrm{eff}}=0.024$ for the lowest galaxy mass bin considered to $y_{\mathrm{eff}}=0.007$ for the highest galaxy mass bin.
The decreasing effective yield is an indication that the more massive galaxies are more efficient at removing metal-mass from any/all of the gas phases when compared against their lower mass counterparts.
Gas-phase metal removal can be achieved by either (i) locking an increasingly large fraction of metals into stars, or (ii) by ejecting metals from the halo entirely.
To distinguish between these two possibilities, we need to know the total metal yield for the IllustrisTNG simulations.

The global yield for metals in the IllustrisTNG simulations is set by the adopted IMF along with the adopted metal yield tables (both discussed in Section~\ref{sec:Methods}), and is influenced by the age distribution and metallicity distribution of stellar populations in the simulation.
The global metal yield cannot be directly analytically calculated owing to the dependence on the age and metallicity distribution of stellar populations~\citep{Nelson2017}, 
but can be calculated from the global stellar mass density and metal-mass density
\begin{equation}
y_{\mathrm{global}} = \rho_{\mathrm{Z}} / \rho_* \approx 0.05.
\end{equation}
This global metal yield has been derived for redshift $z=0$ and will increase somewhat with decreasing redshift owing to the evolving age and metallicity distribution of the stellar populations.  
The effective yield of $y_{\mathrm{eff}}=0.024$ for all gas in the low-mass galaxies only accounts for half of the total metal yield.\footnote{We note that not every galaxy will have an identical metal yield given their unique metallicity and formation histories.   All quoted retention efficiencies should therefore be treated as approximate.}
The remaining half of the metals produced by these low-mass galaxies are split between being locked into stars (about 20 per cent) and being ejected from the galaxy (about 30 per cent).
In the highest mass bin, the effective yield for all gas drops significantly to $y_{\mathrm{eff}}=0.007$ (roughly $\sim15$ per cent), but the amount of metals locked into stars significantly increases (roughly $\sim40$ per cent), meaning that just over half of the metals produced have been ejected from the galaxy.
High-mass IllustrisTNG galaxies are more efficient at ejecting metals from their systems than their lower mass companions.
The broad picture painted in Figure~\ref{fig:galaxy_phase_diagrams} is that redistributing metals between different gas phases is important in determining the ISM metallicity, and that the ISM metal retention efficiency changes with galaxy mass.

\subsection{Time Evolution of Metal Retention Efficiencies}
Figure~\ref{fig:yield_evo} shows the time evolution of the effective yield for the ISM (left), the hot CGM (center left), the cool-diffuse gas (center right), and stars (right).
As in Figure~\ref{fig:galaxy_phase_diagrams}, the effective yields shown in Figure~\ref{fig:yield_evo} indicate the total effective yield from the 100 galaxies with masses closest to the target value.
Lines are omitted where the 100 closest galaxies span more than 0.1 dex in stellar mass.
We emphasize that these lines shown are for independent galaxy selections at each redshift, and are not tracking individual galaxies or galaxy populations in time.

In general, the effective yield for the stars and the hot CGM remains reasonably static with time.  
The hot CGM retains an effective yield of $\sim0.003-0.008$ which represents roughly $\sim5-15$ per cent of the metal budget for all of the three galaxy mass bins out to redshift $z=3$.  
Stars retain a metal yield that depends clearly on the total stellar mass of the system, but remains reasonably static with time.
The time-independence of the stellar metal yield is an indication that stellar mass is more important in determining stellar metallicity than redshift in the IllustrisTNG model, with higher mass galaxies having higher metallicity stars.

The cool-diffuse gas and ISM phases show more pronounced evolution with time.
The ISM metal yield increases with increasing redshift similarly for all the three galaxy mass bins.
The highest mass bin is offset toward lower values by a factor of $\sim8$, as was discussed in the previous subsection for redshift $z=0$.
The general trend of increasing ISM effective metal yield with increasing redshift indicates that higher redshift galaxies keep a higher fraction of their metals in their ISM compared against their low redshift counterparts.

We return to the discussion of these effective yields (or metal retention efficiencies) when addressing the evolution of the MZR.

\section{Results: Metal Distribution and Evolution in Galaxies}
\label{sec:Results2}

\subsection{Metal Distribution within Galaxies}

\begin{figure}

\centerline{\vbox{\hbox{
\includegraphics[width=0.15\textwidth]{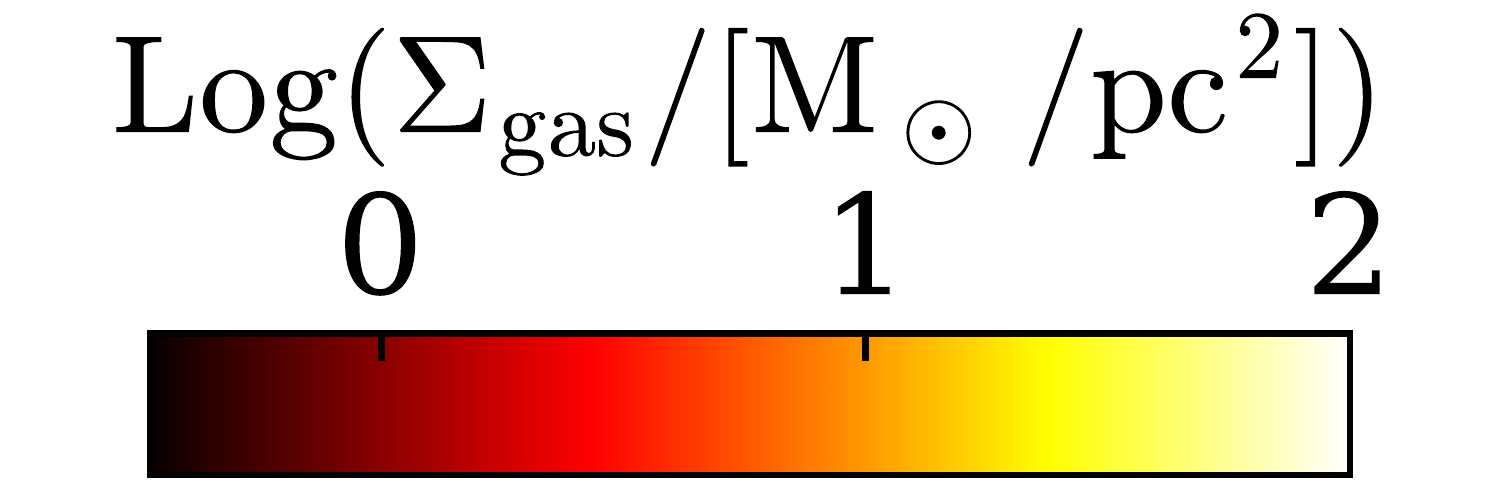}
\includegraphics[width=0.15\textwidth]{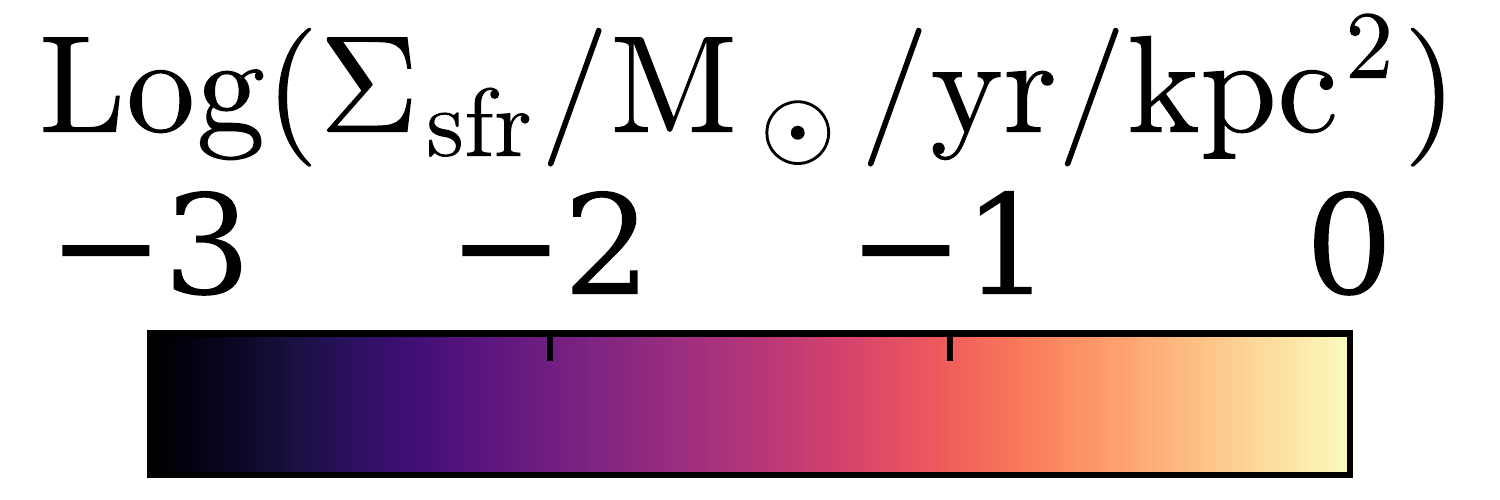}
\includegraphics[width=0.15\textwidth]{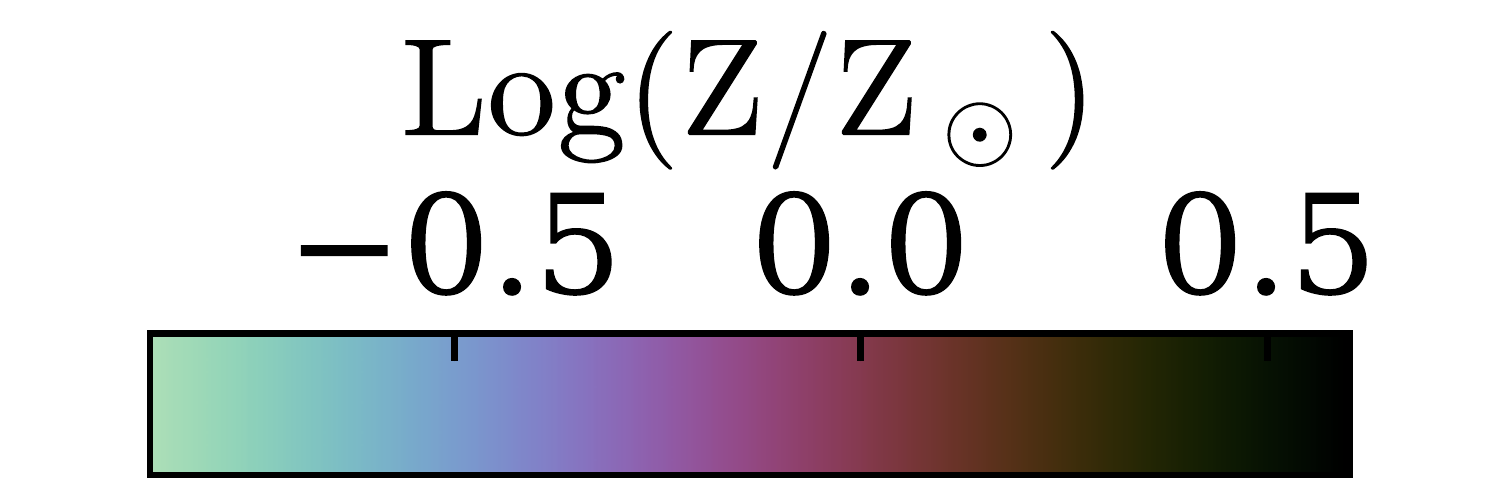}
}}}

\centerline{\vbox{\hbox{
\includegraphics[width=0.15\textwidth]{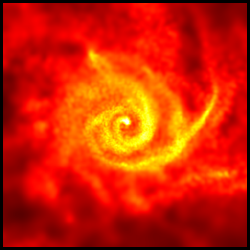}
\includegraphics[width=0.15\textwidth]{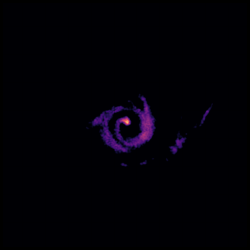}
\includegraphics[width=0.15\textwidth]{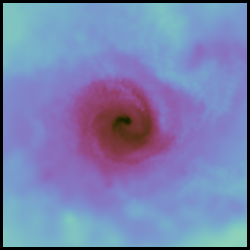}
}}}

\centerline{\vbox{\hbox{
\includegraphics[width=0.15\textwidth]{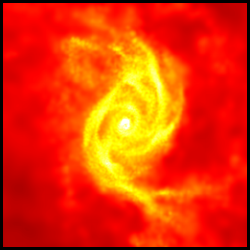}
\includegraphics[width=0.15\textwidth]{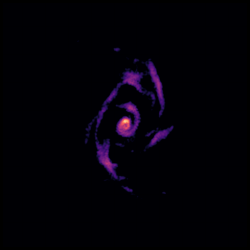}
\includegraphics[width=0.15\textwidth]{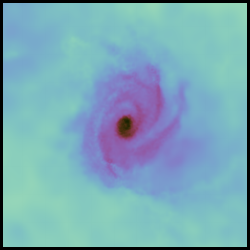}
}}}

\centerline{\vbox{\hbox{
\includegraphics[width=0.15\textwidth]{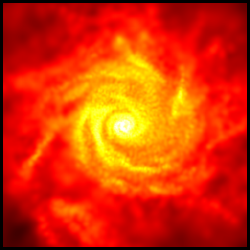}
\includegraphics[width=0.15\textwidth]{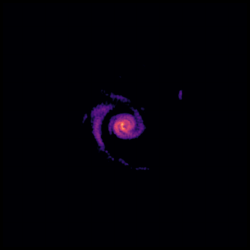}
\includegraphics[width=0.15\textwidth]{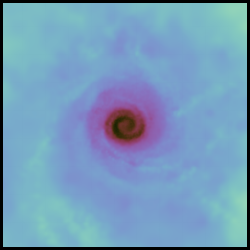}
}}}

\centerline{\vbox{\hbox{
\includegraphics[width=0.15\textwidth]{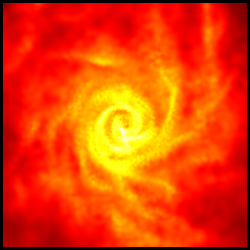}
\includegraphics[width=0.15\textwidth]{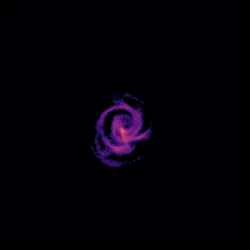}
\includegraphics[width=0.15\textwidth]{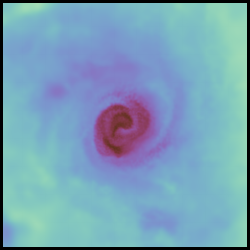}
}}}

\centerline{\vbox{\hbox{
\includegraphics[width=0.15\textwidth]{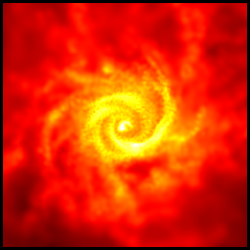}
\includegraphics[width=0.15\textwidth]{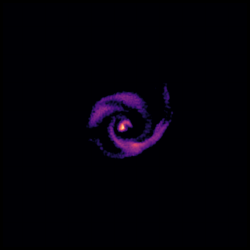}
\includegraphics[width=0.15\textwidth]{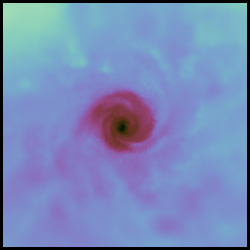}
}}}

\caption{ 
Projections of several gas properties for five galaxies with masses $M_* \approx 10^{10}\mathrm{M}_\odot$ from the TNG100 simulation showing the gas surface density (left), 
the star formation rate surface density (center),  and 
the gas-phase mass-weighted average metallicity (right). 
Associated color bars for each panel are shown at the top of each column.
Face on projections were achieved by projecting the galaxy properties along the net angular momentum axis for gas above the star formation density threshold.
Each image shows a field of view that is 50 kpc on a side.
  }
\label{fig:galaxy_postage_images}
\end{figure}

\begin{figure}

\centerline{\vbox{\hbox{
\includegraphics[width=0.15\textwidth]{./cbar_gas_surface_density.pdf}
\includegraphics[width=0.15\textwidth]{./cbar_sfr_surface_density.pdf}
\includegraphics[width=0.15\textwidth]{./cbar_met.pdf}
}}}

\centerline{\vbox{\hbox{
\includegraphics[width=0.15\textwidth]{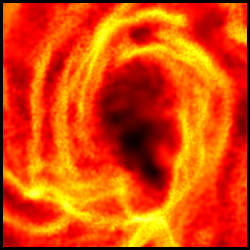}
\includegraphics[width=0.15\textwidth]{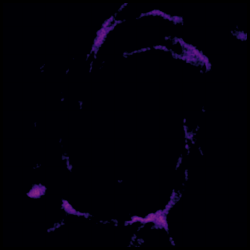}
\includegraphics[width=0.15\textwidth]{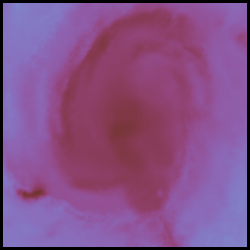}
}}}

\centerline{\vbox{\hbox{
\includegraphics[width=0.15\textwidth]{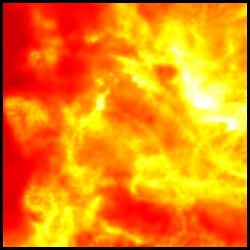}
\includegraphics[width=0.15\textwidth]{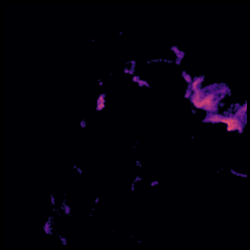}
\includegraphics[width=0.15\textwidth]{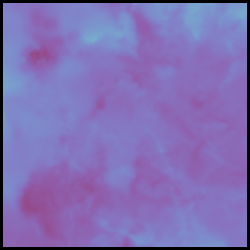}
}}}

\centerline{\vbox{\hbox{
\includegraphics[width=0.15\textwidth]{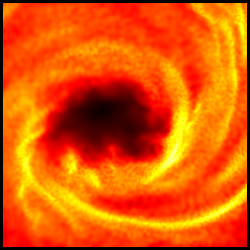}
\includegraphics[width=0.15\textwidth]{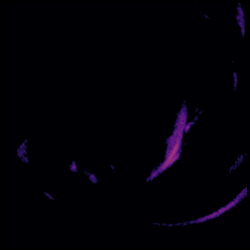}
\includegraphics[width=0.15\textwidth]{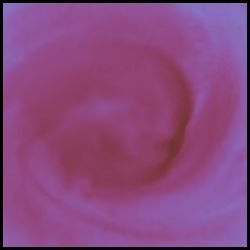}
}}}

\centerline{\vbox{\hbox{
\includegraphics[width=0.15\textwidth]{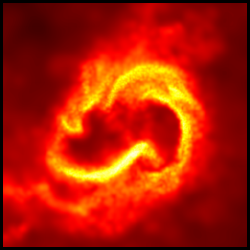}
\includegraphics[width=0.15\textwidth]{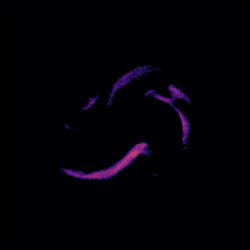}
\includegraphics[width=0.15\textwidth]{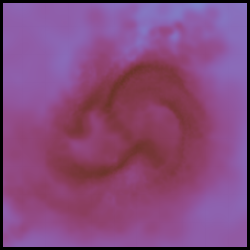}
}}}

\centerline{\vbox{\hbox{
\includegraphics[width=0.15\textwidth]{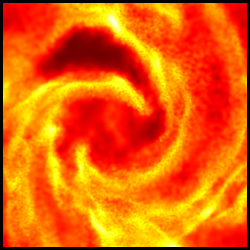}
\includegraphics[width=0.15\textwidth]{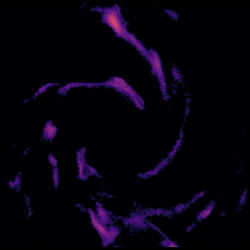}
\includegraphics[width=0.15\textwidth]{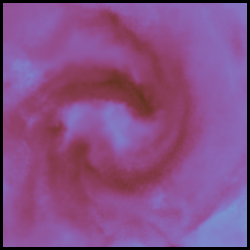}
}}}

\caption{Same as Figure~\ref{fig:galaxy_postage_images}, except for galaxies selected around $M_*\approx 10^{11} \mathrm{M}_\odot$.
The higher mass galaxies considered here have less well ordered gas disks owing to the increased prominence of AGN feedback, and have less clearly identified metallicity gradients.
}
\label{fig:galaxy_postage_images_massive}
\end{figure}

Figure~\ref{fig:galaxy_postage_images} shows face-on projections of galaxies with stellar masses $M_* \approx 10^{10}\mathrm{M}_\odot$ from the TNG100 simulation.
The three rows show 
the gas surface density (left), 
the star formation rate surface density (center), and  
the mass weighted average metallicity (right).
Associated color bars for each panel are shown at the top of each column.
The field of view is 50 kpc wide in each image.
The specific systems shown are the 5 galaxies with masses closest to $M_*=10^{10}\mathrm{M}_\odot$ with star formation rates greater than 1$\mathrm{M}_\odot/\mathrm{yr}$.
The star formation rate cut was employed to ensure star forming galaxies with prominent gas disks.
The projections are taken to be along the angular momentum vector for all star forming gas in the galaxy, leading to roughly face on projections of the central gas disk.

The gas surface density distribution gives a sense for the structure and resolution of galaxies in IllustrisTNG.
The five systems shown in this panel have total gas masses of $M_{\mathrm{gas}}\approx 10^{10.75}-10^{11} \mathrm{M}_\odot$, meaning they are resolved with roughly $N_{\mathrm{gas}} \approx 3-5\times10^4$ gas elements.
This number of gas cells allows for the resolution of a radial surface density profile, as well as some internal structure in the gas disk.
The star formation surface density closely traces the high density gas.  
The star forming gas is confined to the central region of each galaxy, with enhanced star formation activity along the dense spiral arm gas features.
Although we do  not show it here, young stars closely trace the star formation surface density, which serve as sites of rapid heavy element enrichment.

The gas metallicity shown in the third column was calculated as a mass weighted average of the gas phase metallicity.\footnote{We use mass weighted, rather than star formation rate weighted, gas phase metallicity here to provide a continuous map, rather than simply highlighting the star forming gas.}  
The peak central metallicity for this galaxy sample is slightly above solar.
We find there is a metallicity gradient present where the high central metallicities fall off rapidly and nearly monotonically with radial distance.
For the selected galaxies shown, the magnitude of this total metallicity drop is of order $\sim$0.5 dex over the central $\sim$10 kpc.
At larger radii (i.e. toward the edge of the frames) the low metallicity gas outside of the central gas disk shows a less pronounced metallicity gradient.
The metallicity gradient is influenced by two competing effects:  localized enrichment from aging stellar populations and gas mixing.
The metallicity of the gas near the centrally concentrated star forming regions is continually increased as stellar populations enrich the local ISM -- which increases the metallicity gradient prominence.
However, as can be inferred from the surface density plots in the left column, most of the disks have clear non-axisymmetric features, leading to a radial redistribution of metals -- which decreases the metallicity gradient.
The presence of a metallicity gradient is consistent with lower metallicity gas being found in the diffuse-cool gas and hot CGM of the phase diagrams presented in the previous Section.

Figure~\ref{fig:galaxy_postage_images_massive} shows the same quantities as Figure~\ref{fig:galaxy_postage_images}, but now for galaxies with masses $M_* \approx 10^{11}\mathrm{M}_\odot$.
The more massive galaxies share a number of characteristics with their lower mass counterparts, including identifiable gas disks with dense spiral patterns which coincide with ongoing star formation.
However, whereas all five of the lower mass galaxies featured a well-ordered and monotonically decreasing gas surface density profile, the five more massive galaxies show visible distortions, and in some cases significantly depressed central gas densities.
The gas distortions and depressed central gas densities are the result of AGN feedback, which is more impactful to massive galaxies in the IllustrisTNG model.
The presence of strong feedback has a ripple effect on the metallicity measurements, with metallicity gradients being less pronounced.
The same strong feedback which can be seen to influence the gas distributions in Figure~\ref{fig:galaxy_postage_images_massive} is responsible for the diminished mass content and `metal yield' for the ISM and cool-diffuse medium discussed in the previous Section.

For all subsequent portions of this paper, we will quote galaxy metallicities as a single quantity per galaxy. 
We specifically calculate galaxy metallicities as being a star formation rate weighted average metallicity since the majority of observations of gas-phase metallicity rely on nebular emission lines, which probe  gas associated with star forming regions.
Comparing the second and third columns, it can be seen that the star formation rate weighted metallicities will be biased toward the central, dense gas within our galaxy populations.

\begin{figure*}
\centerline{\vbox{\hbox{
\includegraphics[width=0.333\textwidth]{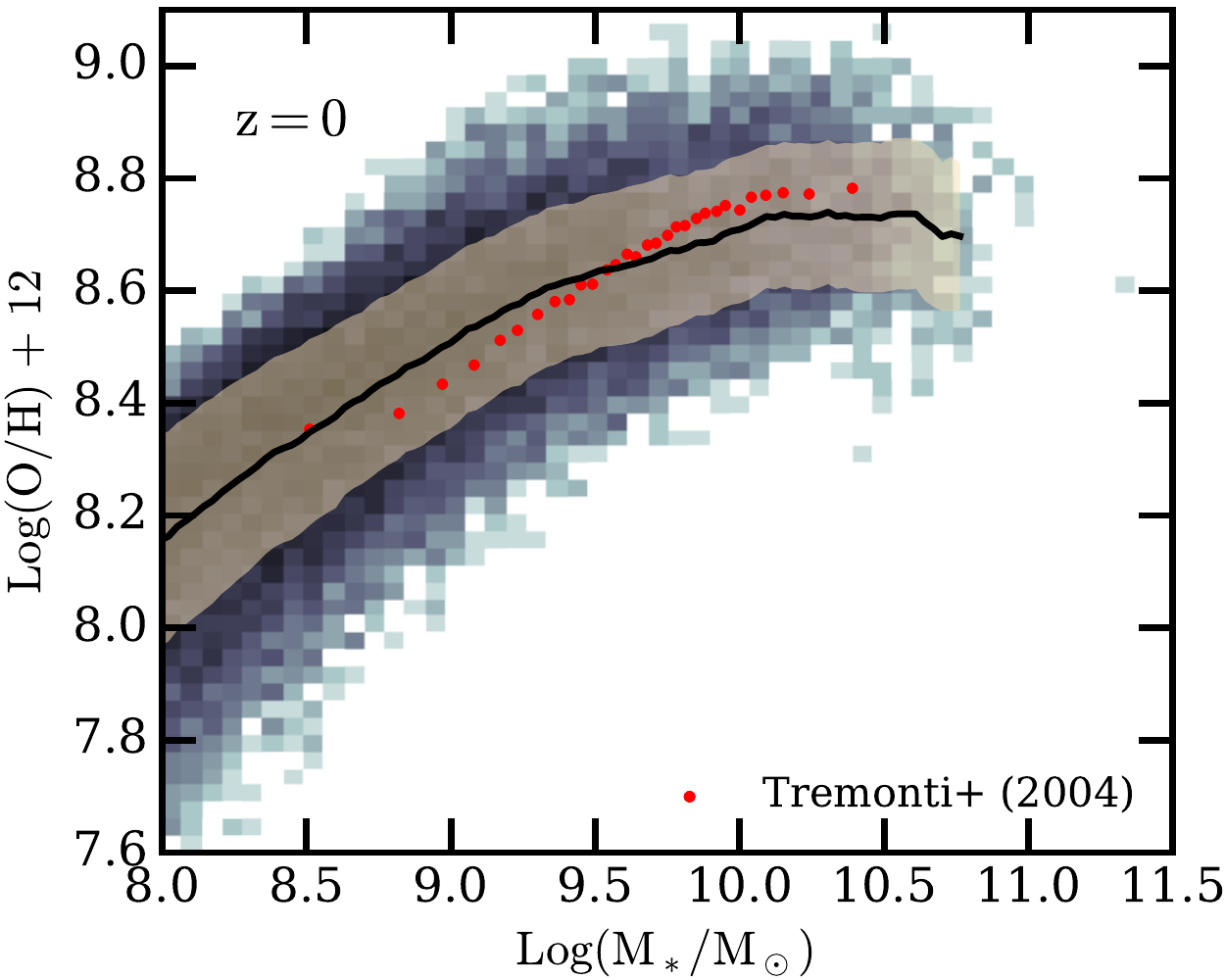}
\includegraphics[width=0.333\textwidth]{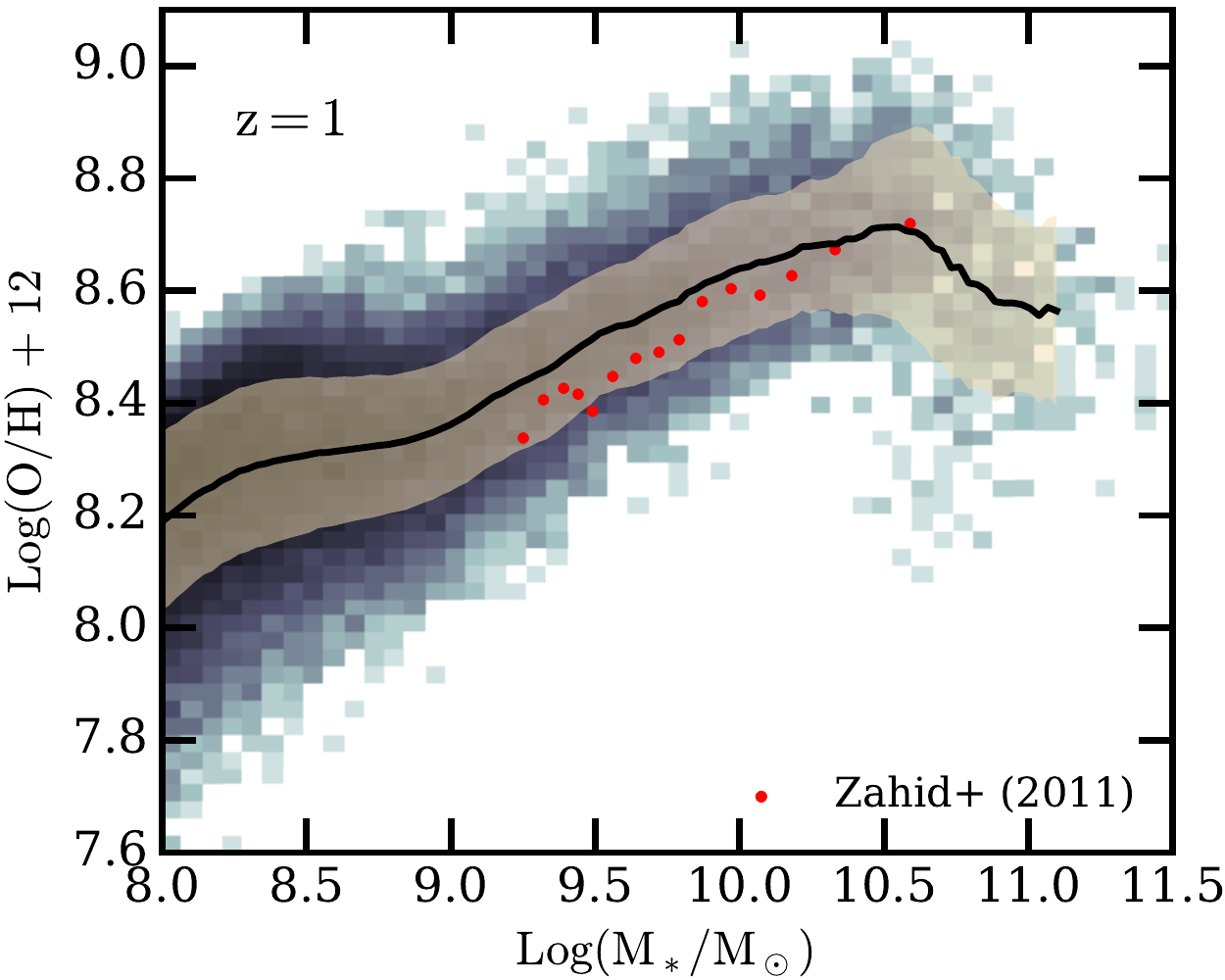}
\includegraphics[width=0.333\textwidth]{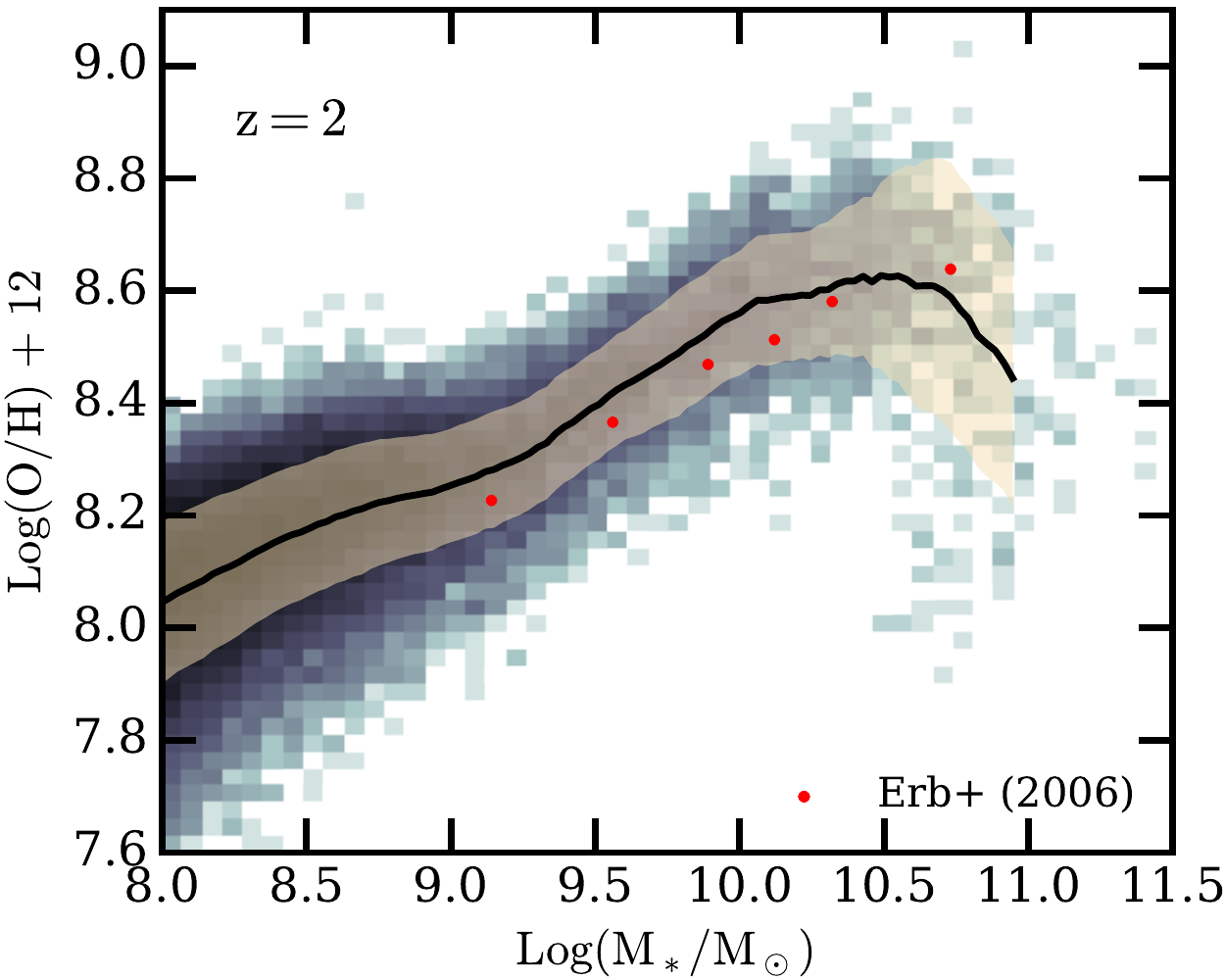} }}}
\centerline{\vbox{\hbox{
\includegraphics[width=0.333\textwidth]{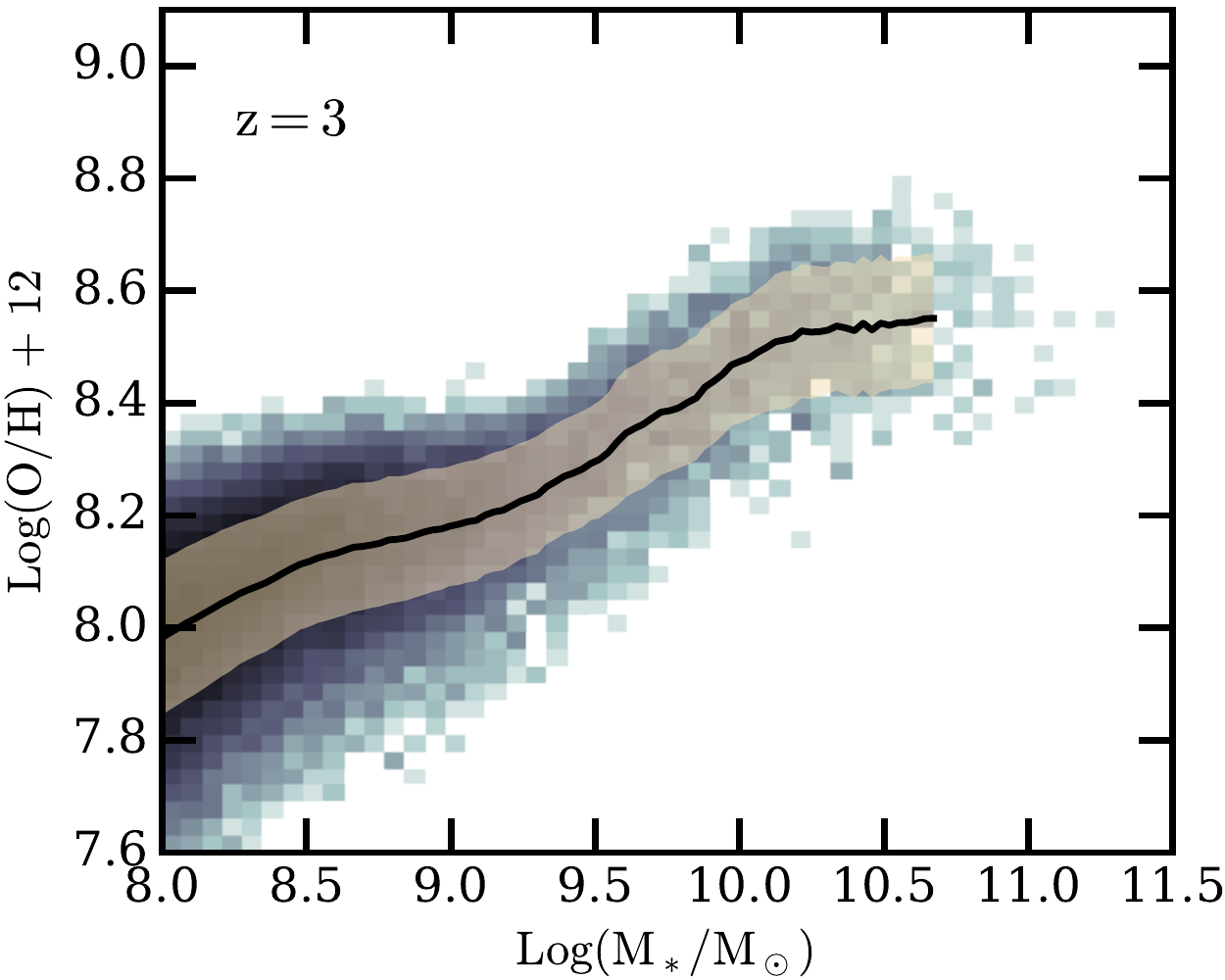}
\includegraphics[width=0.333\textwidth]{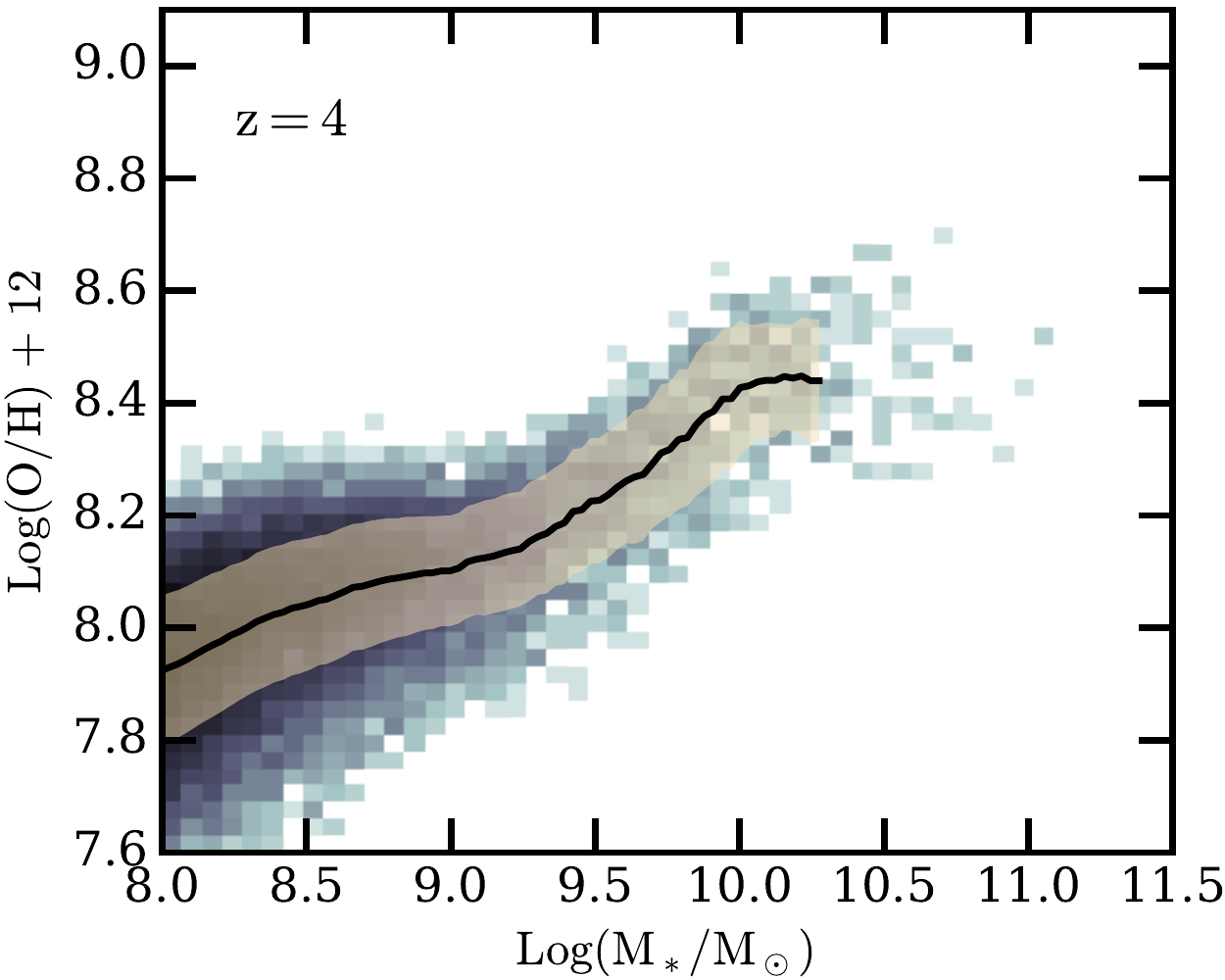}
\includegraphics[width=0.333\textwidth]{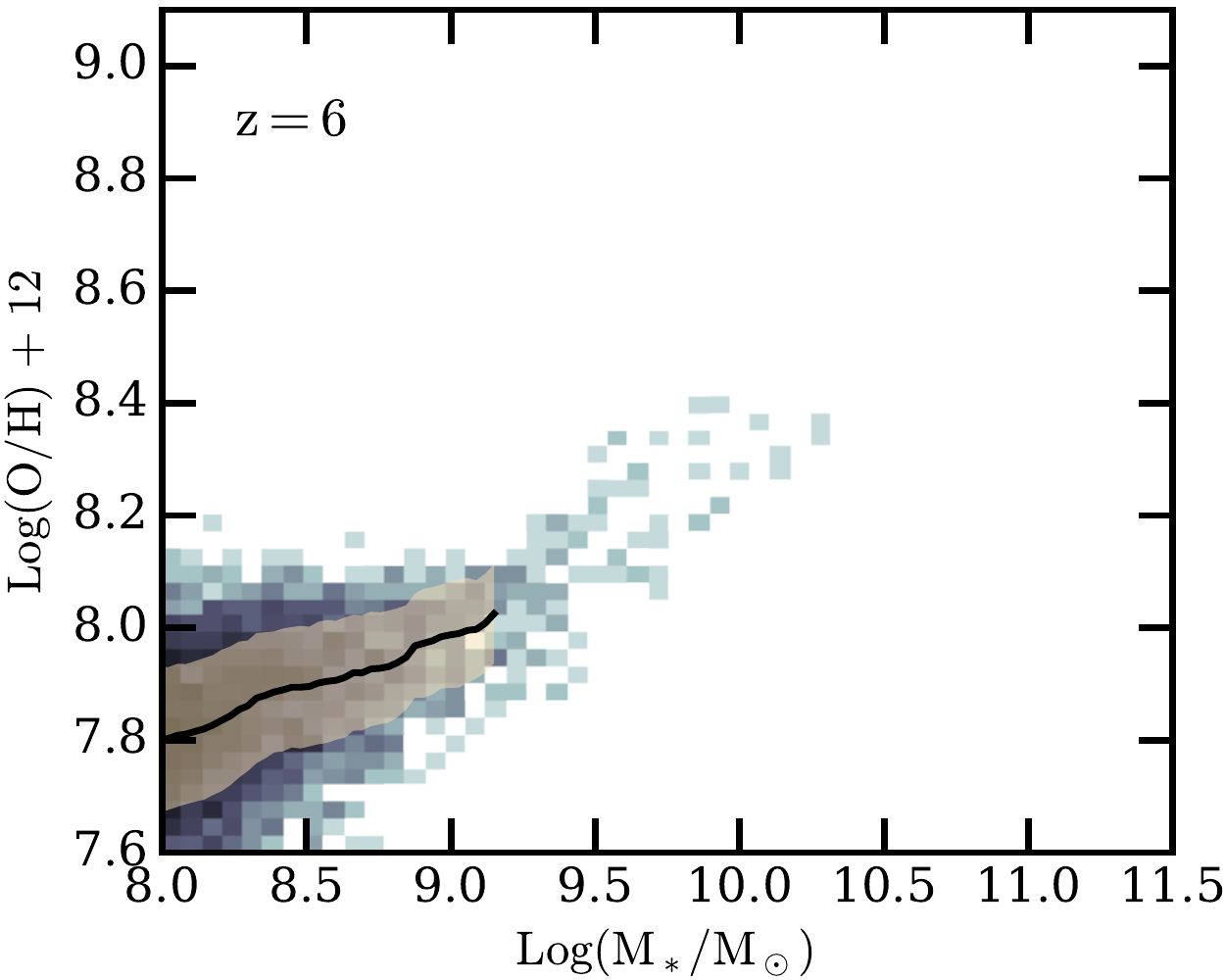} }}}
\caption{
Gas-phase mass-metallicity relation at redshifts $z=\{0,1,2,3,4,6\}$, as labeled in the upper left corner of each panel. 
The colored two-dimensional background histogram indicates the distribution of central galaxies from the TNG100 simulation.  
The solid black line and surrounding shaded band indicate the median MZR and the one sigma scatter, respectively.
Observational data from~\citet{Tremonti2004}, \citet{Zahid2011}, and \citet{Erb2006} have been included in the $z=0$, $z=1$, and $z=2$ panels, respectively.
We do not place any emphasis on the absolute  normalization of the MZR owing to uncertainties in the observed metallicity diagnostics, but instead focus on the slope and normalization evolution.
Our models broadly match the low redshift shape of the MZR and we find there is a gradual decline in the normalization with increasing redshift. 
}
\label{fig:mz_relation}
\end{figure*}

\subsection{MZR and redshift evolution}

Figure~\ref{fig:mz_relation} shows the gas-phase mass-metallicity relation at six distinct redshifts.
The black-and-white colored two-dimensional histograms indicate the distribution of simulated galaxies within this space, with the solid black lines and shaded bands indicating in median and one-sigma scatter of the simulated galaxy distribution.
Where available, we have included median data points for the MZR from the observational literature including \citet{Tremonti2004}, \citet{Zahid2011}, and \citet{Erb2006}.
At redshift $z=0$, the simulated mass metallicity relation shows a clear trend of increasing metallicity with increasing stellar mass, up to roughly $M_*\approx10^{10-10.5} \mathrm{M}_\odot$.
We see a flattening in the MZR beyond this mass scale, which is frequently referred to as the saturation metallicity.
The mass range $M_*\gtrsim10^{10.7} \mathrm{M}_\odot$ is sparsely populated because we only consider galaxies with active/ongoing star formation (i.e. SFR$>0$).
The overall shape of the $z=0$ MZR compares favorably with observations, with the exception that the simulated MZR appears to begin flattening at somewhat lower masses when compared against the observations.  
This metallicity underestimate at the high-mass end may be in part due to contamination with nearly quenched galaxies, as we will discuss later in this Section.

\begin{figure*}
\centerline{\vbox{\hbox{
\includegraphics[width=0.5\textwidth]{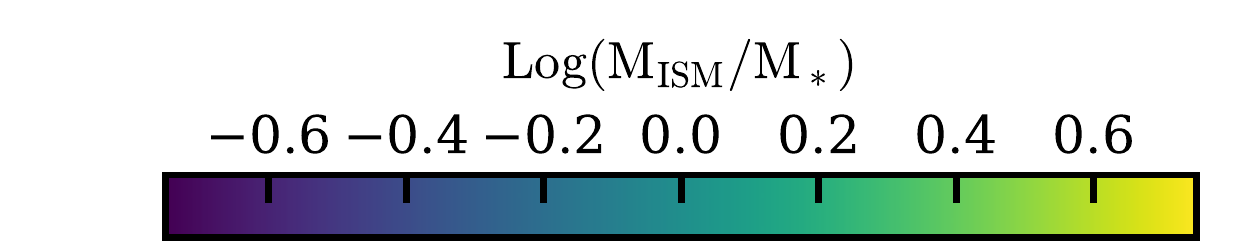}
}}}
\centerline{\vbox{\hbox{
\includegraphics[width=0.333\textwidth]{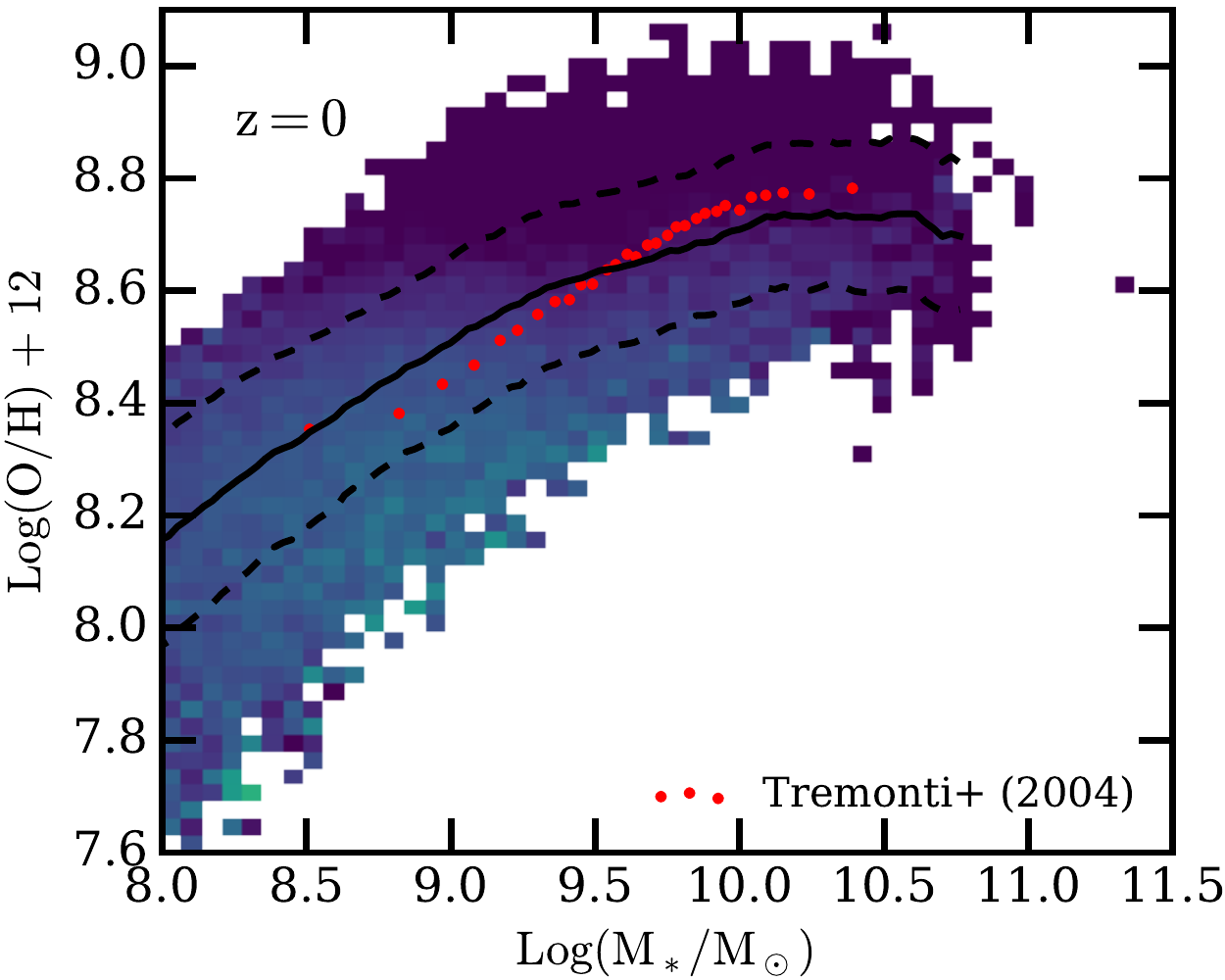}
\includegraphics[width=0.333\textwidth]{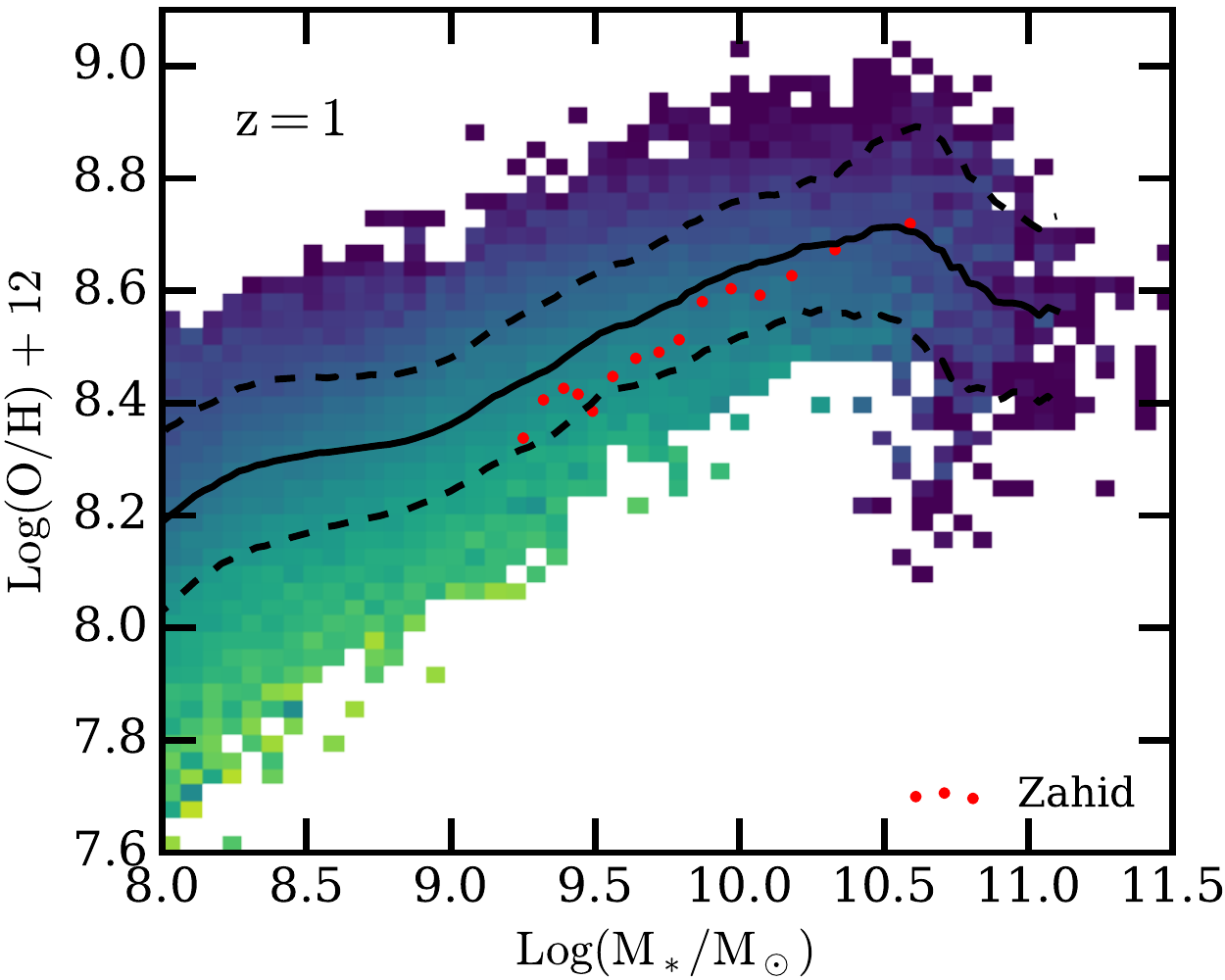}
\includegraphics[width=0.333\textwidth]{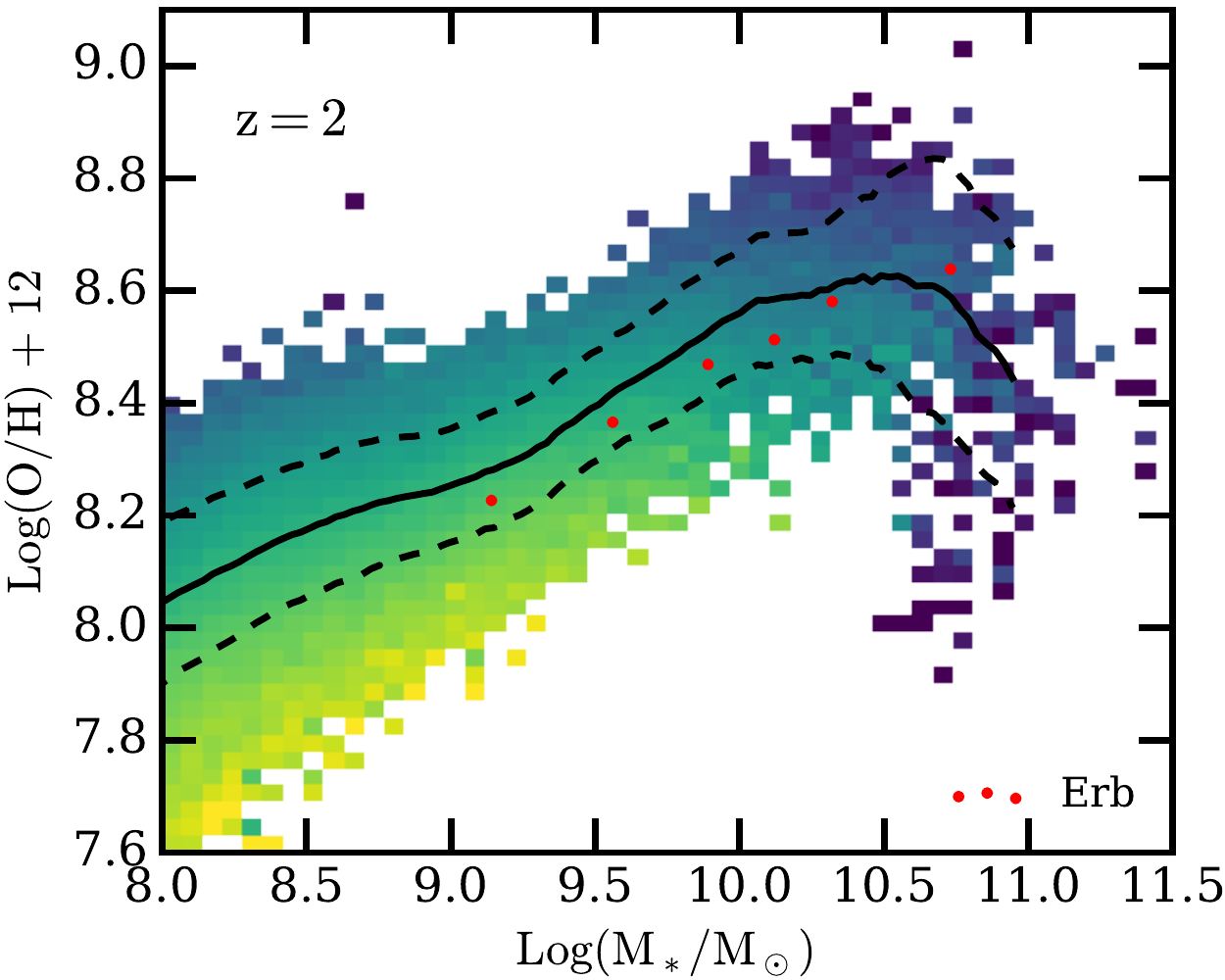} }}}
\centerline{\vbox{\hbox{
\includegraphics[width=0.333\textwidth]{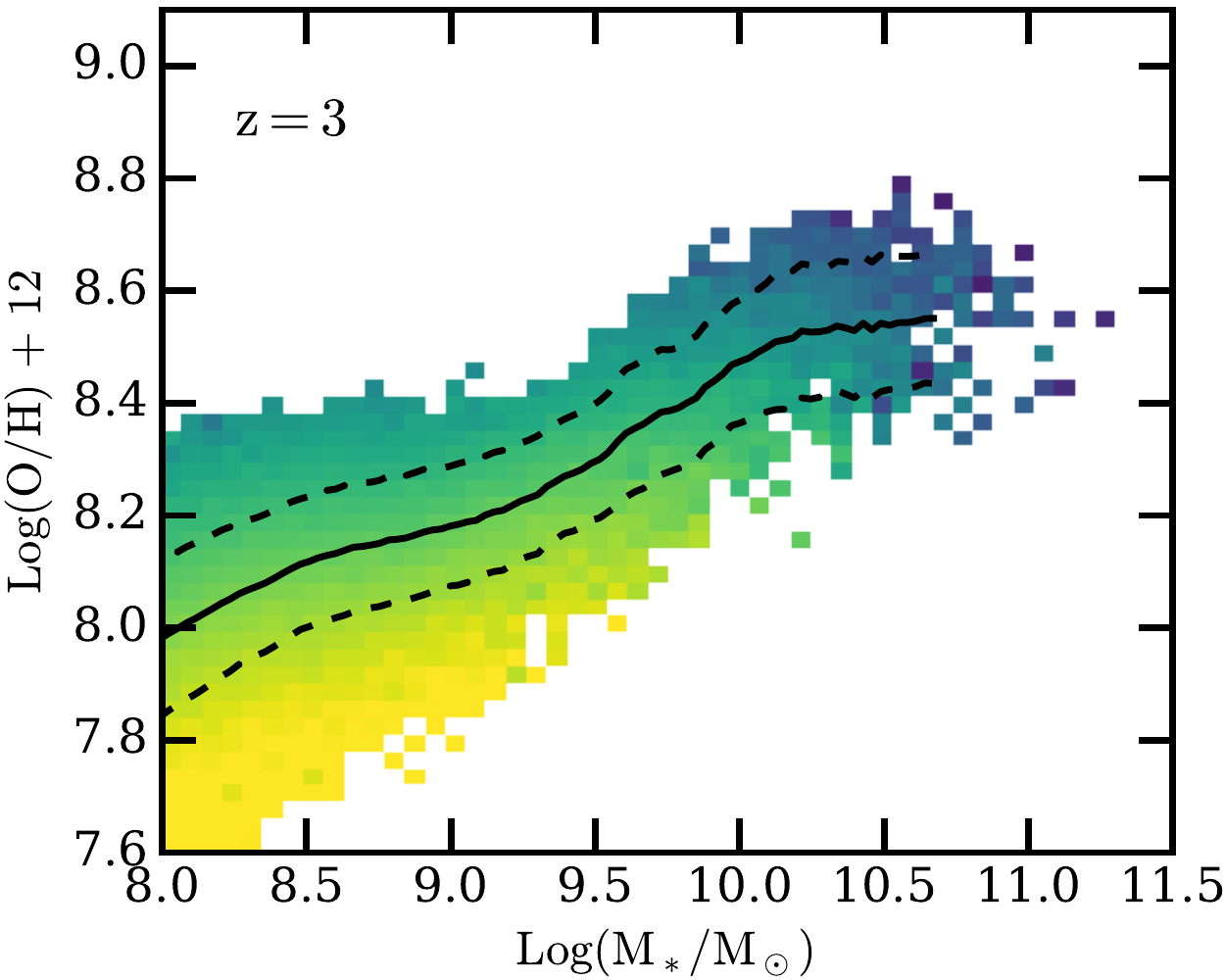}
\includegraphics[width=0.333\textwidth]{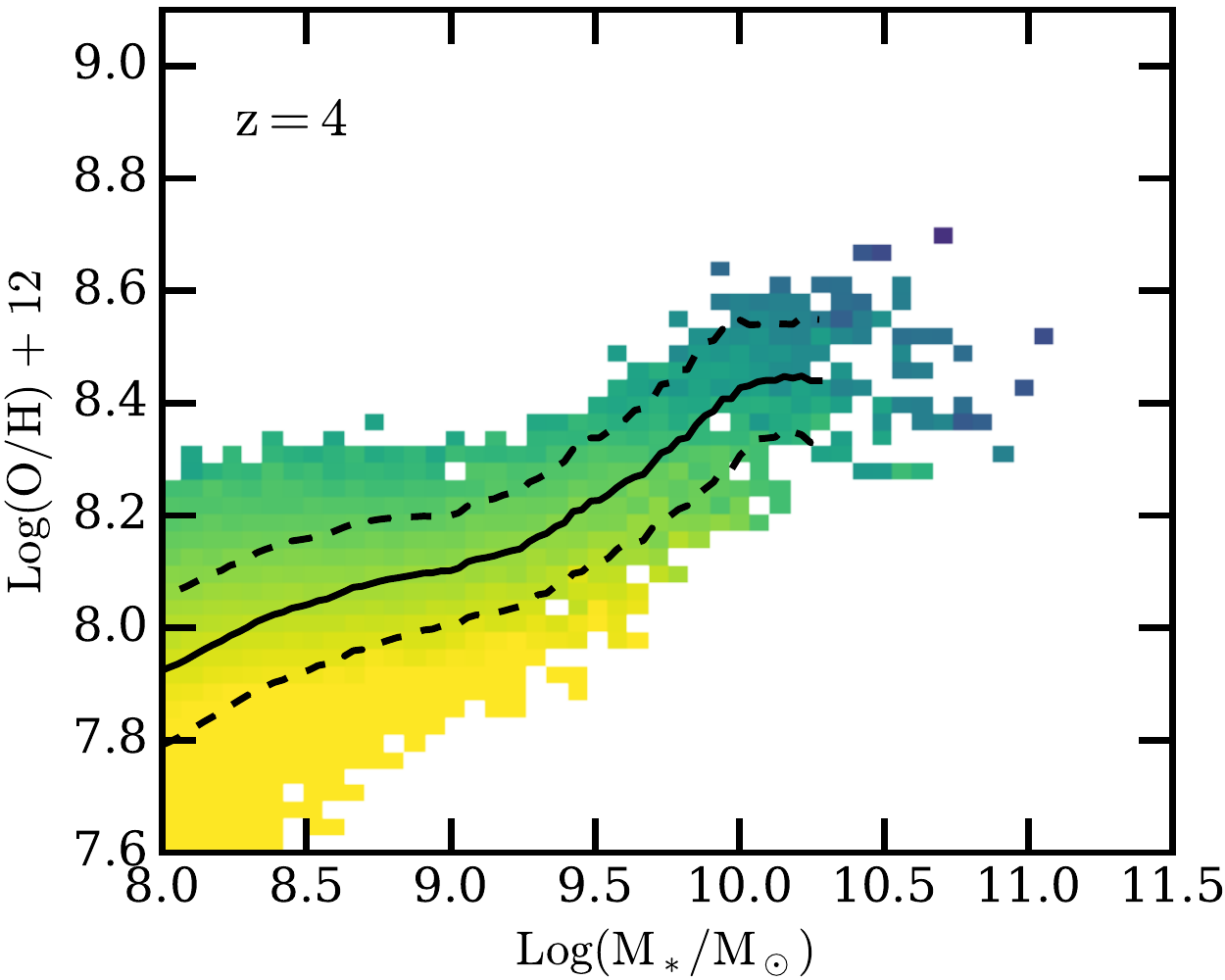}
\includegraphics[width=0.333\textwidth]{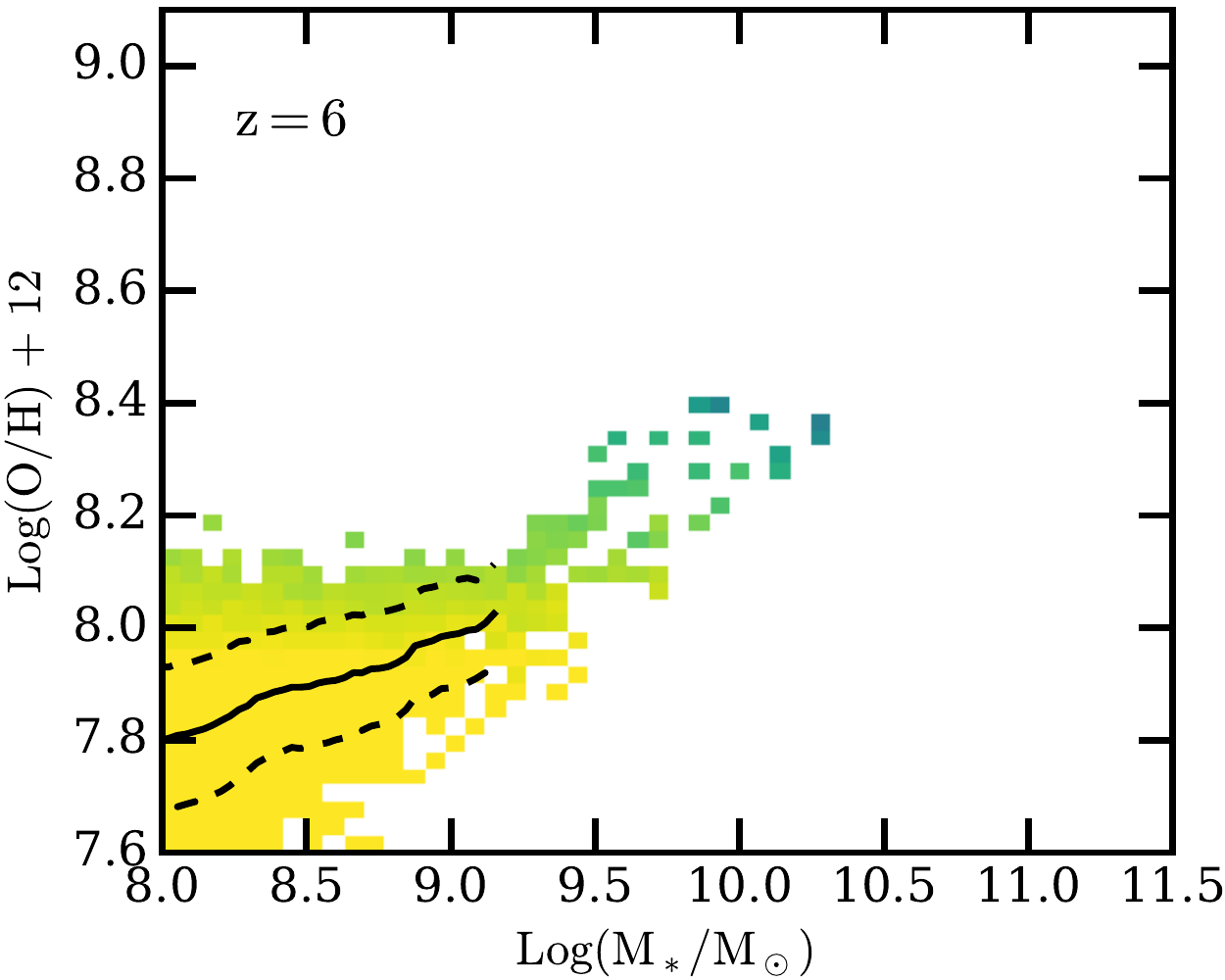} }}}
\caption{
The median gas fraction, defined as the ratio of ISM mass to stellar mass, is shown with respect to the mass metallicity relation for six distinct redshifts as labeled in the upper left corner of each panel.
As in Figure~\ref{fig:mz_relation}, the median MZR is indicated with a solid black line, with the one sigma variation indicated with dashed black lines.
A trend can be identified where galaxies with high (low) gas fractions have low (high) metallicities, when compared against the MZR.
This trend holds across a wide redshift and mass range.  
 }
\label{fig:mz_relation_ism_mass}
\end{figure*}

Comparing the MZRs at higher redshifts to the $z=0$ MZR reveals a continuous evolution in the normalization, with higher redshift MZRs having lower normalizations.
In contrast to the $z=0$ MZR, the high redshift MZRs also contain a break in the low-mass MZR slope at $M_*\approx10^9 \mathrm{M}_\odot$.
No similar break in the low-mass MZR slope was present in the original Illustris model~\citep[e.g.,][]{Torrey2014}.
These bumps are the product of the minimum wind velocity imposed in IllustrisTNG stellar feedback model.
Wind velocities are assigned within the IllustrisTNG model according to
\begin{equation}
v_{\mathrm{w}} = \mathrm{max} \left[  \kappa_{\mathrm{w}}  \sigma_{\mathrm{DM}}   \left(\frac{H_0}{H(z)} \right)^{1/3}, v_{\mathrm{w,min}}  \right ],
\end{equation}
where $\kappa_{\mathrm{w}}= 7.4$ is the wind velocity normalization factor, $\sigma_{\mathrm{DM}}$ is the local dark matter velocity dispersion, and $v_{\mathrm{w,min}}=350$ km/s is the minimum allowed wind velocity~\citep[see][eqn 1 and Table 1]{Pillepich2017}.
Low-mass galaxies in the IllustrisTNG model will have low dark matter velocity dispersions, and will therefore be assigned the minimum wind velocity.
The transition scale for the fiducial to constant wind velocity is not directly set by stellar mass, but roughly corresponds to $M_* \sim10^9 \mathrm{M}_\odot$ which is where the MZR bumps are visible.
At the fixed wind velocity of $v_{\mathrm{w,min}}=350$ km/s, the amount of gas ejected in winds (i.e., the wind mass loading factor) is lower than what would have been ejected if the wind velocity were permitted to be lower.
The reduced ejected wind mass carries away less metal content from the galaxy, results in higher ISM gas-phase metal retention, and manifests in somewhat higher metallicities for low-mass galaxies.
MZR data for lower galaxy mass bins becomes more scarce, but there is not currently strong observational evidence in agreement, or disagreement, with the break in the MZR found in our models.
While the constant minimum wind velocity is critical in the IllustrisTNG model toward matching the low end of the galaxy stellar mass function~\citep{Pillepich2017}, it also leaves a clear potentially observable imprint on the MZR.
Further observational data on the low-mass MZR may help constrain the validity of the adopted minimum wind velocity assumption.

In general, the median MZR from the simulations and observations compare favorably where data exists out to $z=2$.
While we place limited emphasis on the absolute normalization of the MZR~\citep{KewleyEllison}, 
there is a normalization evolution between redshifts $z=0$ and $z=2$ that is present in the observed data and which appears to be captured in the simulations.
Our model predicts that this normalization evolution should continue out to higher redshifts yet, which may be observable with JWST and which is discussed further in Section~\ref{sec:Discussion}.

\begin{figure*}
\centerline{\vbox{\hbox{
\includegraphics[width=0.333\textwidth]{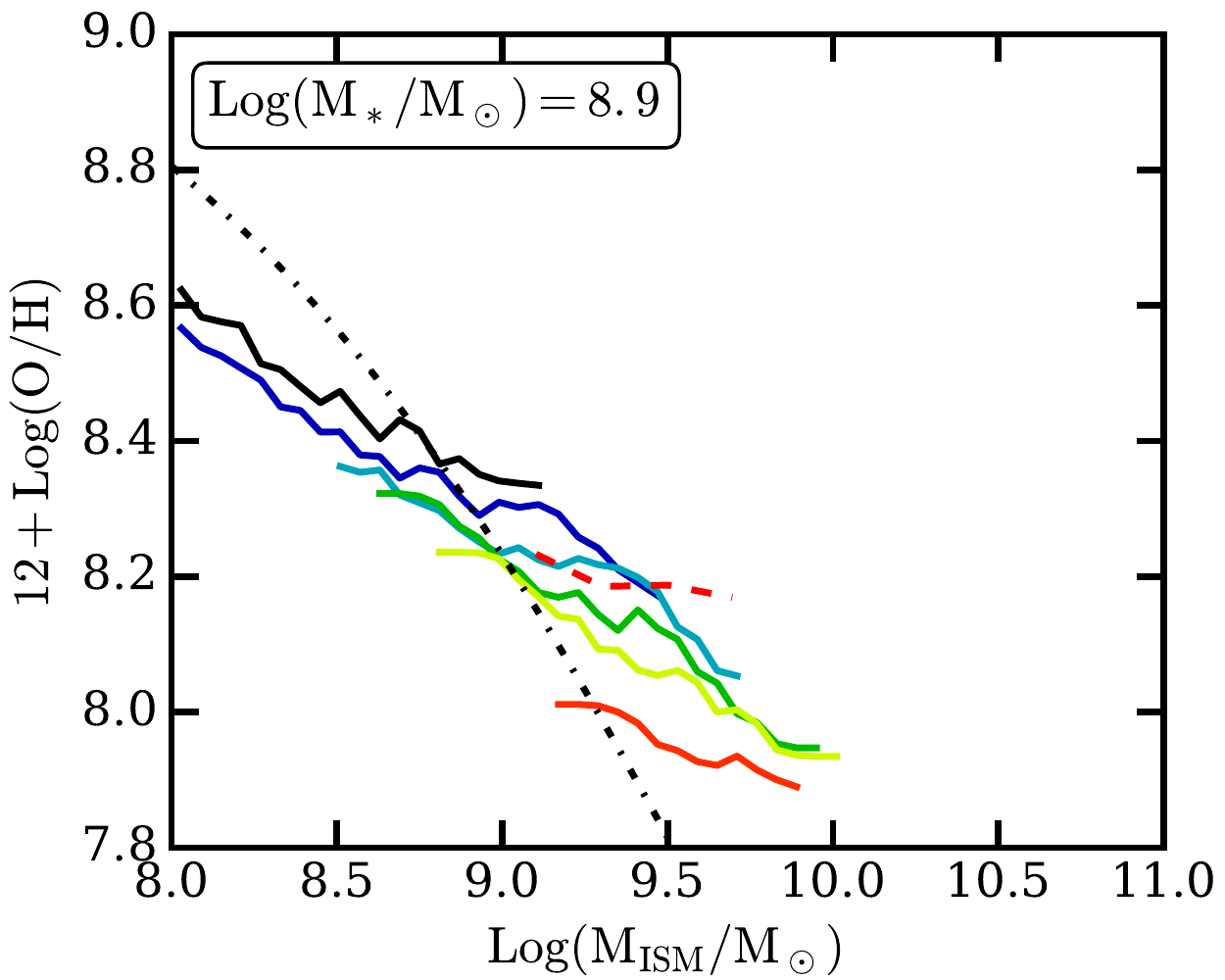}
\includegraphics[width=0.333\textwidth]{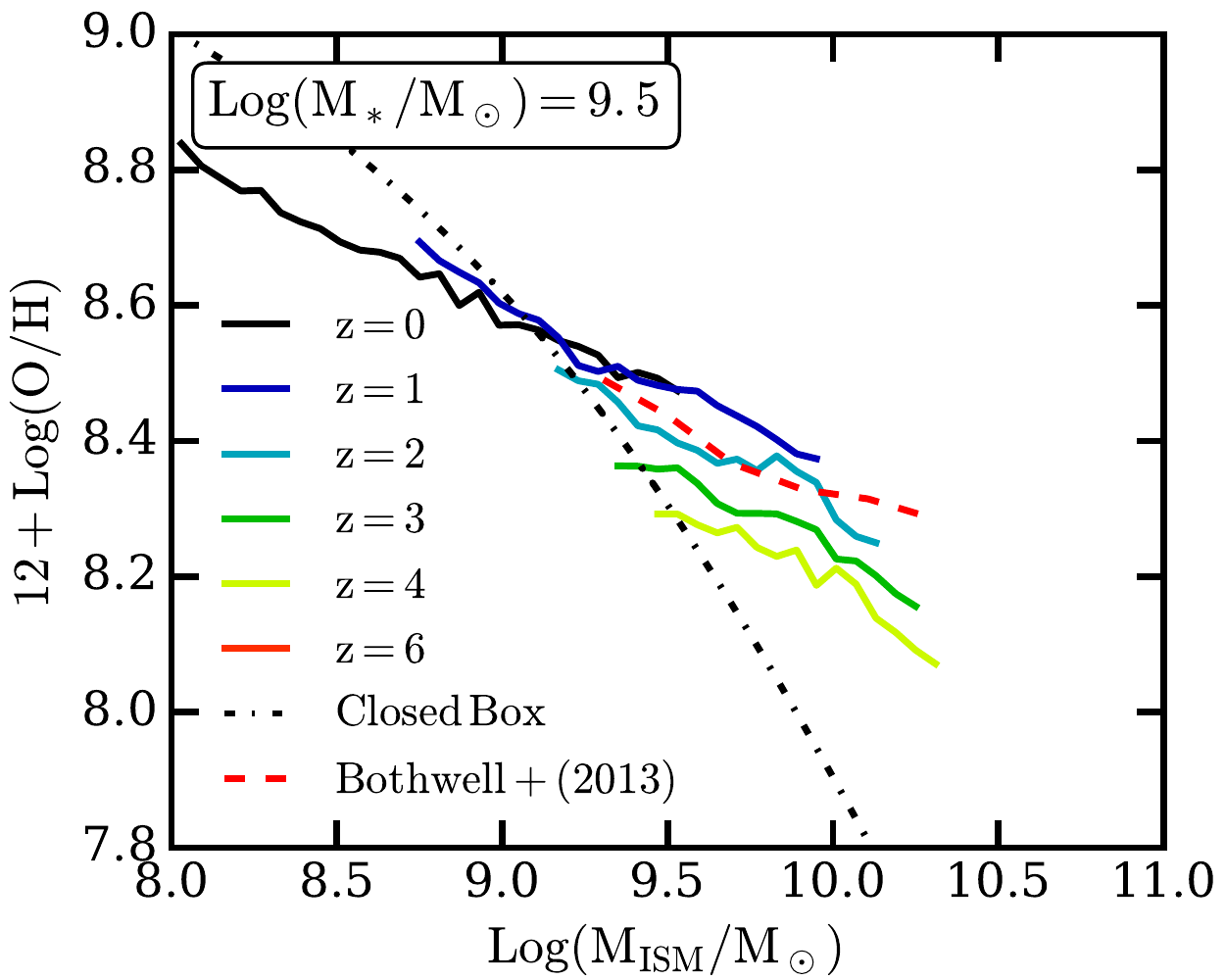}
\includegraphics[width=0.333\textwidth]{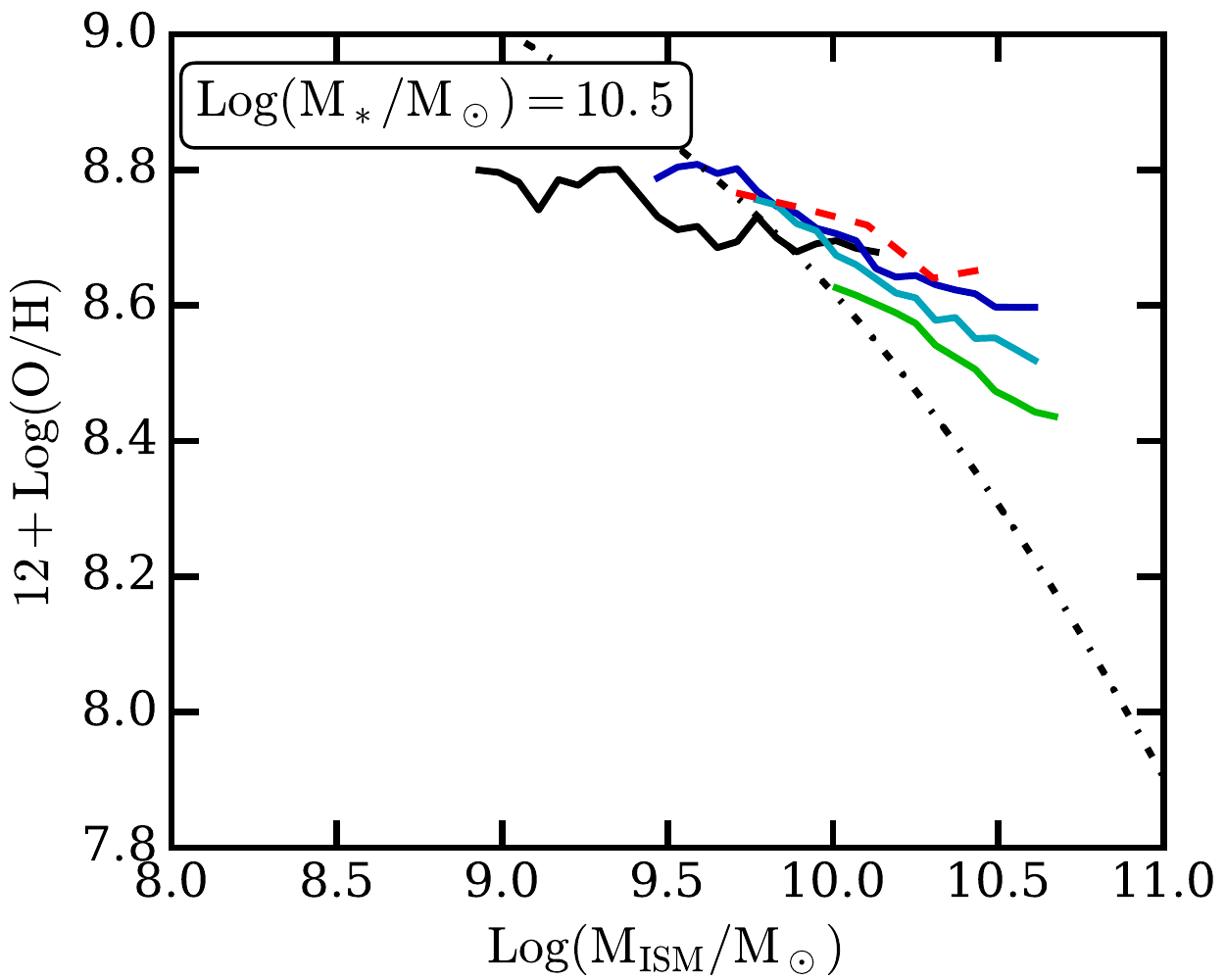} }}}
\caption{ The average metallicity is shown as a function of $M_{\mathrm{ISM}}$ for three thin mass bins (corresponding to the three panels) at several redshifts as indicated within the legend.
There is a clear correlation between galactic metallicity and the ISM mass where galaxies with higher ISM masses have lower metallicities.
The strength of this correlation decreases for higher galaxy mass samples.
The dashed red lines indicate observational data from~\citet{Bothwell2013} which broadly agrees with the simulated trends.
The dot-dashed black lines indicate the closed box model with an effective yield of $y=0.02$.  }
\label{fig:secondary_correlations_mism}
\end{figure*}

\subsection{MZR Dependence on Gas Fraction}
Figure~\ref{fig:mz_relation_ism_mass} shows the median gas fraction of galaxies with respect to the MZR for six distinct redshifts.
We define gas fraction in Figure~\ref{fig:mz_relation_ism_mass} as being the ratio of total gas-mass above the star formation density threshold to the total stellar mass within twice the stellar half mass radius.
As has been found previously~\citep[e.g.,][]{Dave2012}, there is an overall trend with mass where higher mass galaxies have lower gas fractions.
Additionally, higher redshift galaxy populations have higher gas fractions on average than their lower redshift analogs at a fixed stellar mass.
However, perhaps most interesting, Figure~\ref{fig:mz_relation_ism_mass} clearly reveals the presence of a residual trend in the simulated MZR, where galaxies with high metallicities have low gas fractions, and galaxies with low metallicities have high gas fractions.  
The residual trend between offset from the MZR and gas fraction can be seen at all redshifts below $z=6$.
The correlation between offset from the MZR and gas fraction appears notably monotonic, with limited exceptions.	
Variations in the gas fraction at a fixed mass reach up to an order of magnitude in value.
The strength of the trend between offset from the MZR and gas fraction becomes somewhat less clear at very high redshift (i.e. $z=6$) because the volume used in the present study is simply too small to have a sufficient number of well-resolved, massive-galaxies that might allow us to continue the study of this trend.  
However, for the narrow mass range of $8.5 < \mathrm{log(}M_*/\mathrm{M_\odot)} < 9.25$ a trend with gas fraction is still present.

Figure~\ref{fig:secondary_correlations_mism} identifies the residual trend between scatter in the MZR and gas fraction by showing the median metallicity as a function of ISM gas-mass in thin mass bins, at several redshifts.
The three panels each select galaxies in $\pm$0.1 dex stellar mass bins centered around $\mathrm{Log(}M_*/\mathrm{M}_\odot \mathrm{)}=\{8.9, 9.5, 10.5\} $ in the left, center, and right panels, respectively.
Figure~\ref{fig:secondary_correlations_mism} shows that the simulated galaxy population in IllustrisTNG displays an anti-correlation between gas-phase metallicity and ISM gas-mass, at a fixed stellar mass.
A trend of this nature holds across a wide mass range.
Comparing the three panels over the mass range $8.9 < \mathrm{Log(}M_*/\mathrm{M_\odot)} < 10.5$ shows that there is a visible increase in the average metallicity -- as is expected from the MZR -- while the residual trend between metallicity and ISM mass remains present but decreases in strength.

For comparison at redshift $z=0$, the dashed red line in Figure~\ref{fig:secondary_correlations_mism} shows the data from~\citet{Bothwell2013} comparing the gas-phase metallicity versus HI gas-mass.
\citet{Bothwell2013} assembled HI gas masses from the Arecibo Legacy Fast ALFA survey~\citep[ALFALFA;][]{Haynes2011} and metallicity measurements from the Sloan Digital Sky Survey (SDSS) using the R23 parameter and [N II]/H$\alpha$ ratio as metallicity diagnostics~\citep[as applied in][]{Maiolino2008}.
We note that there is significant uncertainty associated with the absolute normalization of {\it both} axes in this comparison.
Metallicity diagnostics are known to be systematically uncertain at the $\gtrsim$0.3 dex level~\citep{KewleyEllison} and making a fair comparison between the observed HI gas masses  employed in \citet{Bothwell2013} and gas masses from our simulation is non-trivial~\citep{Bahe2016, Marinacci2017b}.
Here we simply take the total mass of all star forming gas in each galaxy, which likely scales with the HI gas-mass, but may underestimate the total HI gas-mass of the system.
Owing to these two normalization concerns, we focus our attention on the slope of the residual correlation between metallicity and gas-mass.

In general, the slope between metallicity and HI gas-mass found in the \citet{Bothwell2013} data is very similar to the slope found in the majority of the IllustrisTNG data between metallicity and ISM mass.
In more detail, the redshift $z=0$, low-mass IllustrisTNG galaxy metallicity (black line in the left panel, in Figure~\ref{fig:secondary_correlations_mism}) scales as $Z \propto  M_{\mathrm{ISM}}^{-0.3}$, which is very similar to the \citet{Bothwell2013} trend.  
Moving from the lowest mass bin (left panel) to the intermediate mass bin (center) and highest mass bin (right) we find the continued presence of a correlation between metallicity and ISM gas-mass.
However, the slope gradually flattens to $Z \propto  M_{\mathrm{ISM}}^{-0.1}$ for the highest mass bin.
This slope flattening is consistent with the observed mass dependent trend.
We emphasize again, however, that some caution should be used when interpreting the exact quantitative slope comparisons between these two datasets owing to non-trivial conversions from $M_{\mathrm{ISM}}$ to $M_{\mathrm{HI}}$.

The colored solid lines in the three panels of Figure~\ref{fig:secondary_correlations_mism} indicate the metallicity versus gas-mass trend found at redshifts $z>0$.
There is a gradual normalization evolution, where higher redshift galaxy populations have lower metallicities and higher ISM gas masses at a fixed stellar and gas-mass~\citep[see also][]{DeRossi2017, Dave2017}.
The slopes that describe the metallicity versus gas-mass relation are, however, nearly unchanged.
Our models predict that a residual correlation between offset from the MZR and ISM gas-mass should persist in high redshift galaxy observations.

The dot-dashed black line in Figure~\ref{fig:secondary_correlations_mism} indicates the line $Z \propto - y\; \mathrm{ln}(M_{\mathrm{ISM}} / (M_{\mathrm{ISM}} + M_*))$, which corresponds a closed-box-model type evolution track.
In these plots, we have adopted an effective yield of $y=0.02$ rather than the total yield of $y=0.05$ to improve the normalization agreement between the model and simulated data.
In general, the closed box model trend is steeper than the tracks at any fixed redshift.
We return to this point when discussing the connection between individual galaxy evolution tracks and net galaxy population trends in Section~\ref{sec:Results3}.

A number of observational papers had previously identified residual correlation between offset from the MZR and galaxy SFR~\citep{Ellison_FMR, LaraLopez2010, Mannucci_FMR}.
Although we do not show it here, a very similar residual trend between ISM metallicity and star formation rate is also found in our simulations (see, e.g., Figure~\ref{fig:full_page} and Torrey et al. in prep.).
We omit showing this result because it is qualitatively identical to that presented in Figure~\ref{fig:secondary_correlations_mism} and the ISM gas-mass is a more direct driver of ISM metallicity since it directly enters into the metallicity determination.

The existence of residual trends in the scatter in the MZR is a critical piece to understanding how the MZR emerges and how galaxies evolve with respect to the MZR.
In the following Section, we explore the evolution of individual galaxies with respect to the MZR to consider how halo scale gas inflows, disk scale gas dynamics, and star formation activity all act to shape the MZR and the emergence of these residual trends.

\section{Results: Time evolution of galaxies}
\label{sec:Results3}

\begin{figure*}
\centerline{\vbox{\hbox{
\includegraphics[width=0.5\textwidth]{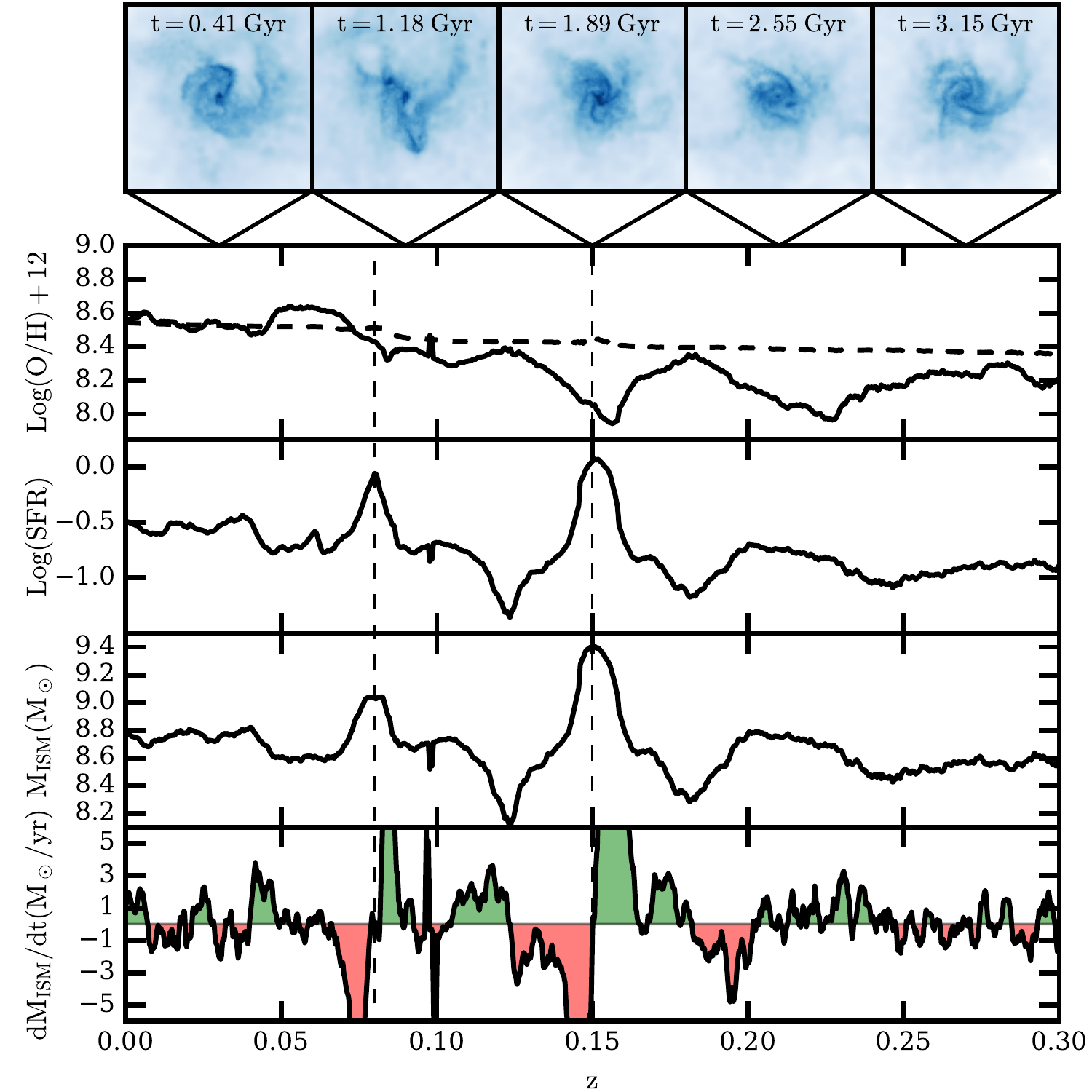}  
\includegraphics[width=0.5\textwidth]{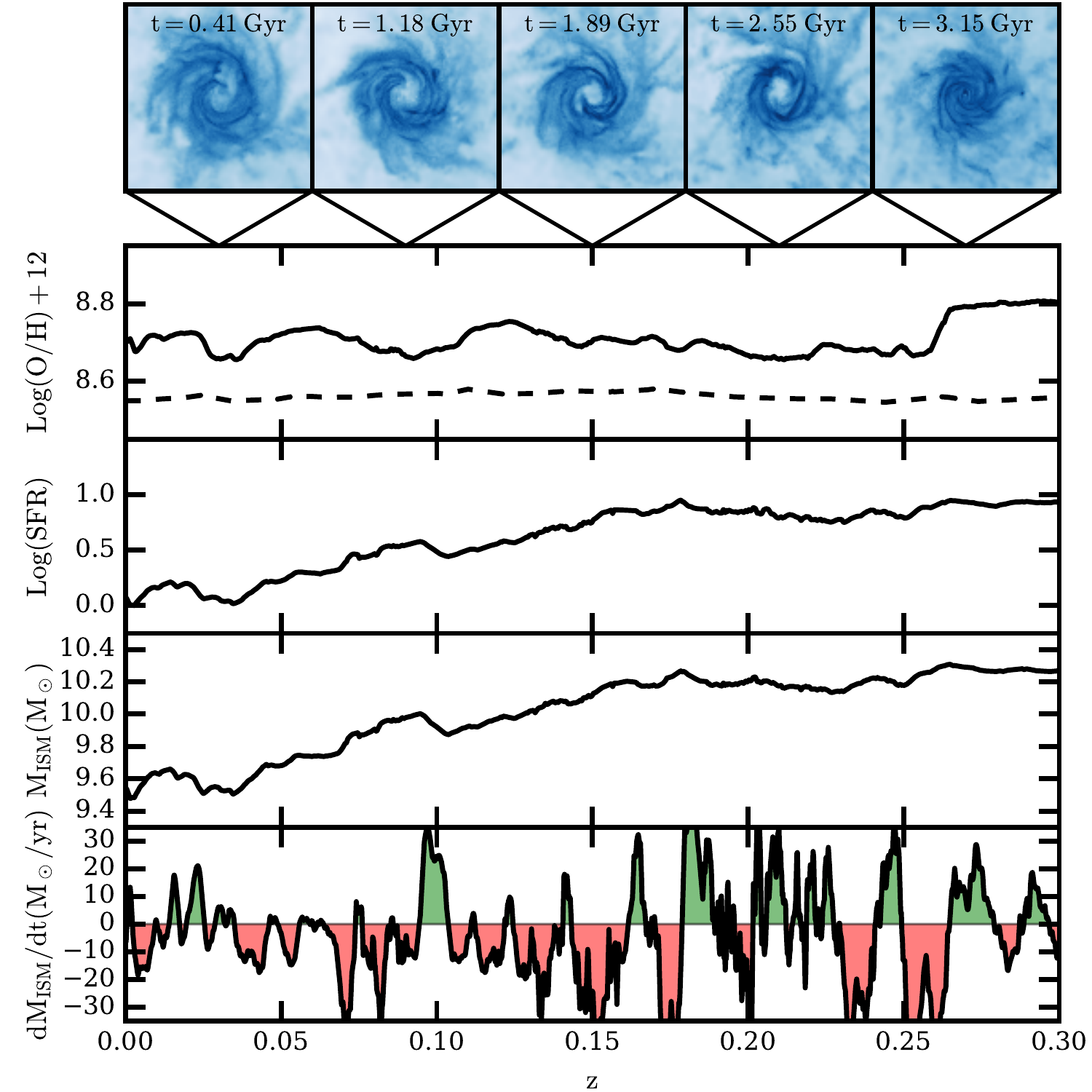}  
}}}
\caption{ 
Example evolution tracks for two galaxies are shown.  
The top postage stamp images indicate the gas-mass distribution at 5 times.
The bottom time series indicates the evolution of the ISM metallicity, 
SFR,
ISM gas-mass,
and net rate of change of ISM gas-mass.
The dashed line in the metallicity evolution time series indicates the mean value for the MZR for a galaxy at that mass and redshift.
The system in the left panel is lower mass ($M_*\approx10^{9} \mathrm{M}_\odot$) than the system on the right ($M_*\approx10^{11} \mathrm{M}_\odot$).
The system on the left undergoes two mergers which are indicated with vertical dashed lines.
   }
\label{fig:full_page}
\end{figure*}

In this Section we explore how galaxies evolve with respect to the MZR.
Figure~\ref{fig:full_page} shows two  detailed examples of how galaxies evolve in time including their gas distribution (top postage stamps), metallicity (top time series), star formation rate (second row), ISM mass (third row), and the rate of change of ISM mass (bottom row) over a period of a few ($\sim$3) Gyrs starting at low redshift.
The evolution tracks have been shaded with red and green to indicate periods of time when the ISM is net increasing or decreasing in mass, respectively.
Changes to the ISM mass can be driven by inflows, outflows, star formation, or stellar mass return.
The dashed line in the metallicity evolution time series indicates the mean value for the MZR for a galaxy at that mass and redshift.
The two systems presented in Figure~\ref{fig:full_page} are not particularly special, but were selected because 
(i) they reside within one of the `subboxes' in the IllustrisTNG simulation that has high time frequency snapshot output and 
(ii) they highlight some of the diverse physical mechanisms that can drive galaxy metallicity evolution.

The left panel of Figure~\ref{fig:full_page} features a relatively low-mass system, with a stellar mass of $M_*\approx10^{9} \mathrm{M}_\odot$.
This system spends most of its early evolution with a metallicity at or below the MZR, which changes very close to redshift $z=0$.
There are two notable galaxy merger events which occur for this system at $z\approx 0.08$ and $z\approx0.15$ (these times have been marked in the figure using vertical thin black dashed lines) that both have easily identifiable impacts on the metallicity, star formation rate, and ISM mass.
Leading up to the merger event, there is an easily identifiable influx or increase of ISM mass which coincides with the decreases in the ISM metallicity owing to metallicity dilution, and increases in the SFR.
This is the same physical picture that has been found in idealized galaxy merger simulations~\citep{Torrey2012a, Scudder2012, Patton2013} and which is supported with observations of close pair galaxies~\citep{SloanClosePairs, Rich2012, Scudder2015}.
Outside of these two discrete events, the metallicity evolution of this system is not described by a single coherent physical picture.
The system undergoes periods of time of perpetuated metallicity increase (e.g., from $0.175 < z < 0.225$) which are broadly associated with sustained high levels of star formation activity, 
but the system also undergoes periods of time of sustained metallicity decrease (e.g., from $0.225 < z < 0.275$) which are not immediately qualitatively distinct from periods of metallicity increase.

The right panel of Figure~\ref{fig:full_page} is similar to the left, but features a relatively massive galaxy, with a stellar mass of $M_*\approx10^{11} \mathrm{M}_\odot$.
In contrast to the left panel, this system does not undergo any significant merger activity over the observed period but is actively experiencing AGN feedback.
The impact of AGN feedback can be seen clearly through the change in the gas distribution between the rightmost and second-to-right gas distribution where a central gas cavity is created and sustained.
The central low density gas cavity is created by removing the central -- highest metallicity -- gas from the ISM.
This drives a sharp drop in the central gas metallicity that can be observed around $z\approx0.26$~\citep{Nelson2017}.  
After this drop, the metallicity evolution of this system is reasonably constant, while the ISM mass and star formation rates drop rather continuously from redshift $z=0.2$ to $z=0$.

One of the main points to take away from Figure~\ref{fig:full_page} is that the metallicity evolution of galaxies is governed by a somewhat subtle competition between gas inflows, gas outflows, and metal enrichment.
We can assert that periods of metallicity decrease are associated with dominant contributions of influxes of lower metallicity gas to the ISM or ejections of high metallicity ISM.
However, outside of merger events the magnitude of gas inflows onto the ISM is set by subtle changes to the ambient CGM properties which itself can be impacted by recent accretion activity, recent outflow activity, feedback from the AGN, or environmental interactions.
This demonstrates the core difficulty associated with modeling galaxy metallicity evolution analytically: galactic metallicity evolution is highly variable and dependent on a number of rapidly varying factors.

\subsection{Characterizing the Nature of ISM Metallicity Evolution Tracks}

\begin{figure*}
\centerline{\vbox{\hbox{
\includegraphics[width=0.3\textwidth]{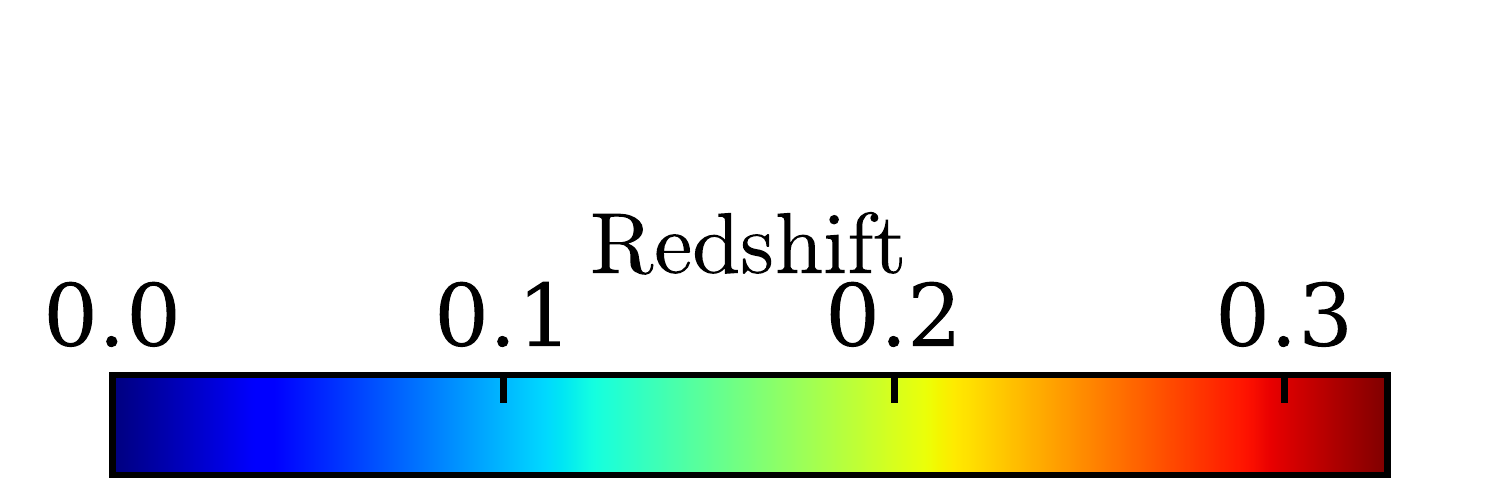}
}}}
\centerline{\vbox{\hbox{
\includegraphics[width=0.2\textwidth]{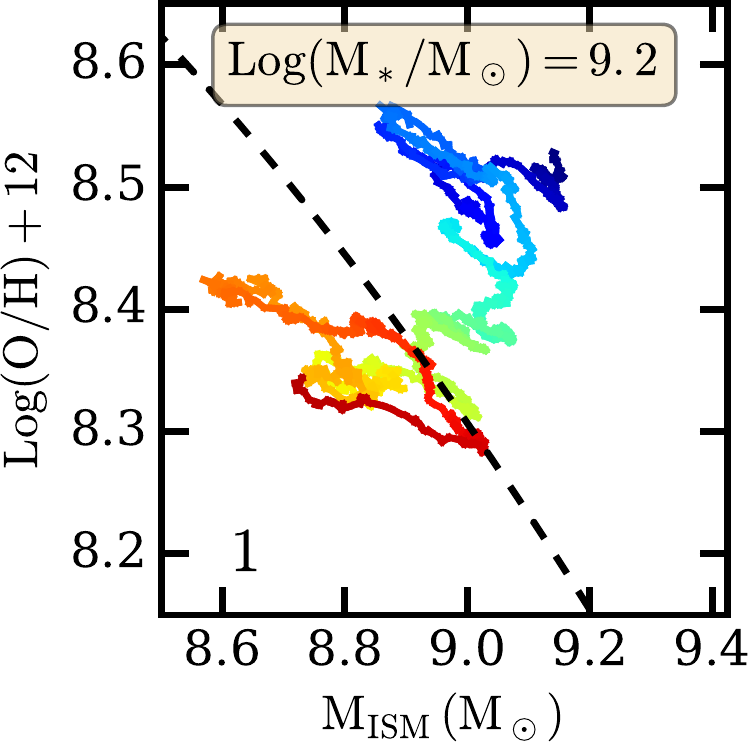}  
\includegraphics[width=0.2\textwidth]{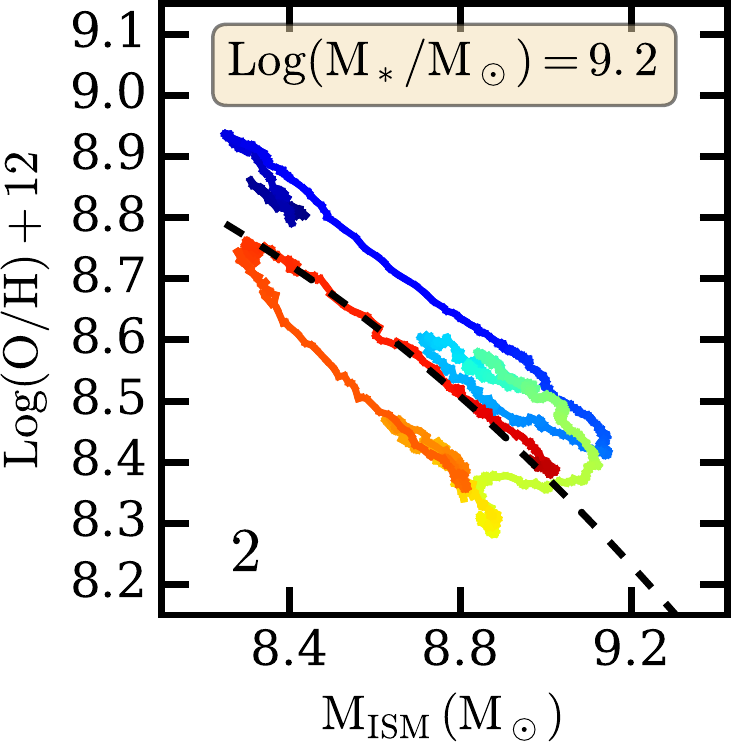}  
\includegraphics[width=0.2\textwidth]{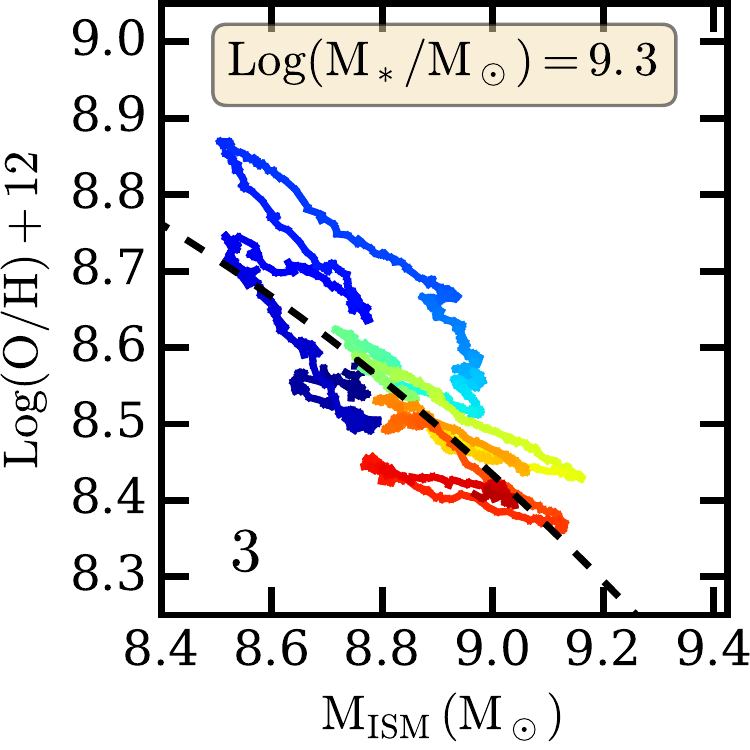}  
\includegraphics[width=0.2\textwidth]{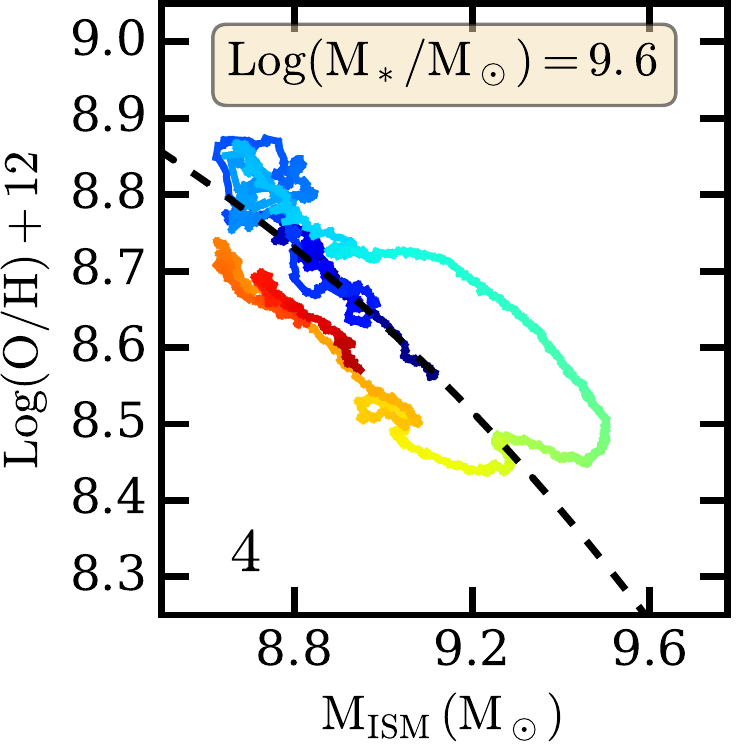}  
\includegraphics[width=0.2\textwidth]{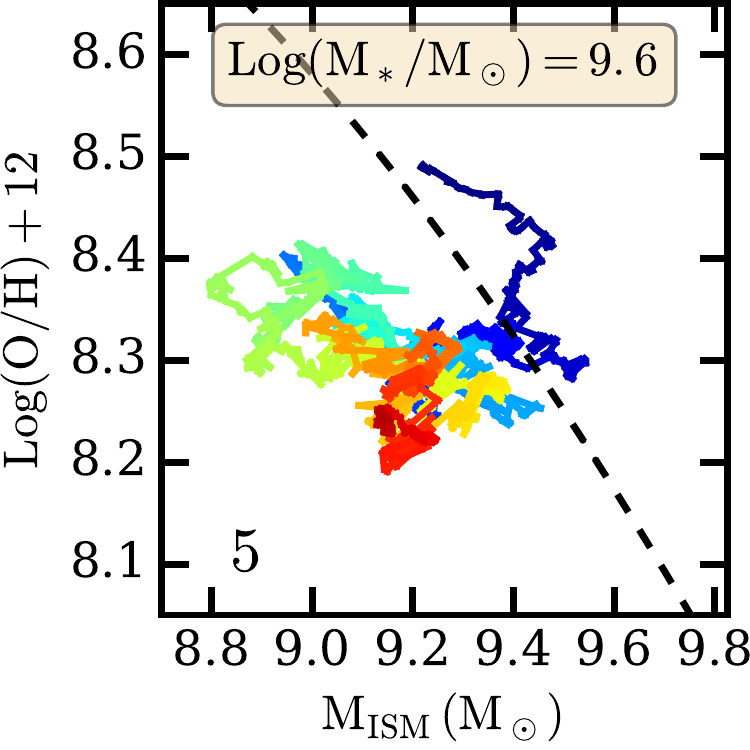}  
}}}

\centerline{\vbox{\hbox{
\includegraphics[width=0.2\textwidth]{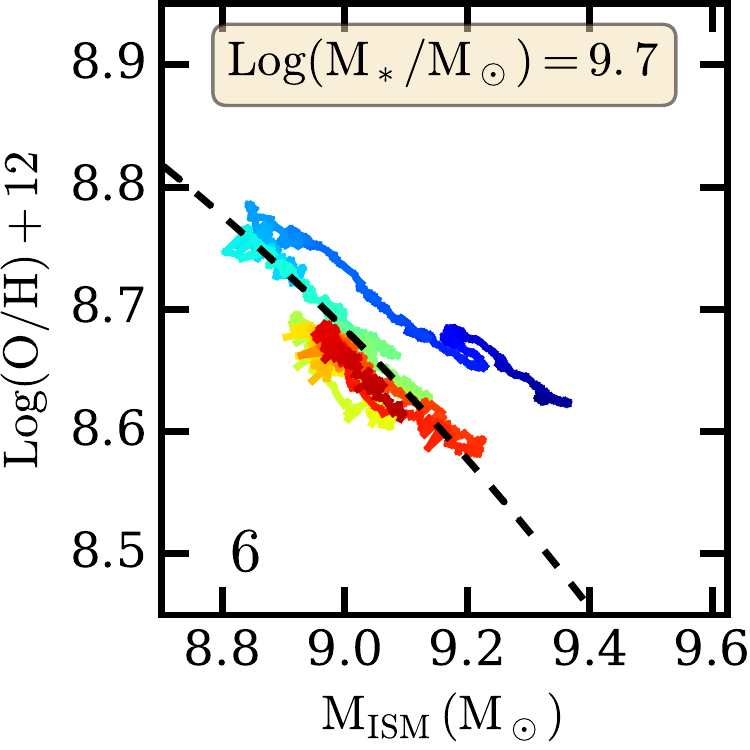}  
\includegraphics[width=0.2\textwidth]{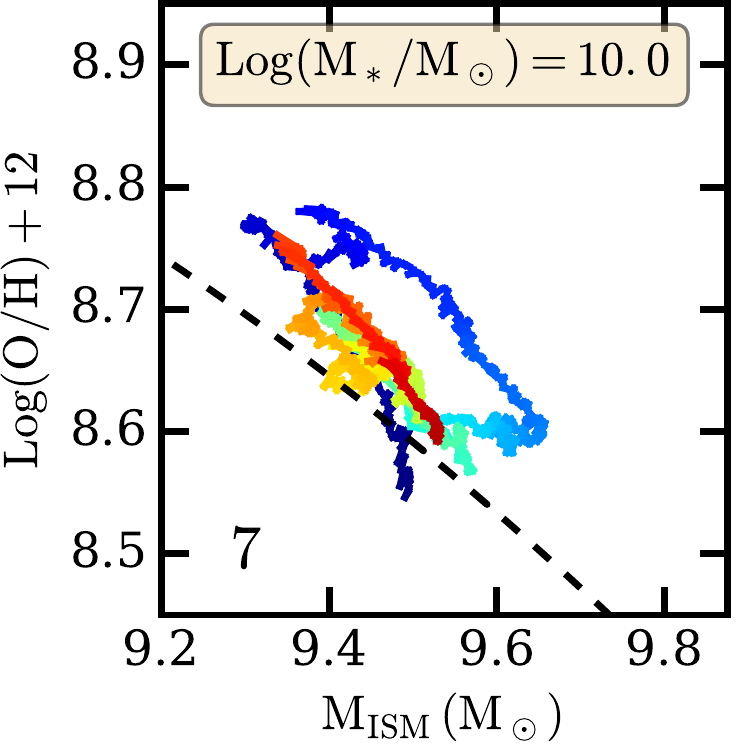}  
\includegraphics[width=0.2\textwidth]{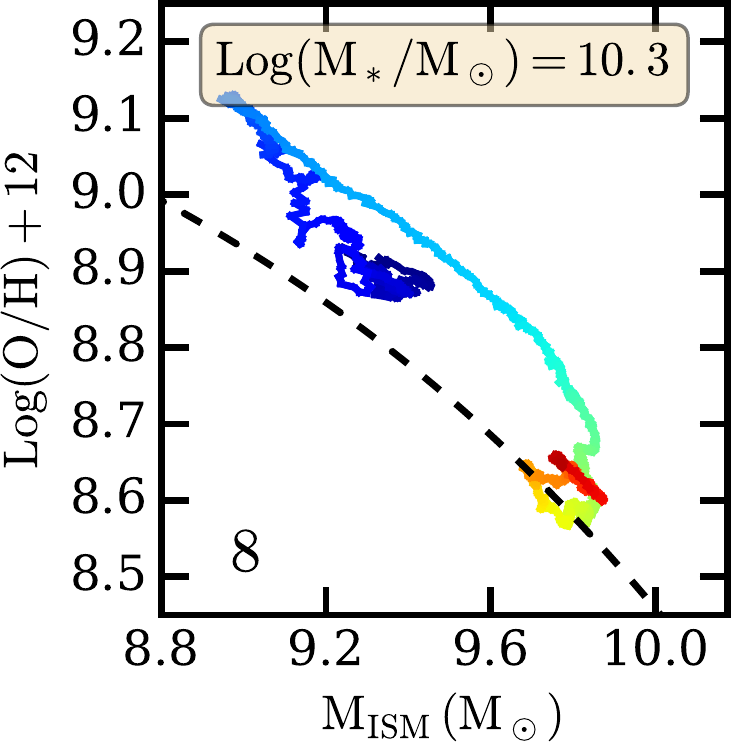}  
\includegraphics[width=0.2\textwidth]{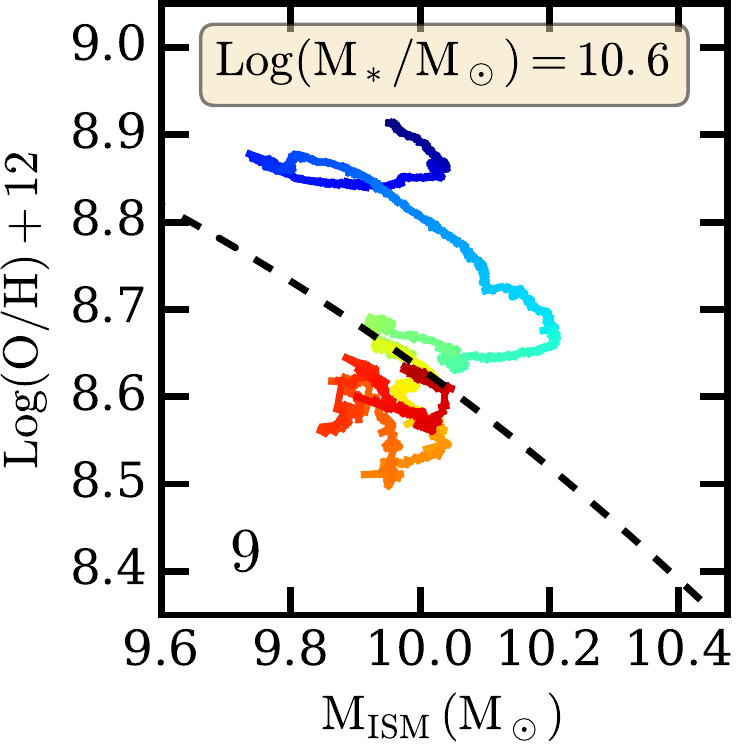}  
\includegraphics[width=0.2\textwidth]{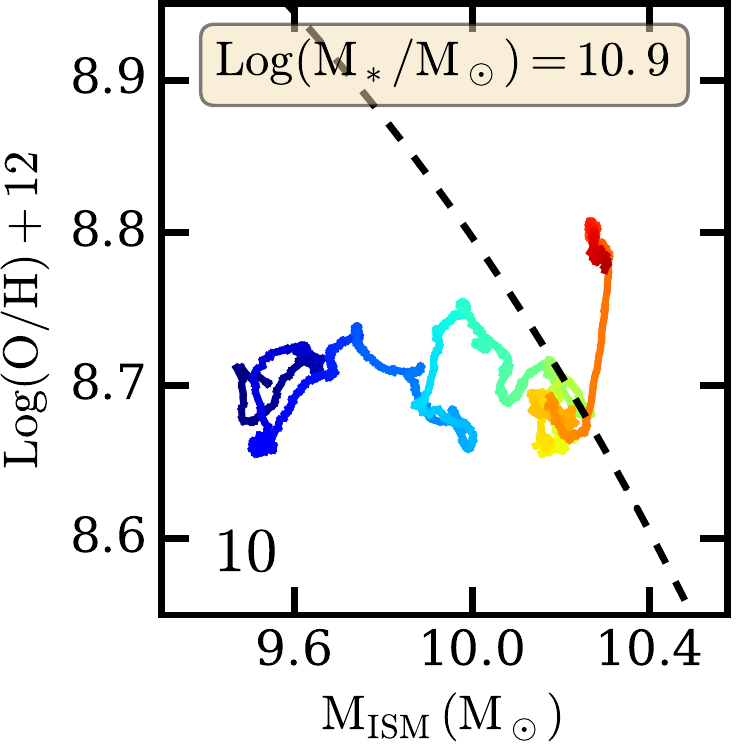}  
}}}
\caption{ Evolution tracks of metallicity and ISM mass are shown as a function of time for ten individual galaxies.  
Tracks in each panel show the evolution from $z=0.35$ (in red) to $z=0$ (in blue).
The black dashed lines indicate closed box model type evolution tracks where we have used an effective yield of $y=0.02$.
Redshift $z=0$ stellar masses for each system are indicated within each plot.
The number in the lower left corner of each panel is assigned for ease of identification.  }
\label{fig:cmb_tracks}
\end{figure*}

To further probe the nature of the metallicity evolution tracks that galaxies follow, 
Figure~\ref{fig:cmb_tracks} shows evolution tracks in metallicity versus ISM mass space for ten individual galaxies tracked in time using {\small SUBLINK}~\citep{RodriguezGomez2015}.  
For clarity of discussion, each panel is labeled with a number in the lower left corner.
Evolution tracks in Figure~\ref{fig:cmb_tracks} can help diagnose the relative importance of inflows/outflows -- which will have a direct impact on the ISM mass -- versus star formation and chemical enrichment.
The evolutionary tracks within each panel are colored to go from red at $z=0.35$ to blue at $z=0$.  
The dashed black line indicates a closed box model line of $Z = - y \; \mathrm{ln(} M_{\mathrm{ISM}} / (M_{\mathrm{ISM}} + M_*) \mathrm{)}$ where we adopt a reduced yield of $y=0.02$ and use a fixed stellar mass of $M_* = M_*(z=0)$ for each system -- which is a reasonably good approximation for most of the systems shown over the plotted redshift range.

The first point to take away from Figure~\ref{fig:cmb_tracks} is that galaxies can move in any direction in metallicity versus ISM space.  
Each galaxy has a truly unique track which is, in general, non-monotonic both in terms of ISM mass evolution and metallicity evolution.
While the overall galaxy population shows a strong correlation between metallicity and ISM mass (see, e.g. Figure~\ref{fig:secondary_correlations_mism}), individual galaxy evolution tracks do not follow this trend with any obvious regularity.

However, Figure~\ref{fig:cmb_tracks} does allow us to identify some regular features in the individual galaxy metallicity evolution tracks.  
One such trend is that there are identifiable periods of time where individual galaxies closely follow closed box model evolution tracks.
For example, panels 2, 8, and 9 all feature galaxies that evolve along, or directly parallel to, the closed box model for a Gyr or longer.
Closer inspection of these systems reveals that they are evolving like the closed box model tracks because they are in relative isolation, with modest or low accretion rates onto the ISM, and strong ongoing star formation.  

Systems can be driven off closed box model tracks by rapidly changing the properties of the ISM gas reservoir.
The ISM gas reservoir can be perturbed by merger events (as was demonstrated in Figure~\ref{fig:full_page}). 
However, we find that most of the non closed box model evolution is more subtle in nature and not associated with distinct merger events.
We find that perturbations to the central gas disk and outflows associated with feedback are the two primary drivers of non closed-box-model behavior.
Bars or clumps that form in the central gas disk have the ability to perturb an otherwise stable system, which can cause the central gas disk to torque down the gas around it.
When this occurs, the central gas density rises and the ISM mass increases, forcing systems to the right in Figure~\ref{fig:cmb_tracks}.
At the same time, this process of gas inflow/accretion can dilute the ISM metallicity depending on the metal content of the newly acquired ISM material.
The behavior is demonstrated, e.g., at early times in panel 6, intermediate times in panel 2, or late times in panel 4.

Outflows, on the other hand, predominately function to drive down the ISM mass.
Panel 10 of Figure~\ref{fig:cmb_tracks} shows the same system featured in the right panel of Figure~\ref{fig:full_page} which is dominated by AGN feedback over the time period shown.
As a result of the AGN feedback driving out ISM material, this system gradually moves toward lower ISM masses without very systematic or strong changes to the ISM metallicity.

The central complication with deciphering the evolution tracks in Figure~\ref{fig:cmb_tracks} is that -- while there are periods of clearly understood evolution driven by distinct physical mechanisms -- much of the evolution of these systems is dominated by a combination of multiple effects.
In general, systems are experiencing accretion onto the ISM, star formation, enrichment, and outflows simultaneously.
It can be broadly argued that in many cases galaxies appear to oscillate between accretion driven movement to the lower right, followed by star formation and enrichment dominated movement to the upper right.
However, any claim along these lines requires a more statistical characterization of the distribution and movement of galaxies in this space.

\begin{figure*}
\centerline{\vbox{\hbox{
\includegraphics[width=0.5\textwidth]{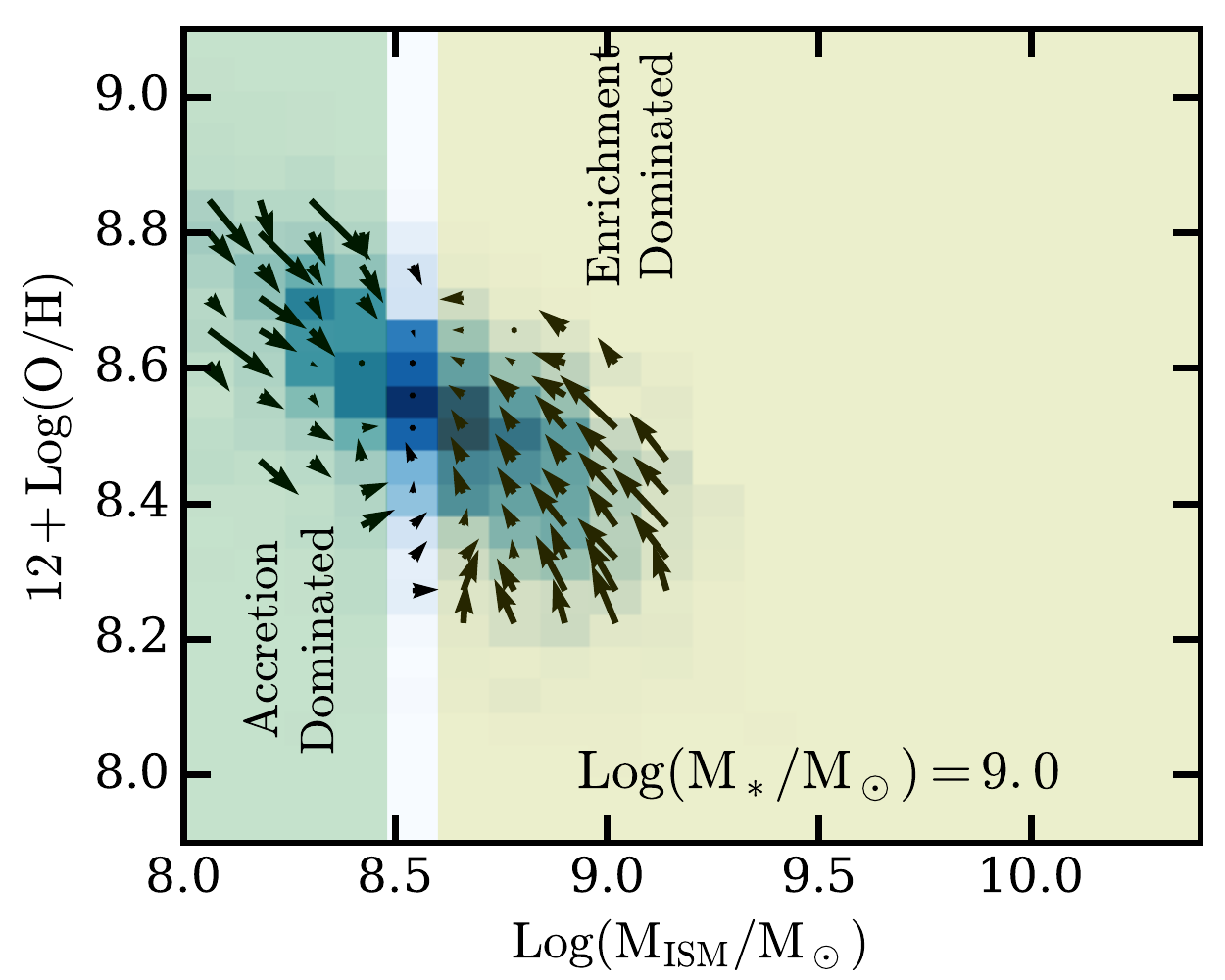}  
\includegraphics[width=0.5\textwidth]{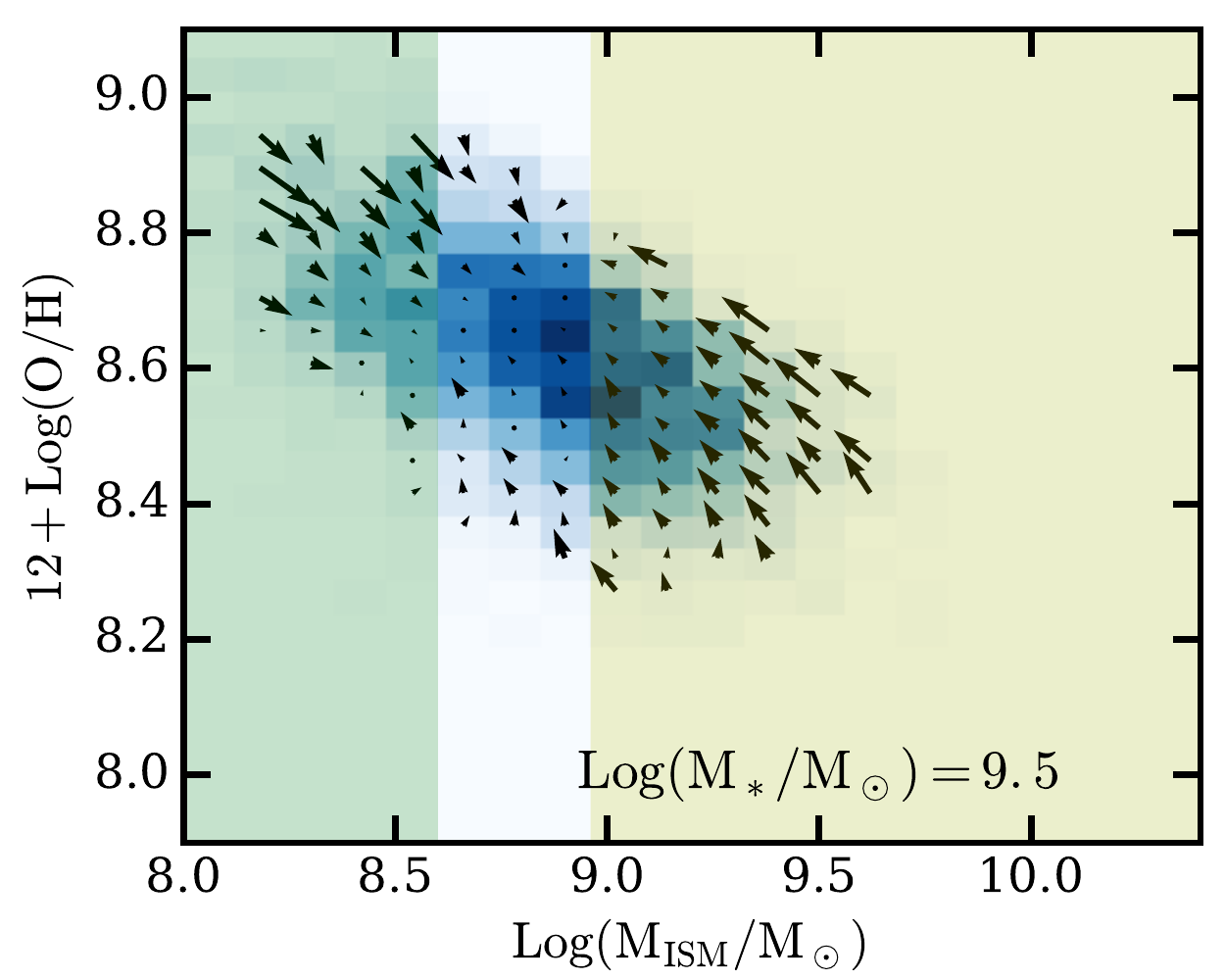}  
}}}
\centerline{\vbox{\hbox{
\includegraphics[width=0.5\textwidth]{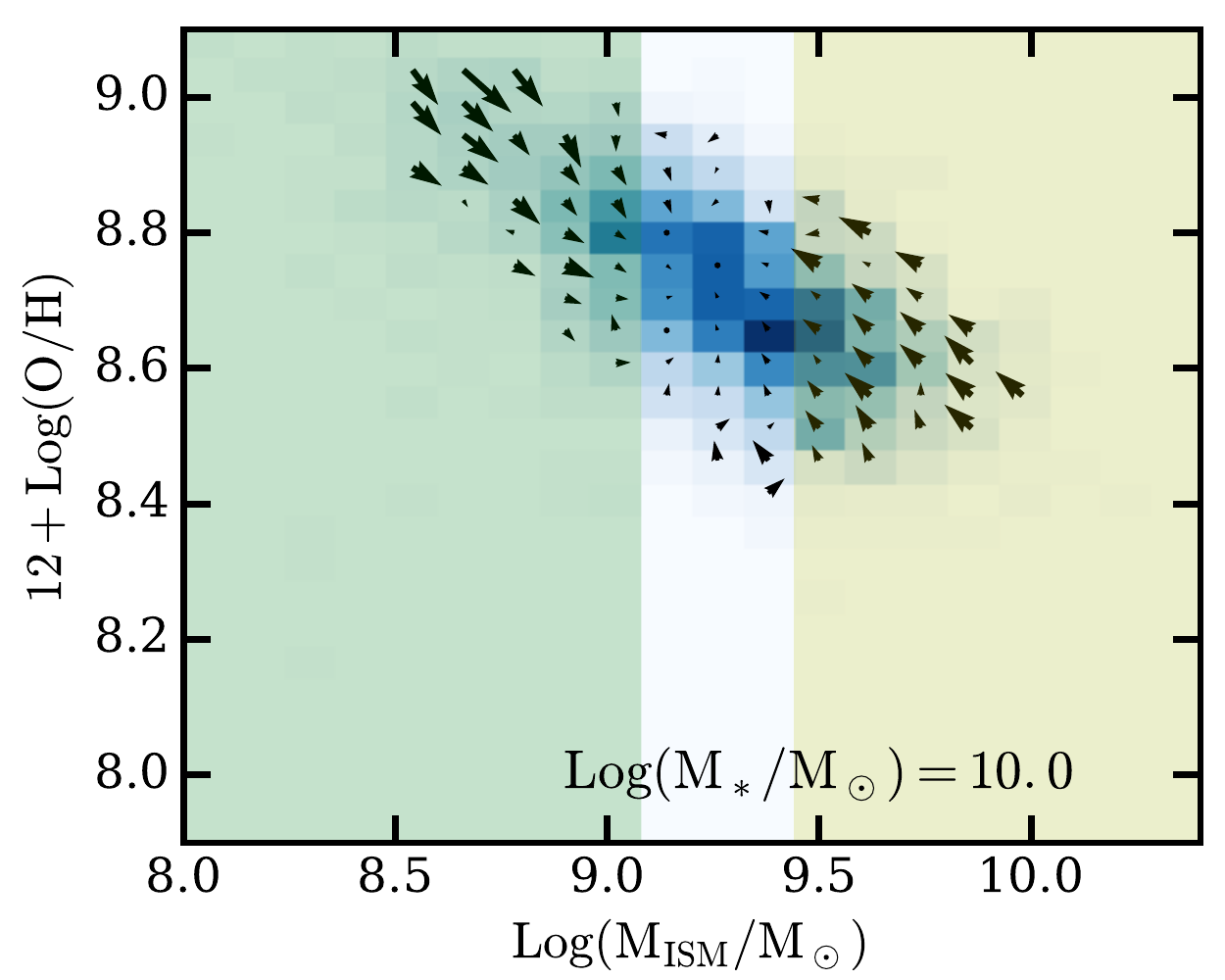}  
\includegraphics[width=0.5\textwidth]{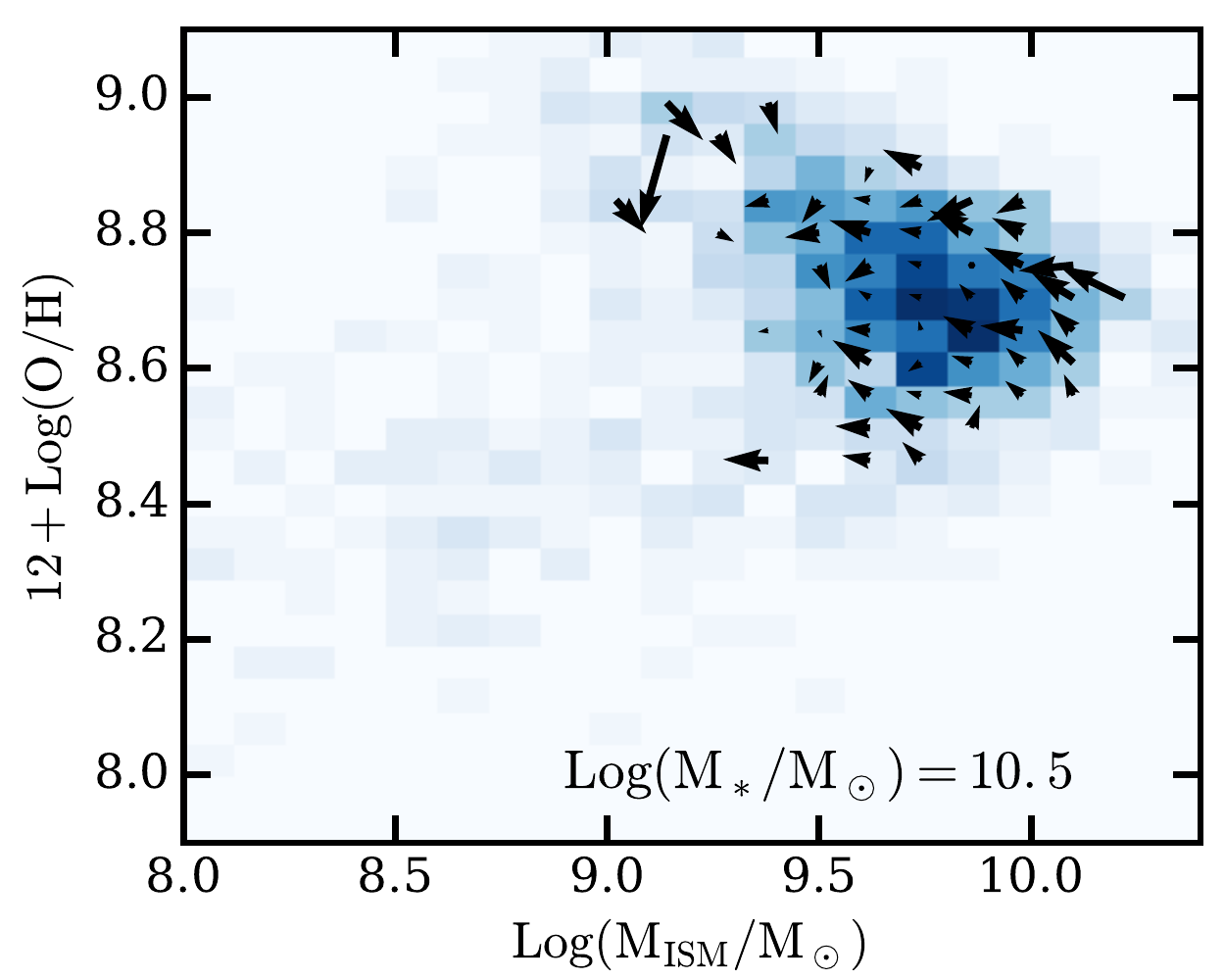}  
}}}
\centerline{\vbox{\hbox{
\includegraphics[width=0.5\textwidth]{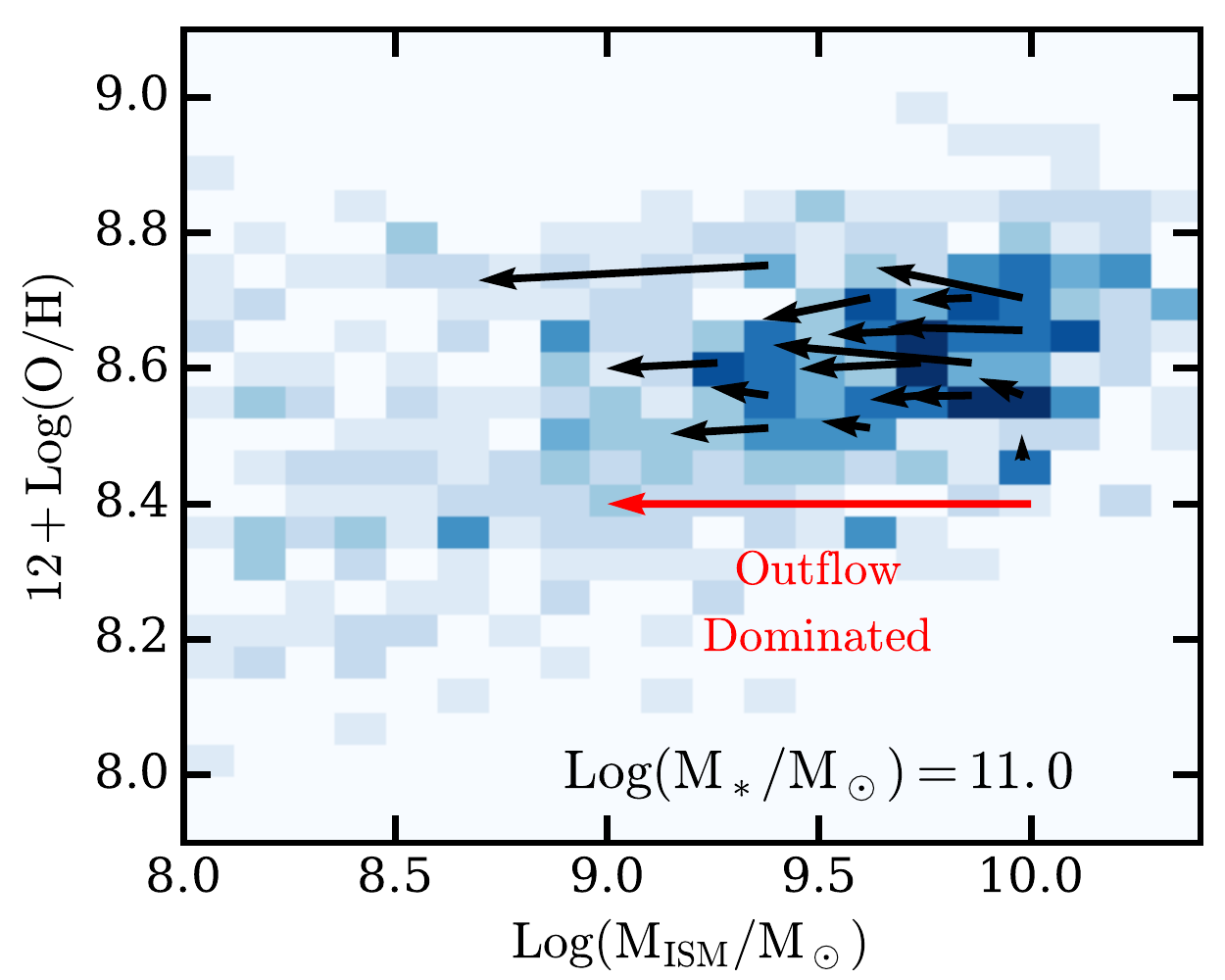}  
}}}
\caption{ 
The distribution and evolutionary trends of galaxies in metallicity versus ISM mass is shown for several stellar mass bins, each of width 
$\pm 0.25$ dex.
The background two-dimensional histograms indicate the distribution of galaxies at $z=0.058$, while the vectors indicate the average evolution direction and speed within this space calculated by tracking galaxy movement over the redshift range $0 < z < 0.058$.
Clear patterns can be identified for the movement of galaxies in the lowest three mass slices.
The first three panels (lowest three mass slices) include green and yellow color coded regions which indicate regions in this space where galaxies are predominately evolving owing to fresh gas accretion and star formation driven enrichment.  
In contrast, the bottom panel (highest mass bin) includes only a single arrow pointing to the left which indicates the direction galaxies move when they are outflow dominated.
 }
\label{fig:fmr_slice_evolution}
\end{figure*}

Figure~\ref{fig:fmr_slice_evolution}  shows the net distribution and evolution for an ensemble of galaxies in metallicity versus ISM mass space.
Each panel shows a two-dimensional histogram indicating the distribution of galaxies as well as a series of vectors indicating the average evolution direction and speed for galaxies in this space.
The distribution of galaxies is shown at redshift $z=0.058$, and the evolution direction and speed is calculated by tracking each galaxy forward to $z=0$.  
The average evolution direction and speed is taken to be the mean ISM mass change and mean metallicity change for all galaxies residing in each pixel.
Arrows are only indicated for pixels containing at least 5 galaxies.

For the lowest mass bin (upper left) a clear anti-correlation can be seen between the ISM metallicity and ISM mass with galaxies with higher ISM masses having lower ISM metallicities.
Examining the vector distribution we find that galaxies are moving back toward a central ISM mass and ISM metallicity value from both ends of the distribution when averaged over the full population.
This restoring movement can be interpreted based on the evolutionary tracks explored  in Figures~\ref{fig:full_page} and~\ref{fig:cmb_tracks}.
Galaxies sitting in the lower right portion of this plot which have large ISM gas masses have large star formation rates (see Figure~\ref{fig:full_page}) and are therefore most likely to be enrichment dominated and possibly moving along closed-box-model type trajectories (as demonstrated in Figure~\ref{fig:cmb_tracks}).
The right  portion of the plot has been shaded yellow to indicate this ``enrichment dominated" portion of the galaxy population.
Conversely, galaxies in the upper left portion of this plot have comparatively low ISM masses, comparatively low star formation rates, and therefore comparatively low enrichment rates.
However, the low ISM gas-mass reservoirs in these systems make them particularly susceptible to influence from gas inflow events.
The restoring force for galaxies in the upper left  portion of this plot is accretion of new gas from the CGM, which will simultaneously drive the ISM to lower metallicities and higher ISM masses.
The left portion of this plot has been shaded green to indicate the regions where the ISM mass and metallicity evolution is accretion dominated.

Similar trends can be seen for the upper-right and middle-left panels, which display  higher mass bins.  
There are clearly identifiable trends for galaxies in the enrichment and accretion dominated regimes, although the median ISM mass and metallicity for the overall population shifts to somewhat larger values with increasing galaxy mass.
These trends, however, become less clear as we move to the middle-right and bottom panels, which display the two highest mass bins.  
In the bottom panel, we find that there is not a visible residual trend between ISM mass and metallicity and galaxies are moving horizontally to the left.
This trend for these highest mass galaxies is dominated by AGN driven outflows which drive down the ISM gas mass without significantly diluting or enriching the ISM metallicity.
The middle-right panel is a transitionary population of galaxies, which shows some indication of accretion and enrichment, but where the trends are less pronounced owing to the influence of an increasingly impactful AGN contribution.

\section{Discussion}
\label{sec:Discussion}

In this paper we have provided a top-down examination of the metal distribution and evolution for the TNG100 galaxy formation simulation.
As in the original Illustris simulation, the IllustrisTNG feedback model was refined predominately based on matching stellar mass functions and the cosmic star formation rate density~\citep[black hole mass to galaxy or halo mass relation, halo gas fractions, and galaxy sizes were also considered;][]{Pillepich2017, Weinberger2017}.
The stellar and AGN feedback that was employed was tuned to help galaxies regulate their stellar mass growth accordingly. 
However, the same feedback that is used to regulate the stellar mass growth of the simulated galaxy population has a distinct impact on the distribution of metals in and around galaxies.
Examining the metal distribution is therefore an interesting test of the model as it is both independent of the core constraints used to tune the IllustrisTNG model, and sensitive to the feedback strength and implementation.

\subsection{High Redshift Evolution of the MZR}
\begin{figure}
\centerline{\vbox{\hbox{
\includegraphics[width=0.5\textwidth]{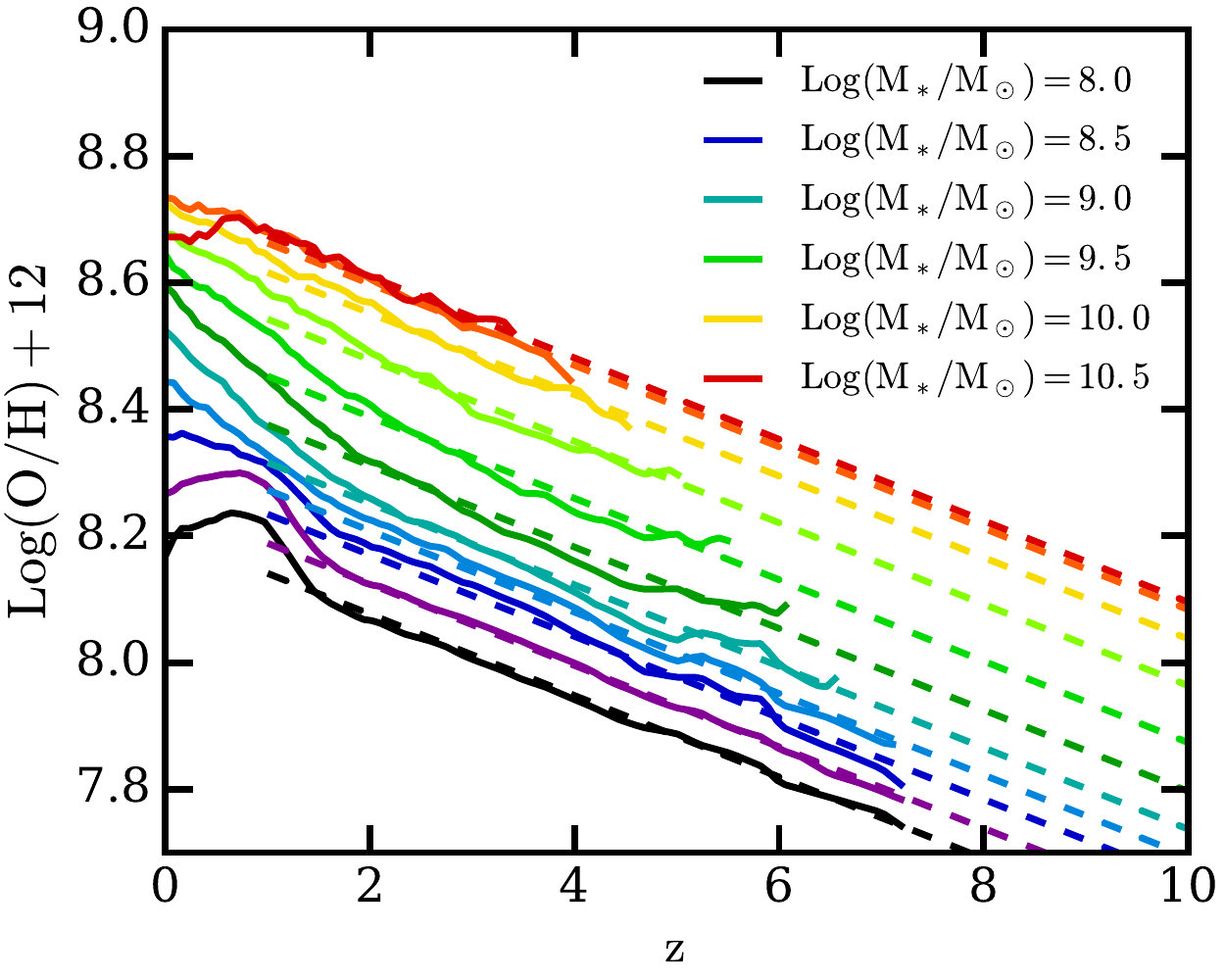}  
}}}
\caption{ Evolution of galactic metallicity as a function of redshift for several mass bins, as indicated in the legend.  
Solid curves indicate the metallicity as determined from the simulation, while dashed lines indicate best fits and extrapolations to the data.
We find a best fit slope of $Z \propto  - 0.064 z$ which applies reasonably well for $z>2$ across the full mass range resolved in our simulations.  }
\label{fig:predictions}
\end{figure}

Our simulations make clear predictions for the continued evolution of the MZR out to high redshift which will likely be testable in the near future with JWST.
The OIII, NII, H$\alpha$ nebular emission lines will shift into JWST NIRSPEC instrument for $ z \gtrsim 4$, which will likely lead to a large influx of metallicity data for these high redshift galaxy populations.
Our models predict that the gradual decrease in MZR normalization that has currently been observed out to redshift $z\sim2$ will continue out to higher redshift yet.
Figure~\ref{fig:predictions} provides some guidance on the magnitude of this expected continued evolution.
The solid curves in Figure~\ref{fig:predictions} show the metallicity evolution as a function of redshift for several mass values.
The dashed lines show the best fit lines determined via a chi-squared minimization on the full dataset for $z>1$ of $Z = Z_0 - 0.064 z$.
Predictions for metallicity of galaxies out past $z>6$ are dependent on the nature of our adopted extrapolation, but clearly point toward continued decline in the metallicity for all galaxy mass bins.
By redshift $z=8$ we expect that galactic metallicity will be $\sim0.5$ dex lower than redshift $z=0$ galaxies of the same mass.

We note that our prediction of a decreasing metallicity for the highest mass bins discussed here is already somewhat in tension with claims in the literature that the saturation metallicity for the MZR is redshift independent~\citep{Zahid2014}.
\citet{Zahid2014} argue that the existing observational MZR data out to $z=1.55$ is accurately fit with a single, unchanging saturation metallicity.
The simulated saturation metallicity, on the other hand, clearly evolves significantly as can be seen in Figure~\ref{fig:predictions} (or in Figure~\ref{fig:mz_relation}).
Importantly, however, a direct comparison of the simulated MZR relations and the data explored in \citet{Zahid2014} shows that the two datasets agree reasonably well, with the strongest point of tension being whether and where one defines the saturation metallicity based on sparsely populated high-mass high-redshift data points.
Our simulations agree with the \citet{Zahid2014} data in the sense that neither the simulated MZR nor the observed MZR saturation metallicity evolve significantly over the redshift range $0<z<1.5$, but our models do predict that there is an evolution of the MZR saturation metallicity and that this evolution will become more pronounced at increasingly high redshift.
We expect that this prediction can be (in)validated in the coming years.

The driving force behind the evolution in the normalization in the MZR has been subject to widespread debate.
Galactic metallicity trends with redshift or mass have been explained with competing scenarios based on changes to the metal retention efficiency or changes to the gas fractions of galaxies.
Our models indicate that the ISM metal retention efficiency (which is driven by feedback) is \textit{not} responsible for the redshift evolution of the MZR in our models.
While our models indicate a decrease in the normalization of the MZR with increasing redshift, 
the ISM metal retention efficiency (as presented in the ``effective yields" in Figure~\ref{fig:yield_evo}) \textit{increases} with increasing redshift which should give rise to an increasing metallicity.
Changes in the ISM effective yields or metal retention efficiency are not responsible for the lowered MZR normalization at high redshift in our model.

\begin{figure}
\centerline{\vbox{\hbox{
\includegraphics[width=0.5\textwidth]{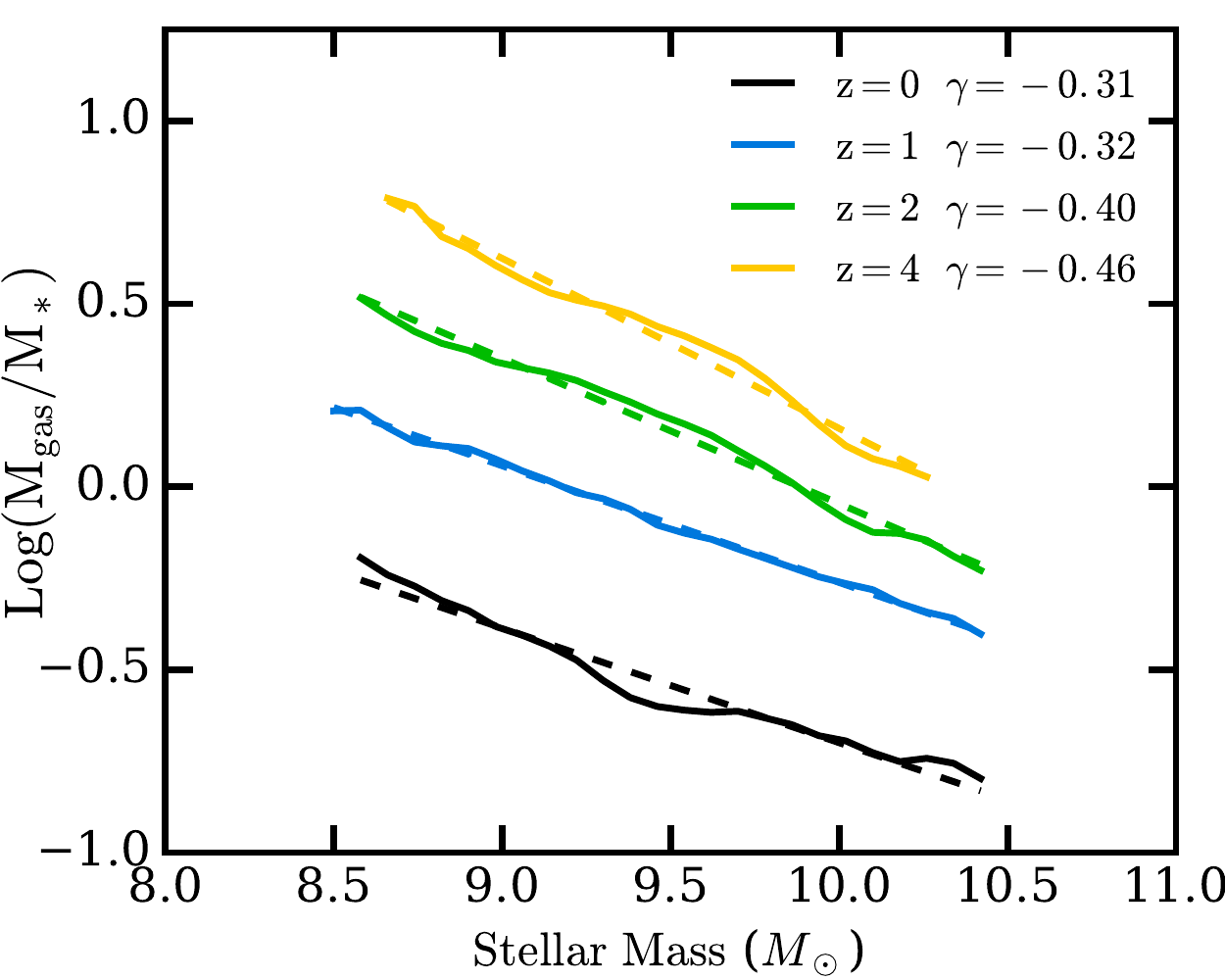}  
}}}
\caption{ Gas fractions are shown as a function of stellar mass for four distinct redshifts for the simulated IllustrisTNG galaxies.
We fit each redshift independently and show the slopes of the best fit $M_{\mathrm{gas}}/M_* \propto  M_* ^\gamma$ line in the legend.
Higher redshift galaxies have significantly higher gas fractions which impacts the MZR normalization evolution.  }
\label{fig:gas_fractions}
\end{figure}

Instead, the increasing gas fractions of high redshift galaxies are the primary driver behind the redshift evolution of the MZR.
In addition to demonstrating that part of the scatter about the MZR is driven by gas fraction, 
Figure~\ref{fig:mz_relation_ism_mass} indicated that higher redshift galaxy populations have on average higher gas fractions and
Figure~\ref{fig:gas_fractions} shows directly the gas fraction (defined here as $M_{\mathrm{gas}}/M_*$) as a function of stellar mass for several redshifts.
There are two trends present:  (i) At a fixed redshift, gas fraction decays with stellar mass and (ii) at a fixed mass, higher redshift galaxies have higher gas fractions.
We have found best fits to the gas fraction dependence on stellar mass using a power law form $M_{\mathrm{gas}}/M_* \propto  M_* ^\gamma$~\citep[same as in][]{Peeples2011} where the best fit values of $\gamma$ fall in the range $\gamma = -0.3$ to $\gamma = -0.5$.
In general the relation steepens with increasing redshift.
The redshift $z=0$ slope of $\gamma=-0.31$ is shallower than the published observed gas fraction values of $\gamma\approx-0.5$~\citep[e.g.,][]{Peeples2011}.
However, this offset in slope values should be regarded with caution since the observed gas fractions are determined from, e.g., HI gas-mass measurements, which is expected to scale in the same way but is not directly comparable to the gas-mass values derived from the simulation shown in Figure~\ref{fig:gas_fractions}.

The main focus of Figure~\ref{fig:gas_fractions} for this paper, however, is the evolution with redshift.
The evolution in the gas fraction at a fixed stellar mass is significant, with gas fractions for the redshift $z=1$ ($z=4$) simulated galaxy population being $\sim0.5$ dex ($\sim1$ dex) higher than the redshift $z=0$ galaxy population.
While the effective yields (i.e. the metal retention efficiency) is increasing toward high redshift, the ISM metallicity is dropping with time, which is explained by the increasing gas fractions.
This point is important to stress because some models that explain the shape of the MZR as an instantaneous competition between metal production and gas inflows do not include a dependence on gas-mass or gas fraction~\citep[e.g.,][]{Finlator2008, Dave2011_met}.
For example, in the~\citet{Finlator2008} ``equilibrium" MZR model the gas inflow rates are assumed to exactly balance the gas consumption/outflow rates such that
$\dot M_{\mathrm{acc}} = (1 + \eta) \dot M_*$.
This assumption implicitly removes the gas-mass dependence associated with ISM metallicity.
Equilibrium MZR models are inviting because they can successfully explain the shape of the MZR and even capture modifications to the shape of the MZR under varied simulation feedback conditions~\citep{Dave2011_met}.
However, they do not yield a clear mechanism for the redshift evolution of the MZR, nor do they explain the role that gas fraction or SFR might play in driving scatter about the MZR.
Our simulations indicate that the evolution of gas-mass (or gas fraction) with time, and even variations in the gas fraction at a fixed redshift are important shaping factors in the MZR that need to be accounted for in MZR models.

A natural extension of the~\citet{Finlator2008} equilibrium model is the ``regulator" style models~\citep[e.g.,][]{Lilly2013, Forbes2014} where the gas-mass of the system is time-variable.
In these models, it is not required that the inflow rates and gas consumption/outflow rates are exactly equal.  
Instead, the gas-mass in the system is assumed to evolve according to
\begin{equation}
\frac{\mathrm{d}M_{\mathrm{ISM}}}{\mathrm{d}t} = \dot M_{\mathrm{acc}} - (1 + \eta) \dot M_*
\end{equation}
where the accretion rate is set by external factors and the star formation rates are set by the gas-mass available for star formation and a star formation consumption timescale.
While increases in the accretion rate can drive up the gas reservoir which will result in higher star formation rates, the gas reservoir mass is allowed to change for galaxies at a fixed redshift and mass, as well as for galaxies as a function of redshift.
The resulting general predicted equilibrium metallicity within the regulator models is
\begin{equation}
\mathrm{Z_{eq}}  = Z_{\mathrm{acc}} + \frac{ y }{1 + 2 \eta  +  f_{g} + 2 \frac{\mathrm{d} \mathrm{ln} f_g}{\mathrm{d}t} \tau_{\mathrm{ISM}}  }
\label{eqn:regulator_solution}
\end{equation}
where $Z_{\mathrm{acc}}$ is the metallicity of accreted gas, $ f_{g} = M_{\mathrm{ISM}}/M_*$ is the gas fraction, and $\tau_{\mathrm{ISM}} =  M_{\mathrm{ISM}} / \dot M_* $ is the  gas consumption timescale.\footnote{In comparison to~\citet{Lilly2013} we have assumed the fraction of mass returned from stellar populations is a fixed value of $R=0.5$.}
There are several factors in Equation~(\ref{eqn:regulator_solution}) that change as a function of mass and redshift including the accreted metallicity, mass loading factor, and gas fraction.
The explicit dependence on gas fraction is consistent with the redshift evolution trend found in our models.
The last term in the denominator of Equation~(\ref{eqn:regulator_solution}) governs non-equilibrium changes in a galaxies gas fraction which can explain residual correlations between metallicity and gas fraction about the MZR.

\subsection{Testing Galaxy Feedback Models with Metallicity}
Although metals are produced via stars that formed from dense gas, the majority of metals live outside of the ISM.
We demonstrated the strong redistribution of metals through global (Figure~\ref{fig:global_phase_diagrams}) and galactic (Figure~\ref{fig:galaxy_phase_diagrams}) phase diagrams, as well as by tabulating the ``effective yield" for the ISM, CGM, and stars.
Some of the metal redistribution can be driven by naturally occurring galaxy dynamics including galaxy mergers and interactions, but the majority of the redistribution of metals is driven by the feedback we have adopted in our simulations.
Feedback pumps metal enriched gas out of the dense ISM into the hot/diffuse CGM around galaxies and is able to eject some fraction of metals out of their haloes all together.
The metal yields quoted in Table~\ref{table:effective_yields} show a significant fraction ($\sim10-20$ per cent) of a galaxy's metal content has been ejected from the halo, across a wide mass range.
The fraction of ejected metal-mass increases for massive haloes to $\sim50$ per cent, which is driven in part by the increased impact of AGN feedback~\citep{Weinberger2017b, Nelson2017}.
In our model, high-mass galaxies are more efficient at ejecting their metal-mass compared with their low-mass counterparts.

Although the central focus of this paper has been on the MZR and metal content of the ISM of galaxies, 
the widespread distribution of metals make observations of \textit{both} the ISM metallicity (e.g. through nebular emission line measurements) \textit{and} CGM metal content (e.g. through quasar absorption line studies) critical toward understand the coevolution of galaxies and their metal content.  
The MZR found in our simulations scales reasonably well with current observational constraints. 
However, this does not imply that the IllustrisTNG model is a unique ``correct solution" to how the MZR is shaped and evolves with redshift.
As we have argued previously, there is a degeneracy between the ISM metal retention efficiency and changes in galactic gas fractions as a function of mass and redshift.
This degeneracy can be in part broken by 
(i) considering more careful comparisons of the galactic gas fractions and/or
(ii) comparing the total metal reservoir found in the CGM against observations.
We leave further consideration of these points to a future exploration~(Nelson et al. in prep).

We note the trend found in our models as a function of redshift for the hot CGM metal retention efficiencies found in Figure~\ref{fig:yield_evo} indicate a slow redshift evolution, changing only by a factor of $\sim2$ out to high redshift.  
This is likely a product of our wind feedback modeling, which itself is only a weak function of redshift.
The slowly evolving hot CGM metal retention efficiency does not necessarily imply a slowly evolving metallicity nor slowly evolving covering fraction for any particular species.
Instead, it simply implies that the hot CGM in our models hosts a fraction of the total metal budget that slowly evolves.
Comparisons with metallicity and/or covering fraction calculations would require further knowledge of the total CGM gas-mass (i.e. how much pristine gas the enriched gas is diluted among) and halo size~(Nelson et al. in prep).

The sensitivity of the metal redistribution to the presence of feedback makes the effective yields discussed in Section~\ref{sec:Results1} important predictions that should be checked against other theoretical models and observations, where possible.
We expect that a more careful examination of the global metal budget could reveal areas of more clear tension between the IllustrisTNG results and observations, which may point to areas where feedback modeling may need to be modified.

\subsection{Is there a fundamental metallicity relation?}
The existence, or lack of existence, of a fundamental metallicity relation has been a point of contention in the literature~\citep[e.g.][]{Kashino2016, Telford2016}.
It remains unclear whether the initial claims about correlated scatter about the MZR are a fundamental property of galaxies~\citep{Salim2014, Brown2017}, or whether these correlations are driven by observational systematic biases in metallicity determinations~\citep{Sanchez2013, BarreraBallesteros2017}.
This debate will likely continue as methods of metallicity determination are refined and larger high redshift galaxy datasets are obtained~\citep[e.g.][]{Steidel2014, Maier2014}.
Using our simulations, however, we can discuss the physical motivation for the fundamental metallicity relation.

We find clear evidence in our models for the existence of a correlation between the scatter in the MZR and galactic gas-mass (see Figure~\ref{fig:mz_relation_ism_mass}).
A similar correlation exists between the scatter in the MZR and galactic SFR.
We explored the role that gas-mass -- not star formation rate -- plays in explaining scatter about the MZR because the increased (decreased) SFRs follow as a consequence of the increased (decreased) gas masses.

This existence of correlated scatter about the MZR persists out to high redshift, with limited changes in the slope of the correlated scatter (see Figure~\ref{fig:secondary_correlations_mism}).
The slope of the correlated scatter flattens with increasing galaxy mass, which is driven by a combination of (i) increasing metallicity of accreted gas and (ii) a flattening of the slope of closed box model evolution tracks at low gas fractions.
We stress that galaxies do not evolve along any single ``fundamental" metallicity plane, but instead follow unique tracks which are shaped by the accretion, merger, and star formation history of every galaxy individually (see Figure~\ref{fig:cmb_tracks}).
There are often repeated trends which are identifiable in the individual galaxy evolution tracks including periods of evolution along closed-box-model trajectories and periods of evolution along accretion driven trajectories.

When we average over the full galaxy population (see Figure~\ref{fig:fmr_slice_evolution}), we find that coherent trends emerge among the movement of galaxy populations about the MZR.
Specifically, there is a coherent effective ``restoring force" that constantly operates to bring galaxies back toward an equilibrium metallicity and ISM mass.
We argued that -- based on inspection of a large number of individual galaxy evolution tracks -- this can be physically explained as a competition between periods of gas-rich, enrichment domination and periods of gas-poor, accretion domination.
Galaxies with high gas masses have accordingly high star formation rates, which increases the likelihood that galaxies move along closed-box-model style trajectories.
Conversely, while galaxies with low gas masses are not necessarily more likely to be accreting rapidly, their gas accretion timescales (i.e. $\tau_{\mathrm{acc}} = M_{\mathrm{ISM}} / \dot M_{\mathrm{acc}}$) are shorter than their gas rich counterparts for a fixed accretion rate.
These systems are therefore more likely to be accretion dominated.

In our models, the MZR emerges as a consequence of natural variability in the accretion and enrichment histories of galaxies.
Variability is seeded by natural fluctuations in the accretion history of each galaxy which itself is a result of the cosmic environment in which the galaxy evolves.
\citet{Forbes2014} presented a regulator-style model similar to that of~\citet{Lilly2013}, but with the relaxed assumption that $\mathrm{d}Z/\mathrm{d}t \neq 0$.  
They showed that variability in the galactic gas accretion history can drive correlated scatter about the MZR using a combination of analytic modeling and Monte-Carlo simulations of galaxy growth tracks.
The spirit of the model and results presented in \citet{Forbes2014} is in agreement with our simulations, with one subtle exception.
\citet{Forbes2014} link variability in the ISM accretion rate to variability in the halo accretion rate.
While this is likely in part true, we would argue based on inspection of our simulated galaxy sample that a significant fraction of the ISM mass accretion rate variability is driven by internal galaxy dynamics and the boundary conditions at the disk-halo interface.
This can be important because \citet{Forbes2014} required scatter in the halo accretion rates that was in tension with (smaller than) N-body simulations in order to match the correlated scatter about the MZR.
Effectively, the CGM can act as a variability damper such that the accretion onto the ISM from the CGM can contain lower amplitude and frequency variations compared to the halo accretion rates.

The IllustrisTNG model does contain one variable (the wind metal loading factor) which can impact the overall normalization of the ISM metallicity.  
However, there are not any \textit{direct} parameters that would allow us to impact or tune the presence or strength of the residual correlation between the scatter about the MZR and galactic gas fractions.
Given the clear presence and persistence of this residual correlated scatter in our models as well as the presence of a clear and simple physical picture for the driving of the correlated scatter, 
we expect that this trend should be observable.  
Based on our simulations and analysis, we expect that as methods for calculating metallicity from nebular emissions lines continue to be refined and as high redshift galaxy metallicity data is increased that evidence for the presence of correlated scatter about the MZR will solidify.
Failure to find this trend observationally will indicate a serious point of tension in the IllustrisTNG feedback model.

\section{Conclusions}
\label{sec:Conclusions}
In this paper we have analyzed the metal distribution within the TNG100 simulation, part of the IllustrisTNG simulation suite, with a focus on the properties and evolution of the mass metallicity relation.
The IllustrisTNG simulations contain a comprehensive feedback model that is aimed to regulate the stellar mass growth of galaxies, but this same feedback model has the impact of widely redistributing the metal budget into different gas phases (Figure~\ref{fig:global_phase_diagrams}).
While star formation -- and therefore metal production -- is associated with dense gas, the majority of gas-phase metals ($\sim$85 per cent at $z=0$) in our simulated galaxies are found outside of the dense ISM (Figure~\ref{fig:galaxy_phase_diagrams}).
Understanding the properties and evolution of the MZR therefore requires a combined understanding of the ISM metal retention efficiency and gas-mass evolution as a function of time and galaxy mass.

Our primary conclusions in this paper are as follows:
\begin{itemize}
\item The IllustrisTNG simulated MZR is in broad agreement with observations over the redshift range $0 < z < 2$ (Figure~\ref{fig:mz_relation}).   
We find a gradual decline in the normalization of the MZR which is consistent with observations out to $z=2$ and which we predict to continue out to $z=6$ and beyond (Figure~\ref{fig:predictions}).
\item However, there is a break in the simulated MZR for $z>0$ associated with the minimum wind velocity used in the supernova wind model.
Low galaxy mass MZR data is still somewhat sparse and so it is unclear if the break in the MZR is in tension with current observations. 
If so, it would imply that the minimum wind velocity which is needed to shape the galaxy stellar mass function in the IllustrisTNG model requires modification.
\item We showed that a majority of metals live outside of the ISM, and further that high-mass IllustrisTNG galaxies are more efficient at ejecting metals from their haloes compared to their lower mass companions.   
We calculated the redshift evolution of the metal retention efficiencies to identify the relative partitioning of metals between the ISM, CGM, and stars (Figure~\ref{fig:yield_evo}).
\item We argued that the primary driver of the MZR normalization evolution is not an evolution of metal retention efficiencies, but rather the evolving galactic gas fractions.  
The ISM metal retention efficiency \textit{increases} toward high redshift (Figure~\ref{fig:yield_evo}) which should lead to an increase in the ISM metallicity.
Instead, the decrease in the ISM metallicity toward high redshift is a result of high redshift galaxies having higher gas fractions (Figures~\ref{fig:mz_relation_ism_mass} and~\ref{fig:gas_fractions}).
The higher gas fractions result in diluted/lower metallicities, even with high ISM metal retention efficiencies.
\item There is a clear correlation between the scatter in the MZR and galactic ISM gas-mass or star formation rate (Figures~\ref{fig:mz_relation_ism_mass} and~\ref{fig:secondary_correlations_mism}).
The existence of correlated scatter about the MZR has been observationally postulated to constitute a ``fundamental metallicity relation".
Our models recover a similar residual relationship between metallicity and ISM gas-mass as is found observationally (Figure~\ref{fig:secondary_correlations_mism}).
While there remains observational uncertainty on the existence of the fundamental metallicity relation, our models clearly support the existence of such a relation.
\item Despite the existence of the fundamental metallicity relation in our simulated data, we find that galaxies do \textit{not} move along the fundamental metallicity relation (Figure~\ref{fig:cmb_tracks}).
Instead, galaxies oscillate between periods of being enrichment and accretion dominated.  
The net trend is a very clear effective ``restoring force" that drives galaxies back to equilibrium ISM masses and metallicities (Figure~\ref{fig:fmr_slice_evolution}).
\end{itemize}

Complementary constraints on our models and the results presented in this paper can be obtained through closer examination of the metal budget in the CGM.
While feedback has shaped the MZR in our simulations by expelling a significant fraction of the metal content from the star forming ISM, much of this metal content remains in the CGM.
Further comparisons on the metal budgets for non-ISM phases, including dust~\citep{McKinnon2016, McKinnon2017}, will be important for further constraining the realism or need for improvement in the feedback modeling.

 \section*{ACKNOWLEDGEMENTS}
 
PT acknowledges helpful conversations with Rob Simcoe.
PT acknowledges support from NASA through Hubble Fellowship grants HST-HF2-51384.001-A awarded by the STScI, which is operated by the Association of Universities for Research in Astronomy, Inc., for NASA, under contract NAS5-26555. 
RW, VS and RP acknowledge support through the European Research Council under ERCStG grant EXAGAL-308037, and would like to thank the Klaus Tschira Foundation. 
RW acknowledges support by the IMPRS for Astronomy and Cosmic Physics at the University of Heidelberg. 
VS acknowledges support through subproject EXAMAG of the Priority Programme 1648 ``Software for Exascale Computing" of the German Science Foundation. 
MV acknowledges support through an MIT RSC award, the support of the Alfred P. Sloan Foundation, and support by NASA ATP grant NNX17AG29G. 
JPN acknowledges support of NSF AARF award AST-1402480. 
The Flatiron Institute is supported by the Simons Foundation. 
The flagship simulations of the IllustrisTNG project used in this work have been run on the HazelHen Cray XC40-system at the High Performance Computing Center Stuttgart as part of project GCS-ILLU of the Gauss Centre for Supercomputing (GCS). 
Ancillary and test runs of the project were also run on the Stampede supercomputer at TACC/XSEDE (allocation AST140063), at the Hydra and Draco supercomputers at the Max Planck Computing and Data Facility, and on the MIT/Harvard computing facilities supported by FAS and MIT MKI.


\end{document}